\documentclass[10pt, fleqn, A4, twoside]{article}

\usepackage{amsmath} 
\usepackage{amssymb} 
\usepackage{bm}
\usepackage{graphicx} 
\usepackage{mathtools}
\usepackage{amsthm}
\usepackage{appendix}
\numberwithin{equation}{section}
\usepackage{fullpage}
\usepackage[margin=.8in]{geometry}
\usepackage[]{hyperref}
\usepackage{subcaption}
\usepackage{tikz}
\usepackage{pgfplots}\pgfplotsset{compat=newest}
\usepgfplotslibrary{groupplots}
\usetikzlibrary{matrix,positioning,calc,patterns,angles,quotes,arrows,external}
\usepackage{comment}

\graphicspath{ {PdfImages/} }

\newcommand{\Ha}{\mathcal H} 
\newcommand{\GravC}{\mathcal G} 
\newcommand{\D}[1]{\ensuremath{\operatorname{d}\!{#1}}} 

\newcommand{\E}{E} 
\newcommand{\M}{\ell} 
\newcommand{\N}{n} 
\newcommand{\VARPI}{\varpi} 
\newcommand{\LAMBDA}{\lambda} 
\newcommand{\GAMMA}{\gamma}

\newcommand{\LAMBDONA}{\Lambda}
\newcommand{\GAMMONA}{\Gamma}

\newcommand{\KAPPONA}{\mathcal K}
\newcommand{\KAPPA}{\kappa}
\newcommand{\THETA}{\theta}
\newcommand{\THETAONA}{\Theta}

\newcommand{\PSI}{\psi}
\newcommand{\PSIONA}{\Psi}
\newcommand{\DELTAGAMMA}{\delta\GAMMA}
\newcommand{\THETAP}{\theta'}
\newcommand{\OMEGONA}{\Omega}
\newcommand{\ANGMOM}{\mathcal L}

\newcommand{\Z}{\mathbb Z} 

\newcommand{\bfx}{\mathbf u} 
\newcommand{\dbfx}{\dot\bfx} 

\newcommand{\bfr}{\mathbf r} 

\newcommand{\bfp}{\mathbf p} 
\newcommand{\bfv}{\mathbf v} 

\begin{document}
\font\courier=pcrr at 12pt
\font\codefont=pcrr

\title{Capture into first-order resonances and long-term stability of pairs of equal-mass planets}
\author{
Gabriele Pichierri$^1$, Alessandro Morbidelli$^1$, Aur\'elien Crida$^{1,2}$
\\
{\small ${}^1$CNRS, Laboratoire Lagrange, Universit\'e C\^ote d'Azur, Observatoire de la  C\^ote d'Azur, Nice, France}\\
{\small ${}^{2}$Institut Universitaire de France, 103 Boulevard Saint-Michel, F-75005 
Paris, France}}
\date{\today}
\maketitle

\begin{abstract}
Massive planets form within the lifetime of protoplanetary disks and 
therefore they are subject to orbital migration due to planet-disk 
interactions. When the first planet reaches the inner edge of the disk 
its migration stops and consequently the second planet ends up locked in 
resonance with the first one. We detail how the resonant trapping works 
comparing semi-analytical formulae and numerical simulations. We 
restrict to the case of two equal-mass coplanar planets trapped in first 
order resonances but the method can be easily generalised. We first 
describe the family of resonant stable equilibrium points 
(zero-amplitude libration orbits) using series expansions up to 
different orders in eccentricity as well as a non-expanded Hamiltonian. 
Then we show that during convergent migration the planets evolve 
along these families of equilibrium points. Eccentricity 
damping from the disk leads to a final equilibrium configuration that we 
predict precisely analytically. The fact that observed 
multi-exoplanetary systems are rarely seen in resonances suggests that 
in most cases the resonant configurations achieved by migration become 
unstable after the removal of the protoplanetary disk. Here we probe the 
stability of the resonances as a function of planetary mass. For this 
purpose, we fictitiously increase the masses of resonant planets, 
adiabatically maintaining the low-amplitude libration regime until 
instability occurs.
We discuss two hypotheses for the instability, that of a low-order 
secondary resonance of the libration frequency with a fast synodic 
frequency of the system, and that of minimal approach distance between 
planets.
We show that secondary resonances do not seem to impact resonant systems 
at low-amplitude of libration.
Resonant systems are more stable than non-resonant ones for a given 
minimal distance at close encounters, but we show that the latter nevertheless 
play the decisive role in the destabilisation of resonant pairs.
We show evidence that, as the planetary mass increases and the minimal 
distance between planets gets smaller in terms of mutual Hill radius, 
the region of stability around the resonance center shrinks, until the 
equilibrium point itself becomes unstable.
\end{abstract}

\section{Introduction}

Super-Earths (SE) are planets with a mass between 1 and $\sim$20 Earth masses 
or a radius between 1 and $\sim$4 Earth radii, and so-far discovered 
with orbital period typically shorter than $\sim 100$ days. They are 
estimated to orbit 30 -- 50\% of Sun-like stars 
(\cite{2011arXiv1109.2497M}; \cite{2012ApJS..201...15H}; 
\cite{2013ApJ...766...81F}; \cite{2013PNAS..11019273P}) and 
multi-planetary systems are not rare.
The fact that close-in SE systems form so frequently around stars, but 
not always (for instance not in the Solar System) is an interesting 
constraint on planetary formation models.

It is generally expected that SEs form (mostly) within the lifetime of 
the protoplanetary disk of gas and therefore, regardless of whether they 
form in the inner or outer part of the disk, they should undergo radial 
migration towards the central star, as a result of planet-disk 
interactions (\cite{2015A&A...578A..36O}; \cite{2017MNRAS.470.1750I}). 
Migration brings the SEs to the inner edge of the disk, where inward 
migration stops \cite{2006ApJ...642..478M}. For this reason, the SEs are 
captured into mutual mean motion resonances, where the ratios of orbital 
periods are equal to the ratios of integer numbers. This is observed in 
all simulations (e.g.{ \cite{2007ApJ...654.1110T}; 
\cite{2008A&A...482..677C}; \cite{2008A&A...478..929M}, and the 
aforementioned \cite{2015A&A...578A..36O} and \cite{2017MNRAS.470.1750I}).

Due to this renewed interest in resonant captures, in the first part of 
this paper we revisit the problem of capture in first order resonances 
of two equal-mass coplanar planets in convergent migration, using a 
semi-analytical approach and numerical simulations.  In Section 
\ref{sec:AnalyticalSetup} we compute analytically the locus of 
equilibrium points of first-order resonances, where both the resonant 
and secular oscillations of the planetary orbits have a null amplitude. 
Our calculations are developed for unexpanded Hamiltonians, which allows 
to follow the dynamics up to arbitrarily large eccentricities (e.g. 
\cite{2006MNRAS.365.1160B}, \cite{2006CeMDA..94..411M}). We compare the 
results with those obtained with first and second order expansions of 
the Hamiltonian in the eccentricity, showing qualitative and 
quantitative disagreements. The quantitative accuracy of our results is 
validated with simulations in which planets are forced to migrate 
towards each-other, without any eccentricity damping. These simulations 
have to follow the loci of the equilibrium points, and show perfect 
agreement with the unexpanded model. Moreover, we calculate the two 
frequencies of libration around the equilibrium points, therefore 
obtaining a complete understanding of the system; we again check the 
validity of the analytical calculations against numerical simulations in 
which the amplitudes of resonant and secular librations are slightly 
excited and the frequencies of oscillation of the semi major axis and 
the eccentricity are measured. In Section 
\ref{sec:ConvergentInwardMigration} we introduce the eccentricity 
damping exerted by the disk onto the planets. This leads to a final 
equilibrium configuration where convergent migration stops. The analytic 
calculation of the equilibrium eccentricities and semi major axes ratio 
is presented in the Appendix. We check against numerical simulations the 
validity of these analytical predictions, showing excellent agreement.

Despite resonant capture is typical of migration simulations,
the observed SEs systems show little preference for near-integer period 
ratios and their orbital separations are usually much wider than  those 
characterising planets in resonant chains. \cite{2017MNRAS.470.1750I} 
showed that this observation is not inconsistent with the 
migration/resonant trapping paradigm. In fact, simulations show that, 
after the removal of the disk of gas, the resonant planetary systems 
often become unstable. \cite{2017MNRAS.470.1750I} showed that the 
observations are very well reproduced if the fraction of the resonant 
systems that eventually become unstable exceeds 90\%.  The reasons for 
these instabilities, however, are unexplained.

\cite{2012Icar..221..624M} studied numerically the stability of resonant 
multi-planetary systems for high-integer first-order mean motion 
resonances. They built the desired resonant configuration by  simulating 
the Type-I migration phase in a protoplanetary disk of gas; then they 
slowly depleted the disk. They observed that there is a critical number 
of planets above which the resonant systems go naturally unstable, with 
a crossing time comparable to that of non-resonant systems, and studied 
how this number changes with the planetary masses and resonant order. In 
other words, they demonstrated that, given the planetary masses, there 
is a limit number of planets that can form a stable resonant chain or, 
given the number of planets, there is a limit mass for stability. The 
reason of the instability, however, was not discussed.

Thus, in the second part of this paper we address why resonant planets 
become unstable if they are too massive. We focus here on a system of 
two coplanar planets and study the stability of the resonant center as a 
function of the planets' masses (assumed to be equal for simplicity). In 
a subsequent work, we will generalise this study to more populated 
resonant chains.

Again, we follow a double approach: analytic and numeric. In 
Section \ref{sec:MassIncrease}  we start from a pair of small-mass planets 
deep in resonance and we slowly increase their masses. The mass growth 
preserves, by the adiabatic principle, the original small libration 
amplitude. In this way we can explore the stability of the resonance 
center up the threshold mass for instability. At the same time, we 
detail how one can follow analytically the evolution of the system to a 
good approximation up to high value of the planetary masses. To 
understand why the planets ultimately become unstable, we compare the 
numerical evolution of the system with an analytically computed map of 
secondary resonances (resonances between the libration frequencies or 
between a libration frequency and one of the short-periodic harmonic) as 
well as a map of minimum approach distance between the planets, finding 
that one of them matches well the instability limit observed in the 
numerical simulations.
We summarise our results in the final Section \ref{sec:Conclusions}.

\section{Planetary Hamiltonian}\label{sec:AnalyticalSetup}
We start by considering the Hamiltonian for the planar three-body problem, in canonical Poincar\'e coordinates, 
\cite{1892mnmc.book.....P},
$\bfp_i$, $\bfr_i$, $i=1,2$:
\begin{align}\label{eq:FullHamiltonian}
\Ha 		&= \Ha_{kepl} + \Ha_{pert},\nonumber\\
\Ha_{kepl} 	&= 	\frac{M_* + m_1}{2 M_*}\frac{\bfp_1^2}{m_1} -\frac{\GravC M_* m_1}{\|\bfr_1\|} + 
				\frac{M_* + m_2}{2 M_*}\frac{\bfp_2^2}{m_2} -\frac{\GravC M_* m_2}{\|\bfr_2\|}, \\
\Ha_{pert}	&= \frac{\bfp_1 \cdotp \bfp_2}{M_*} - \GravC \frac{m_1 m_2}{\Delta},\nonumber
\end{align}
where $M_*$ is the mass of the central star, $m_1$ and $m_2$ are the masses of the two planets, $\GravC$ is the 
gravitational constant and $\Delta=\|\bfr_1 - \bfr_2\|$ is the distance between the two planets.
Recall that, with respect to the positions and velocities $(\bfx_i, \dbfx_i)$ 
in a barycentric inertial reference frame, the canonical Poincar\'e coordinates
are given by 
$\mathbf r_0=\bfx_0$, $\mathbf r_i=\bfx_i - \bfx_0$, $i=1,2$ for the positions, and 
$\mathbf p_0 = \tilde\bfx_0 + \tilde\bfx_1 + \tilde\bfx_2$, 
$\mathbf p_i=\tilde\bfx_i$, $i=1,2$ for their conjugated momenta, where 
$\tilde\bfx_0 = M_* \dbfx_0$, $\tilde\bfx_i = m_i \dbfx_i$, $i=1,2$ are the linear (barycentric) momenta of the 
bodies.
In Cartesian coordinates, for a given planet, the heliocentric positions $\bfr = (x,y)$ 
and barycentric velocities $\bfv = (v_x, v_y)$ 
are related to the orbital elements by the usual formal relationships
\begin{equation}\label{eq:CartesianComponents}
\begin{split}
x   &= a (\cos{\E} - e) \cos{\VARPI} - 
     a \sqrt{1 - e^2} \sin{\E} \sin{\VARPI}, \\
y   &= a (\cos{\E} - e) \sin{\VARPI} + 
      a \sqrt{1 - e^2} \sin{\E} \cos{\VARPI},\\
v_x &= \left(-a \sin\E \cos\VARPI - 
    a \sqrt{1 - e^2} \cos\E \sin\VARPI\right) \frac{n}{1 - e \cos\E},\\
v_y &= \left(-a \sin\E \sin\VARPI + 
    a \sqrt{1 - e^2} \cos\E \cos\VARPI\right) \frac{n}{1 - e \cos\E}, 
\end{split}
\end{equation} 
where $a$ is the semi-major axis, $e$ is the eccentricity, $\E$ is the eccentric anomaly, $\VARPI$ is 
the longitude of the pericentre and $n = \sqrt{\GravC (M_* + m)/a^{3}}$ is the mean motion.
Note that the orbital elements defined in this way are different from those usually defined by astronomers,
which are built from the same relationships but using the heliocentric velocities.
For simplicity we restrict to coplanar motions for the planets, so there is no $z$ component, 
no inclination and no ascending node.
All quantities relative to the inner and outer planet will be denoted with subscripts 1 and 2 respectively.
In order to make use of the orbital elements defined from canonical Poincar\'e variables and at the same time  maintain the canonical nature of the system, 
we introduce the modified Delaunay action-angle variables 
$(\LAMBDONA, \GAMMONA, \LAMBDA, \GAMMA)$ given by
\begin{alignat}{2}\label{eq:ModifDelaunayVariables}
\LAMBDONA         &=\mu\sqrt{\GravC (M_*+m) a},                                  &&\quad \LAMBDA = \M + \VARPI, \nonumber\\
\GAMMONA          &=\LAMBDONA(1-\sqrt{1-e^2})\simeq\LAMBDONA e^2/2,       &&\quad \GAMMA = -\VARPI,
\end{alignat}
where $\mu = \frac{M_* m}{M_*+m}$ is the reduced mass, $\LAMBDA$ is the mean longitude and $\M = \E - e\sin\E$ 
is the mean anomaly.
In these variables, the Keplerian part $\Ha_{kepl}$ of the Hamiltonian \eqref{eq:FullHamiltonian} takes the form
\begin{equation}\label{eq:KeplerianPartInModifDelaunayVariables}
\Ha_{kepl} = -\frac{\GravC^2 (M_* + m_1)^2 \mu_1^3}{2\LAMBDONA_1^2} 
             -\frac{\GravC^2 (M_* + m_2)^2 \mu_2^3}{2\LAMBDONA_2^2}. 
\end{equation}

We impose a first order mean motion resonance between the two planets, that is we assume that 
the two mean motions $\N_1=\sqrt{\GravC (M_*+m_1)/a_1^3}$ and $\N_2=\sqrt{\GravC (M_*+m_2)/a_2^3}$ 
satisfy the resonance condition $k \N_2 - (k-1) \N_1 \sim 0$, where $k\in\Z$ is a positive integer, $k\geq2$. 
We now average the Hamiltonian over the fast angles. 
In fact, since the Keplerian part $\Ha_{kepl}$ does not depend on the angles, 
only the perturbation Hamiltonian $\Ha_{pert}$ needs averaging. 
We note that we need to integrate $\Ha_{pert}$ e.g.\ with respect to the angle $\LAMBDA_1$ 
over the interval $[0,2 k \pi]$, corresponding to $k$ revolutions of the inner planet around the star 
(which in turn by the resonance condition is equivalent to $(k-1)$ revolutions of the outer planet), 
in order to fully recover the periodicity of the Hamiltonian. 
This leads to a new averaged perturbing Hamiltonian which we denote with $\Ha_{res}$:
\begin{equation}\label{eq:FullAveragedResonantHamiltonian}
\Ha_{res} := \bar{\Ha}_{pert} = \frac{1}{2k\pi}\int_0^{2k\pi} \Ha_{pert} \D\LAMBDA_1;
\end{equation}
the full averaged Hamiltonian is therefore
\begin{equation}\label{eq:FullAveragedHamiltonian}
\bar{\Ha} = \Ha_{kepl} + \Ha_{res}.
\end{equation}
From an analytical perspective, we remark that only certain combinations of the 
angles will appear in the Fourier expansion of the averaged Hamiltonian $\bar \Ha$. 
Indeed, by the d'Alembert rules, after the averaging procedure, of all angles depending explicitly on 
$\LAMBDA_1$ and $\LAMBDA_2$, only those of the form
\begin{equation}\label{eq:GeneralResonantAngle}
j \big(k \LAMBDA_2 - (k-1)\LAMBDA_1\big) + j_1\GAMMA_1 + j_2\GAMMA_2, \quad 
\text{$j, j_1, j_2\in\Z^{+}$, $j_1+j_2=j$},
\end{equation}
will survive.
With this in mind, in order to simplify the expression of the resonant harmonics appearing in the 
Hamiltonian $\Ha_{res}$ one can introduce the following canonical action-angle variables 
(\cite{1984CeMec..32..307S}):
\begin{alignat}{2}\label{eq:ResHarmonicsCoV}
\KAPPONA	&=\LAMBDONA_1 + \frac{k-1}{k}\LAMBDONA_2,	&&\quad \KAPPA = \LAMBDA_1, \nonumber\\
\THETAONA	&=\LAMBDONA_2/k,	       					&&\quad \THETA = k\LAMBDA_2-(k-1)\LAMBDA_1.
\end{alignat}
The newly defined angle $\KAPPA$ does not appear explicitly in the Hamiltonian, making its conjugated action 
$\KAPPONA$ a constant of motion. 
The significance of the conservation of $\KAPPONA$ is already explained in \cite{2013A&A...556A..28B}; in 
particular it yields the location of exact Keplerian resonance from the observed values of semi-major axes. 
As we will see, especially at low eccentricities the semi-major axes of the two planets deviate away from 
the nominal commensurability, by an amount which also depends on the planetary masses. 
Therefore the observed values of $a_1$ and $a_2$ do not alone reveal how close the planets are to resonance, nor 
they represent the nominal values $\bar{a}_1$ and $\bar{a}_2$ of the semi-major axes that satisfy the exact Keplerian 
resonant relationship $\bar{a}_1/\bar{a}_2=((k-1)/k)^{2/3}$. 
However by calculating from their observed values the value of the constant $\KAPPONA$, and imposing in the formula
\begin{equation}
\frac{\KAPPONA}{\LAMBDONA_2} = \frac{\mu_1}{\mu_2} \sqrt{\frac{(M_*+m_1)}{(M_*+m_2)}\frac{a_1}{a_2}} + \frac{k-1}{k},
\end{equation} 
the condition of exact resonance, $\alpha=a_1/a_2=((k-1)/k)^{2/3}$, one can obtain $\bar{a}_2$ from
$\bar{a}_2 = (\bar\LAMBDONA_2/\mu_2)^2 / (\GravC (M_* + m_2))$
and $\bar{a}_1$ from $\bar{a}_1=((k-1)/k)^{2/3} \bar{a}_2$.

Considering now the remaining three pairs of canonical action-angle variables, 
a final canonical transformation is made: 
\begin{alignat}{2}\label{eq:FinalCoV}
\PSIONA_1	&=\GAMMONA_1+\GAMMONA_2,			&&\quad \PSI_1 = \THETA+\GAMMA_1, \nonumber\\
\PSIONA_2	&=-\GAMMONA_2,       				&&\quad \DELTAGAMMA = \GAMMA_1-\GAMMA_2,\\
\OMEGONA	&=\THETAONA-\GAMMONA_1-\GAMMONA_2,	&&\quad \THETAP = \THETA; \nonumber
\end{alignat}
Using again \eqref{eq:GeneralResonantAngle}, it is trivial to see that in the Hamiltonian $\bar\Ha$ only angles 
of the form 
\begin{equation}\label{eq:GeneralResonantAngle2}
j\PSI_1 + j_2\DELTAGAMMA, \quad j,j_2\in\Z^{+}
\end{equation}
will appear, i.e.\ angles in which $\THETAP$ does not enter explicitly, 
making $\OMEGONA$ our second constant of motion and thus reducing to two the degrees of freedom of our system. Note that the two constants of motion $\OMEGONA$ and $\KAPPONA$ are linked to the total angular momentum $\ANGMOM$, which in these mixed variables (orbital elements derived from heliocentric positions and barycentric velocities) is given by
\begin{equation}\label{eq:AngMomIn}
\ANGMOM = m_1\sqrt{\GravC (M_*+m_1)a_1(1-e_1^2)} + m_2\sqrt{\GravC (M_*+m_2)a_2(1-e_2^2)};
\end{equation}
to first order in the masses, we have $\KAPPONA+\OMEGONA=\ANGMOM$.

\subsection{First and higher order expansions of the Hamiltonian}
An analytical treatment of first order resonances making use of an expansion of the Hamiltonian up 
to first order in the eccentricities was presented in \cite{2013A&A...556A..28B}, 
yielding a qualitative description of the resonant dynamical evolution of two planets.
Following this approach, the resonant Hamiltonian $\Ha_{res}$ in variables \eqref{eq:ModifDelaunayVariables} 
takes the form
\begin{equation}\label{eq:FirstOrderExpansion}
\begin{split}
\Ha_{res} 	&= -\frac{\GravC^2 M_* m_1 m_2^3}{\LAMBDONA_2^2}\left(f_{res}^{(1)}\sqrt{\frac{2\GAMMONA_1}{\LAMBDONA_1}} 
				\cos\big(k\LAMBDA_2-(k-1)\LAMBDA_1 + \GAMMA_1\big)\right.\\
			&\quad \left.+ f_{res}^{(2)}\sqrt{\frac{2\GAMMONA_2}{\LAMBDONA_2}} 
				\cos\big(k\LAMBDA_2-(k-1)\LAMBDA_1 + \GAMMA_2\big)\right),
\end{split}
\end{equation}
where the coefficients $f_{res}^{(1)}$ and $f_{res}^{(2)}$ depend (weakly) on the semi-major axis ratio.
Note that, since this is an expansion up to first order in $e$ and the two terms in parenthesis are already of order 
$\sqrt{\GAMMONA}=\mathcal{O}(e)$, we can evaluate $\LAMBDONA$ on the nominal values of the semi-major axis, thus 
fixing them to $\bar\LAMBDONA_1$ and $\bar\LAMBDONA_2$. 
By doing so, the coefficients $f_{res}$ can be truly considered constant; one may find in 
\cite{2000ssd..book.....M} formul\ae\ to obtain their numerical value in the case of different resonances. 
After the change of variable \eqref{eq:ResHarmonicsCoV} the resonant Hamiltonian $\Ha_{res}$ takes the simple form
\begin{equation}
\Ha_{res}=-\frac{\GravC^2 M_* m_1 m_2^3}{\bar\LAMBDONA_2^2}\left(\alpha_1 \sqrt{2\GAMMONA_1}\cos(\THETA+\GAMMA_1) + 
			\alpha_2 \sqrt{2\GAMMONA_2}\cos(\THETA+\GAMMA_2)\right),
\end{equation}
where $\alpha_i = f_{res}^{(i)}/\sqrt{\bar\LAMBDONA_i}$, $i=1,2$. 
The full Hamiltonian still of course retains the form $\bar\Ha = \Ha_{kepl} + \Ha_{res}$ as in \eqref{eq:FullAveragedHamiltonian}.
While this Hamiltonian contains at the moment two harmonics, it is actually integrable, 
since it is possible to carry out a series of canonical changes of variables, following e.g.\ the approach in 
\cite{1984CeMec..32..307S}, which makes it dependent on only one harmonic and extracts another integral of 
motion.
This advantageous reduction can be used to obtain a general description of the dynamics 
(e.g.\ \cite{2013A&A...556A..28B}, \cite{2017A&A...602A.101R}). 
However it is insufficient when one confronts even qualitatively the prediction of this theoretical model with results 
from numerical simulations, as we will see in the next Section, where we compute the locus of equilibrium points 
(i.e.\ periodic orbits of the full problem) as a function of the system's angular momentum.

Higher order expansions are possible. 
However the Hamiltonian can no longer be reduced to one depending on a
single combination of angles, i.e.\ it will not be integrable. 
Moreover, while they represent a more faithful representation of the real dynamics, 
it is still not adequate enough for good quantitative accord with the results of numerical simulations, 
as we will see in the next section.
Therefore we develop below the formalism for un-expanded Hamiltonians, using a semi-analytical approach 
(i.e.\ computing the integral \eqref{eq:FullAveragedResonantHamiltonian} numerically), already employed e.g.\ in 
\cite{1993CeMDA..57...99M,1995Icar..114...33M}, \cite{2006CosRe..44..440S}, \cite{2017A&A...605A..23P}
for the restricted problem and in
\cite{2006MNRAS.365.1160B}, \cite{2006CeMDA..94..411M} for the full three-body problem. 

\subsection{Equilibrium points of the averaged Hamiltonian}\label{subsec:EquilibriumPointsOfAveragedHa}
We now consider the averaged Hamiltonian $\bar\Ha(\PSIONA_1,\PSIONA_2,\PSI_1,\DELTAGAMMA;\OMEGONA)$ 
as a system with two degrees of freedom with parametric dependence on the value of $\OMEGONA$, the 
action defined in \eqref{eq:FinalCoV} 
(note that the symbol $\OMEGONA$ usually denotes the longitude of the node, 
which is not defined in this case given the planar nature of the problem),
and look for its equilibrium points. 
The Hamiltonian also parametrically depends on $\KAPPONA$, but as we have seen this variable encodes the location 
of exact resonance, for which $a_2/a_1 = \bar{a}_2/\bar{a}_1 = (k/(k-1))^{2/3} =: \bar R$, as well as the value of 
the planetary masses. 
Once we have fixed $m_1$, $m_2$ and $k$, we can choose units in which $\bar{a}_2 = 1$, so that $\KAPPONA$ obtains 
a natural value relative to the problem at hand.

Equilibrium points correspond to stationary solutions, and are therefore found by solving simultaneously 
in the variables $(\PSIONA_1,\PSIONA_2,\PSI_1,\DELTAGAMMA) = \mathbf x$ the set of equations
\begin{equation}\label{eq:EquilibriumPointCondition}
\frac{\partial\Ha}{\partial\PSIONA_1}=0,\quad
\frac{\partial\Ha}{\partial\PSIONA_2}=0,\quad
\frac{\partial\Ha}{\partial\PSI_1}=0,\quad
\frac{\partial\Ha}{\partial\DELTAGAMMA}=0,
\end{equation}
for different values of the constant of motion $\OMEGONA$.
Because of the analytical properties of the Hamiltonian $\bar\Ha$, namely the fact that it contains only cosines
of angles of the form \eqref{eq:GeneralResonantAngle2}, any combination of equilibrium values $\PSI_{1,eq} = 0,\pi$ and 
$\DELTAGAMMA_{eq} = 0,\pi$ will satisfy the last two equations in \eqref{eq:EquilibriumPointCondition}.
Taking any of these possible combination, we solve the first two equations in 
\eqref{eq:EquilibriumPointCondition} for $\PSIONA_1$ and $\PSIONA_2$, and we find two values 
$(\PSIONA_{1,eq},\PSIONA_{2,eq})$.
We then have to check that the point $\mathbf x_{eq} = (\PSIONA_{1,eq},\PSIONA_{2,eq},\PSI_{1,eq},\DELTAGAMMA_{eq})$ 
is a \emph{stable} equilibrium point for the Hamiltonian $\bar\Ha$.
In principle, the last two equations in \eqref{eq:EquilibriumPointCondition} could be satisfied for a combination of 
values of $\PSI_1$ and $\DELTAGAMMA$ different from $0, \pi$ (asymmetric equilibria), but this is the case only if 
all symmetric equilibria are unstable. 
This is because in the adiabatic limit in which one takes the second (slower) degree of freedom $(\PSIONA_2,\DELTAGAMMA)$ as fixed, the Hamiltonian can be considered as describing an integrable one degree of freedom system in the pair of (faster) variables $(\PSIONA_1,\PSI_1)$, with slowly varying parameters corresponding to the slow degree of freedom. It is well known that, for a one-degree of freedom system and at the relatively low eccentricities that are obtained in the process of capturing into resonance, asymmetric equilibria are possible only if a bifurcation occurs which changes the nature of the symmetric equilibria (which always exist) from stable to unstable.
Thus, if one finds a stable symmetric equilibrium the search for asymmetric stable equilibria can be avoided.
The condition for stability of the equilibria is discussed in the next Section and is the usual criterion whereby one imposes that the eigenvalues of the matrix which describes the linear approximation of the system around the equilibrium be purely imaginary.

By changing the value of the constant $\OMEGONA$ we obtain different equilibrium configurations,
and once an equilibrium point in the canonical variables $(\PSIONA_1,\PSIONA_2,\PSI_1,\DELTAGAMMA)$ is obtained, 
we can easily work our way back through the canonical transformation and obtain the equilibrium values for 
the semi-major axes and eccentricities of the two planets, which we denote with $a_{1,eq}$, $a_{2,eq}$, 
$e_{1,eq}$, $e_{2,eq}$.
This results in the stable equilibrium curves shown in Figures \ref{fig:EqCurvesForAllResonances-WithExpansions}, 
\ref{fig:EqCurvesForAllResonances-DifferentMasses}, which are found for $\PSI_{1,eq} = 0$ and $\DELTAGAMMA_{eq} = \pi$.

\begin{figure}[!ht]
\centering
\begin{subfigure}[b]{0.32 \textwidth}
\centering
\includegraphics[scale=0.43
]{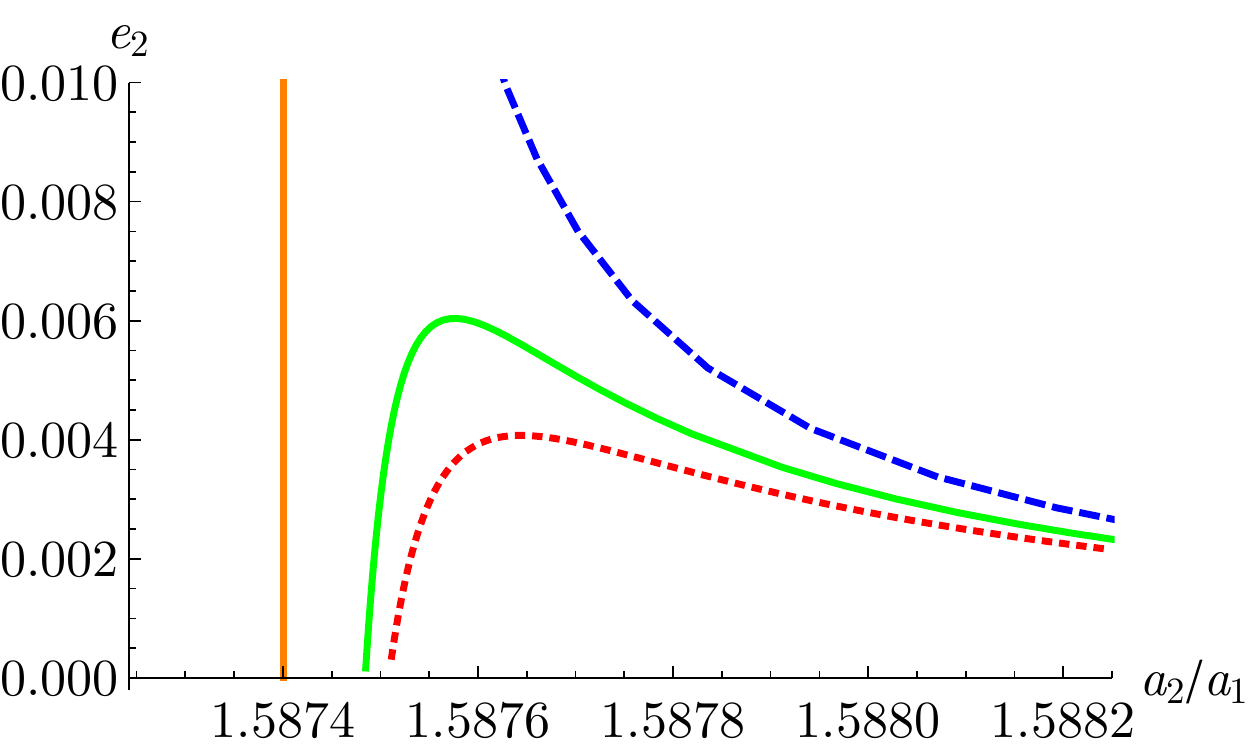}
\caption{2-1 mean motion resonance.}
\label{fig:EqCurvesForAllResonances-WithExpansions.subfig:2-1}
\end{subfigure}
\begin{subfigure}[b]{0.32 \textwidth}
\centering
\includegraphics[scale=0.43
]{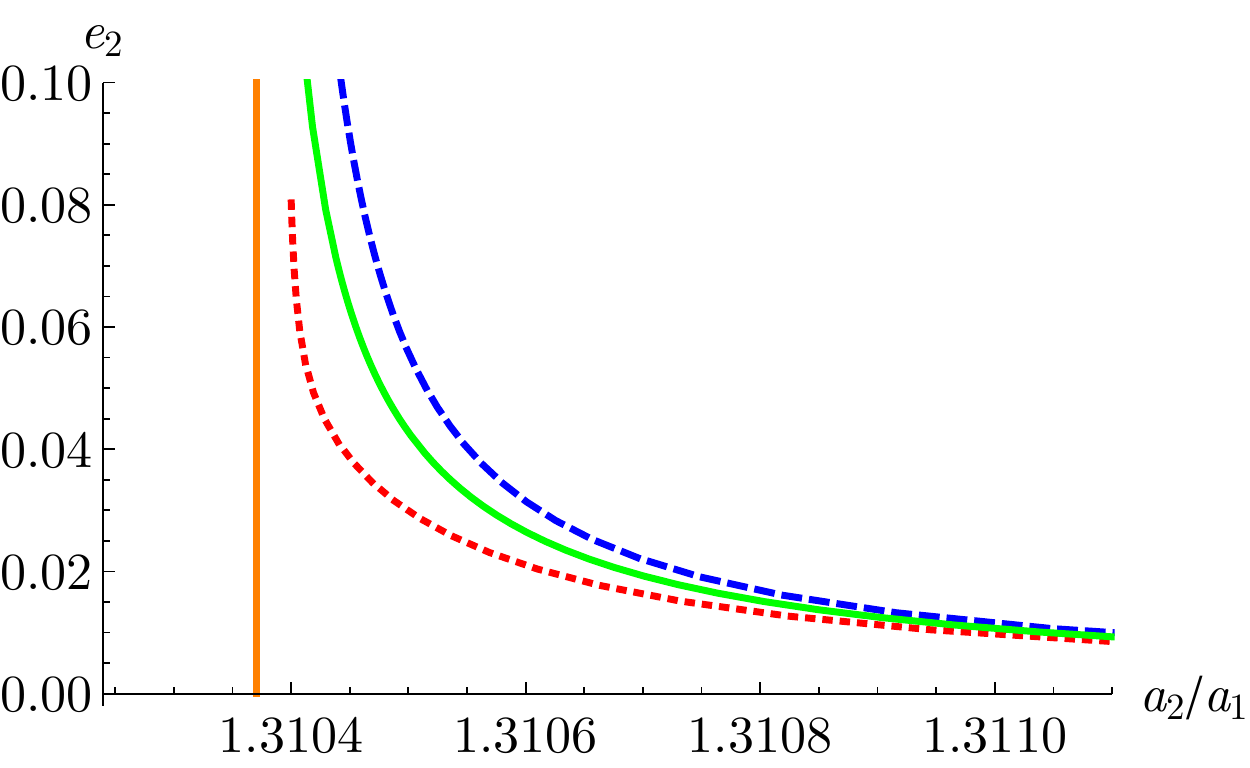}
\caption{3-2 mean motion resonance.}
\label{fig:EqCurvesForAllResonances-WithExpansions.subfig:3-2}
\end{subfigure}
\begin{subfigure}[b]{0.32
 \textwidth}
\centering
\includegraphics[scale=0.43
]{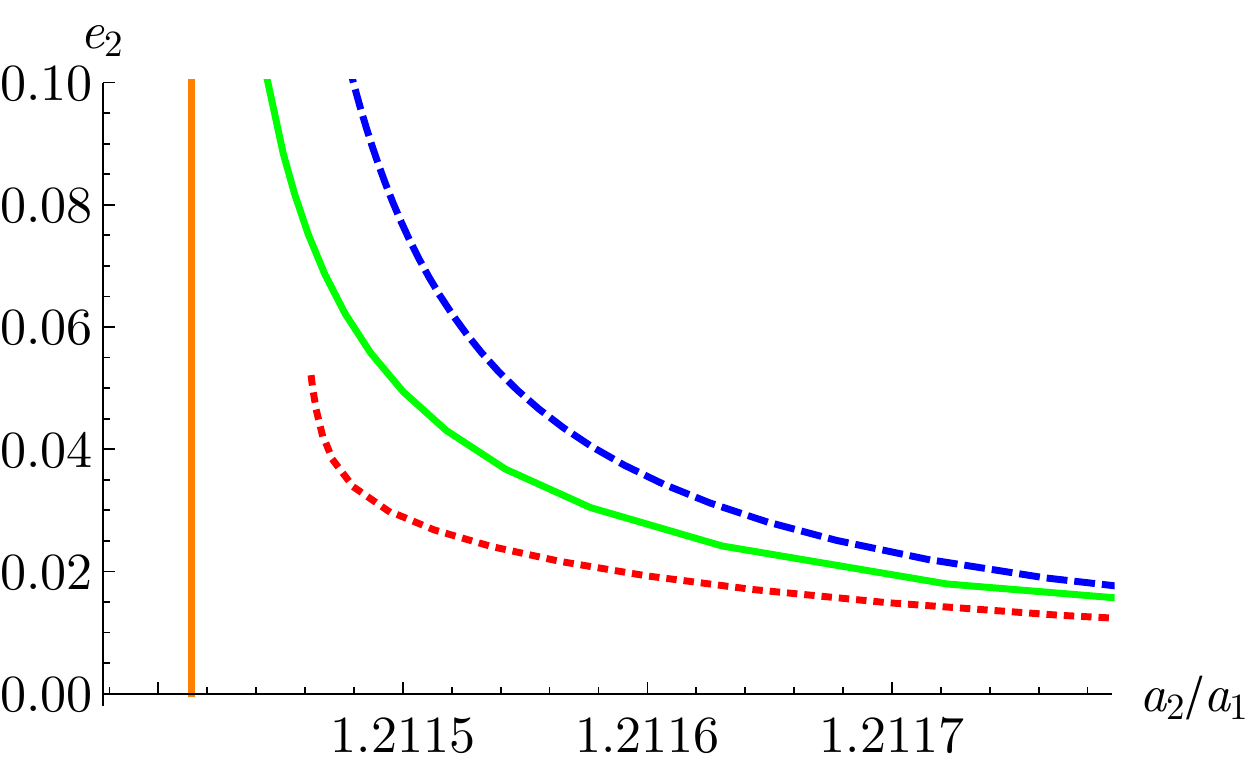}
\caption{4-3 mean motion resonance.}
\label{fig:EqCurvesForAllResonances-WithExpansions.subfig:4-3}
\end{subfigure}
\caption{
Equilibrium curves for three different first-order mean motion resonance, calculated as described in 
the text using the first-order expansion \eqref{eq:FirstOrderExpansion} (dashed blue line), 
a second-order expansion (dotted red line) 
and the full averaged Hamiltonian \eqref{eq:FullAveragedResonantHamiltonian} (continuous green line).
Here we put $m_1 = m_2 = m = 10^{-5} M_*$.
The equilibrium values for the angles are $\PSI_{1,eq} = 0$ and $\DELTAGAMMA_{eq} = \pi$.
The orange vertical line indicates the location of exact Keplerian resonance, 
$a_2/a_1 = \bar{a}_2/\bar{a}_1 = (k/(k-1))^{2/3}$.
Note the discrepancy between the equilibrium curves with and without the expansion of the resonant Hamiltonian, 
due to the presence of higher order harmonics which are not taken into account in the expanded Hamiltonians.
}
\label{fig:EqCurvesForAllResonances-WithExpansions}
\end{figure}

\begin{figure}[!ht]
\centering
\begin{subfigure}[b]{0.32 \textwidth}
\centering
\includegraphics[scale=0.43
]{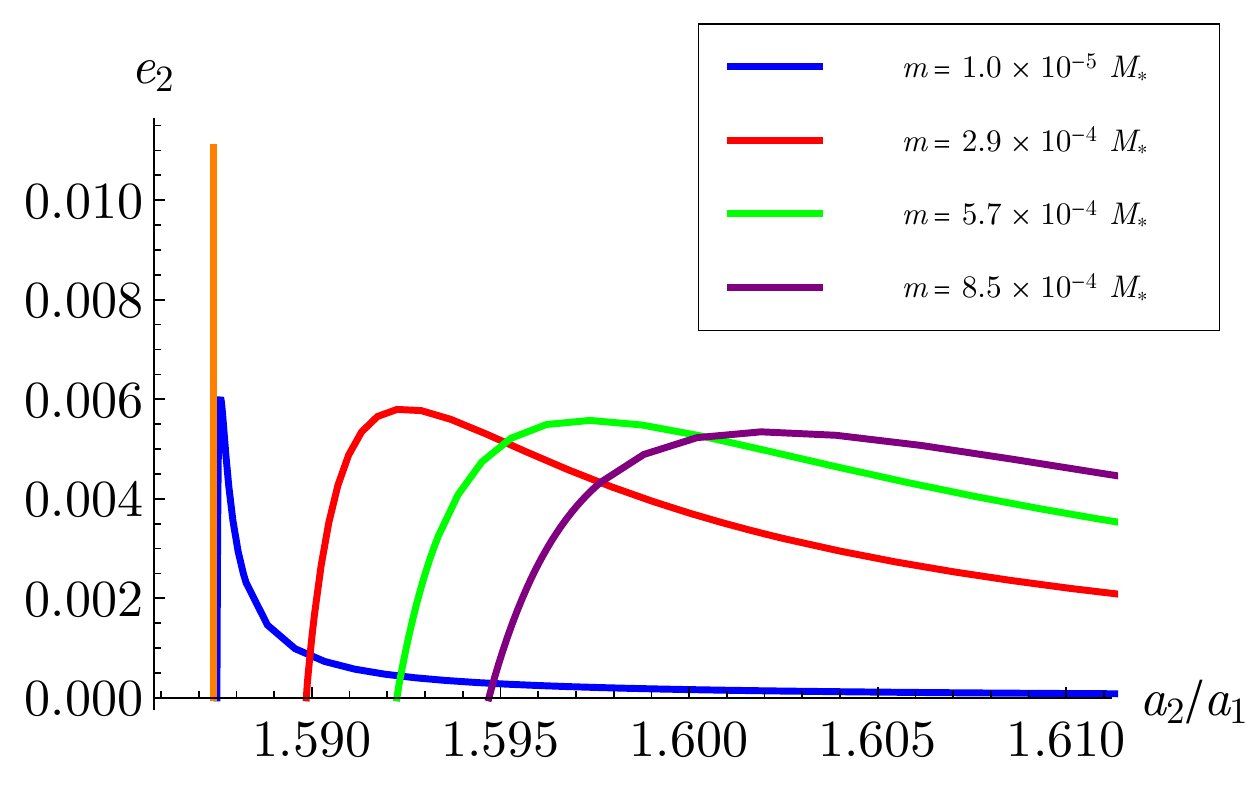}
\caption{2-1 mean motion resonance.}
\label{fig:EqCurvesForAllResonances-DifferentMasses.subfig:2-1}
\end{subfigure}
\begin{subfigure}[b]{0.32 \textwidth}
\centering
\includegraphics[scale=0.43
]{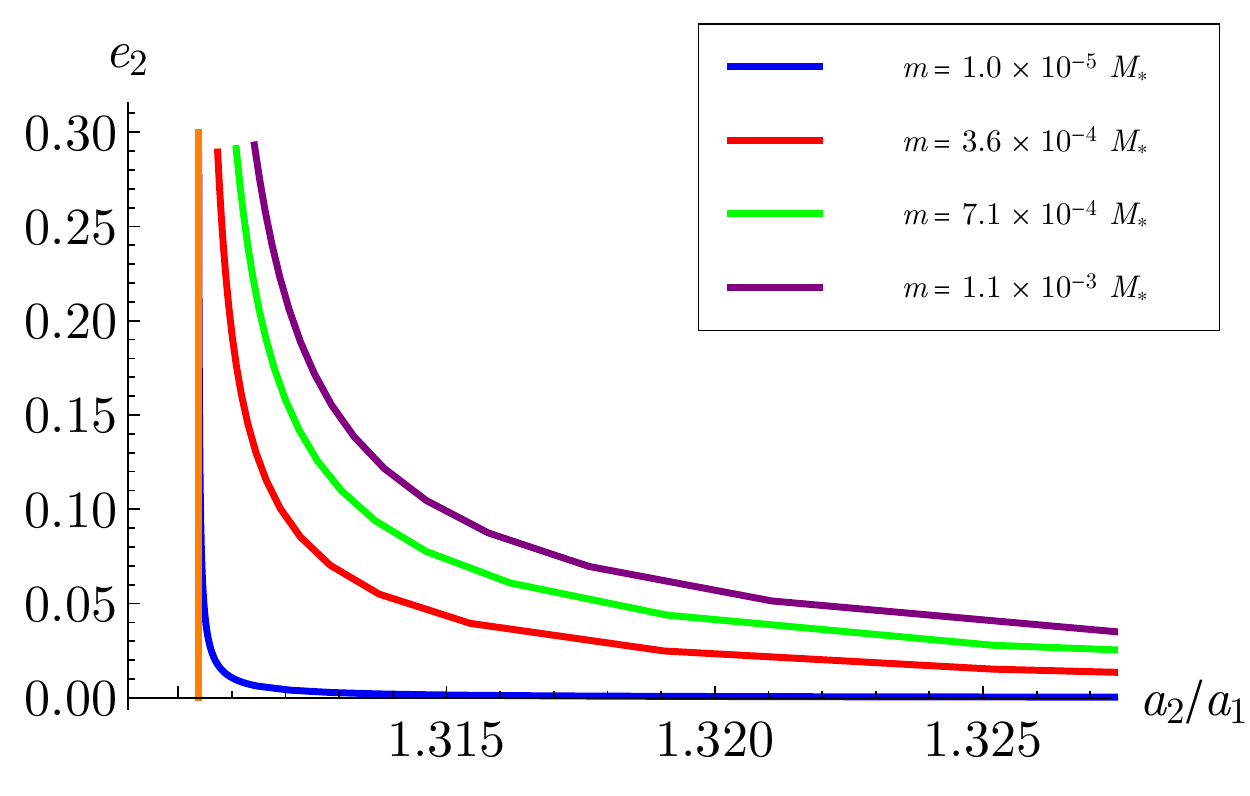}
\caption{3-2 mean motion resonance.}
\label{fig:EqCurvesForAllResonances-DifferentMasses.subfig:3-2}
\end{subfigure}
\begin{subfigure}[b]{0.32 \textwidth}
\centering
\includegraphics[scale=0.43
]{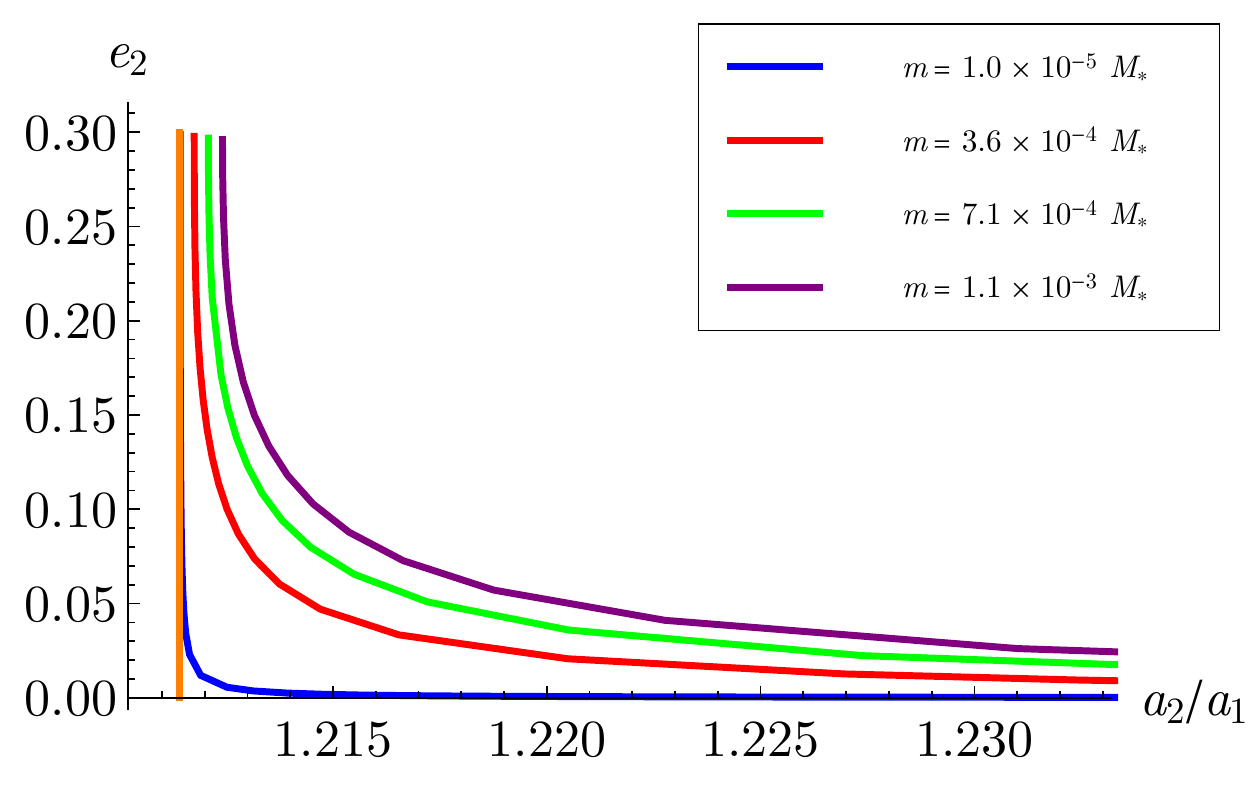}
\caption{4-3 mean motion resonance.}
\label{fig:EqCurvesForAllResonances-DifferentMasses.subfig:4-3}
\end{subfigure}
\caption{
Equilibrium curves for the three different first-order mean motion resonance, calculated as described in 
the text using the full averaged Hamiltonian \eqref{eq:FullAveragedResonantHamiltonian}, with different values for  
the planetary masses $m_1 = m_2 = m$. 
Here again we have fixed $\PSI_{1,eq} = 0$ and $\DELTAGAMMA_{eq} = \pi$.
The orange vertical line indicates the location of exact Keplerian resonance, 
$a_2/a_1 = \bar{a}_2/\bar{a}_1 = (k/(k-1))^{2/3}$.
}
\label{fig:EqCurvesForAllResonances-DifferentMasses}
\end{figure}

We should immediately remark one property of these curves.
As one can see from the first order expansion \eqref{eq:FirstOrderExpansion}, 
the rates of precession of the perihelia are estimated by $\dot\GAMMA \propto 1/\sqrt{\GAMMONA} \sim 1/e$, 
which grows substantially as $e \to 0$.
Therefore, in order to preserve the resonant condition $\dot{(\THETA+\GAMMA)}\sim0$, it is necessary to have 
$\dot\THETA = k\dot\LAMBDA_2 - (k-1)\dot\LAMBDA_1 \nsim 0$, i.e.\ $a_2/a_1 \nsim \bar R = (k/(k-1))^{2/3}$. 
Indeed, we see from Figures \ref{fig:EqCurvesForAllResonances-WithExpansions} 
that as the eccentricities vanish the equilibrium points deviate away 
from exact Keplerian commensurability, in a way that the semi-major axis ratio $a_2/a_1$ grows as $e\searrow 0$.
This effect, as is shown in Figures \ref{fig:EqCurvesForAllResonances-DifferentMasses}, 
is more and more evident as the planetary mass increases, 
since $\dot\GAMMA\propto m$. 
As a consequence, to sample these low-eccentricity equilibrium points with the correct value of $\OMEGONA$, it is 
necessary to plug into its analytical formula values of the semi-major axes such that 
$a_2/a_1 = \bar R + \delta{(a_2/a_1)}$.

We also point out the different equilibrium curves that one obtains using the expanded Hamiltonians and the non-expanded averaged Hamiltonian (Figure \ref{fig:EqCurvesForAllResonances-WithExpansions}).
The case of the 2-1 mean motion resonance is the most striking. Using a first order expansion, as the semi-major axis ratio approaches the exact Keplerian ratio one finds equilibrium points with increasing values of $e_2$ (and $e_1$). This is qualitatively different from the result obtained with higher order expansions and the averaged Hamiltonian: we see that $e_2$ reaches a maximum value and then starts approaching zero again (note that, although $e_2 \sim 0$, $e_1$ is large, so high order terms are important). This fact is known (e.g.\ \cite{2006MNRAS.365.1160B} and \cite{2006CeMDA..94..411M} using the numerical averaging of the Hamiltonian, \cite{2002CeMDA..83..141H} and \cite{2014Ap&SS.349..657A} tracking periodic orbits).
We further note that while the expansion to order 2 in the eccentricities captures this behaviour, it does not agree quantitatively with the averaged Hamiltonian.
On the other hand, the analytical curve obtained with the full averaged Hamiltonian
is in perfect agreement with a simulation in which two planets on initially circular orbits
are subjected to convergent migration resulting in resonant capture (Figure \ref{fig:EvolutionFor2-1}).
These simulations will be detailed in Section \ref{sec:ConvergentInwardMigration}, but they are expected to track the locus of
equilibrium points as the semi-major axis ratio $a_2/a_1$ decreases towards the Keplerian resonant ratio. Because here we apply no
damping on the eccentricities, the latter are {\it a priori} free to grow towards unity.
We observe that at the point in which $e_2$ vanishes, $\delta\VARPI$ flips from $\pi$ to 0, 
which is evident from Figure \ref{fig:EvolutionFor2-1.subfig:deltavarpi}.
Indeed the equilibrium point on the $e_2 \cos(\delta\VARPI)$ axis is initially on the negative side, and as the angular momentum decreases it moves to the positive axis.
This transition from $\delta\VARPI=\pi$ to 0 is smooth, and this is why the planets stay at the equilibrium point, 
without triggering secular oscillations. 

We note that at higher values of $e$ these equilibrium points found for $\DELTAGAMMA=\pi$ 
(or $\DELTAGAMMA=0$ in the case of the 2-1 resonance) might be unstable, and stable asymmetric equilibrium 
points for different values of $\DELTAGAMMA$ are possible (see for example \cite{2003ApJ...593.1124B} and \cite{2006MNRAS.365.1160B}, for a detailed 
study on the 2-1 mean motion resonance); in the case reported here, 
they are unstable for $e_1$ between about $0.28$ and $0.35$ corresponding to $e_2$ between about $0.08$ and $0.11$.  
We should also note that a similar behaviour of the equilibrium curves, where they reach a maximum value in $e$ and then bend 
down to reach 0, is also present in the other resonances that we have considered, but that this happens at much 
higher values of $e$. 
In the case of the 3-2 and 4-3 resonances, it is $e_1$ that reaches a maximum value, of $e_1 \simeq 0.22$ and 
$e_1 \simeq 0.12$ respectively. 
However, these circumstances occur at high values of the eccentricities and are beyond the scope of this work.

\begin{figure}[!ht]
\centering
\begin{subfigure}[b]{0.32 \textwidth}
\centering
\includegraphics[scale=0.43
]{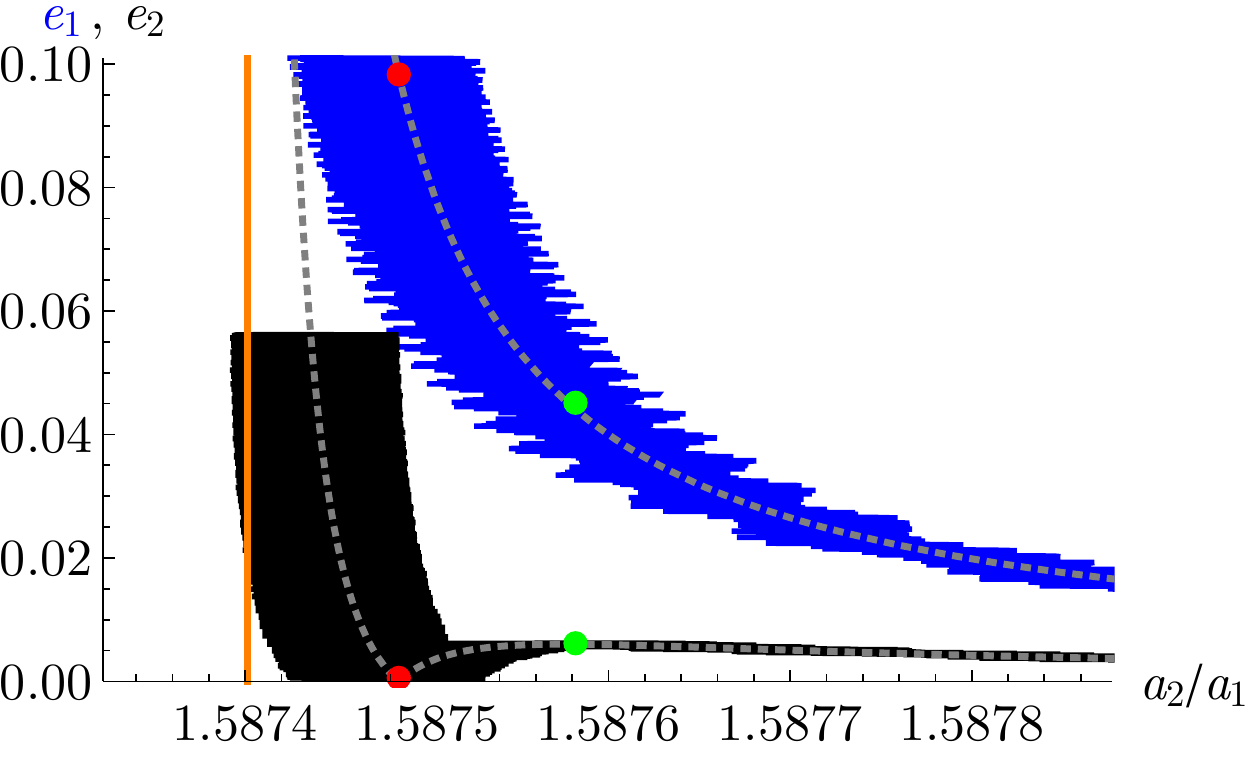}
\caption{$e_1$ and $e_2$.}
\label{fig:EvolutionFor2-1.subfig:e1e2}
\end{subfigure}
\begin{subfigure}[b]{0.32 \textwidth}
\centering
\includegraphics[scale=0.13
]{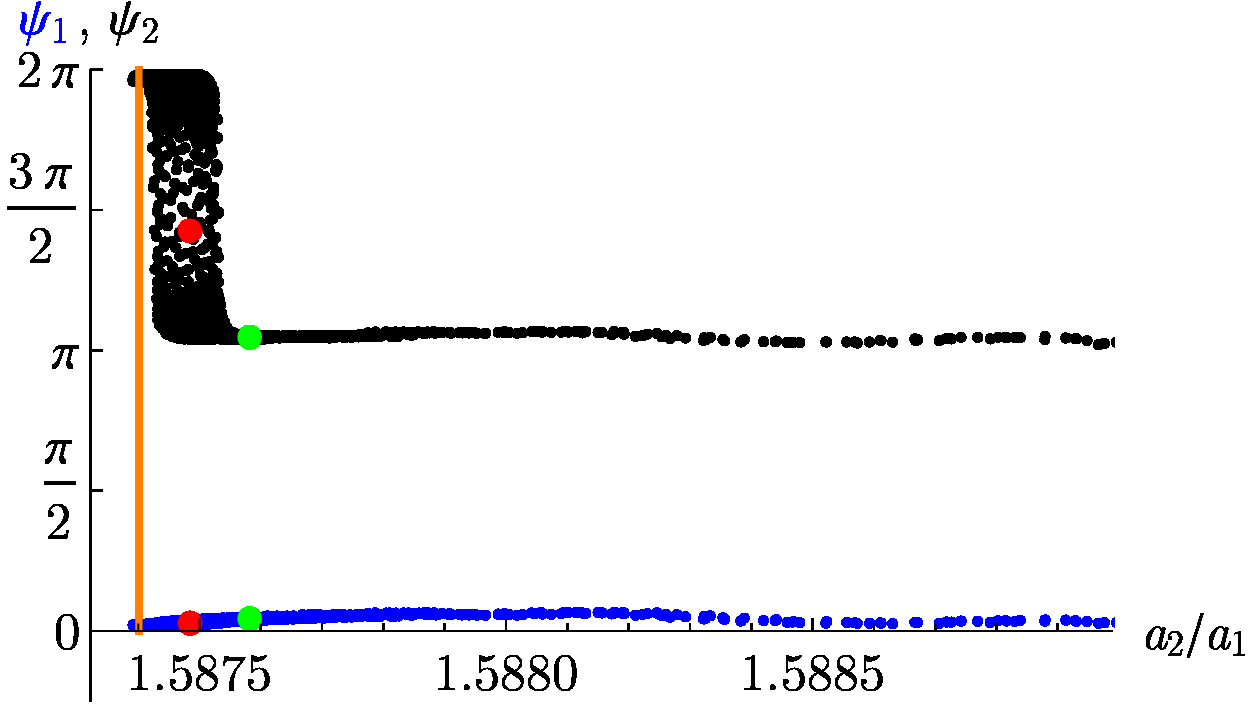}
\caption{Resonant angles $\PSI_i = \THETA+\GAMMA_i$.}
\label{fig:EvolutionFor2-1.subfig:resonantangles}
\end{subfigure}
\begin{subfigure}[b]{0.32 \textwidth}
\centering
\includegraphics[scale=0.12
]{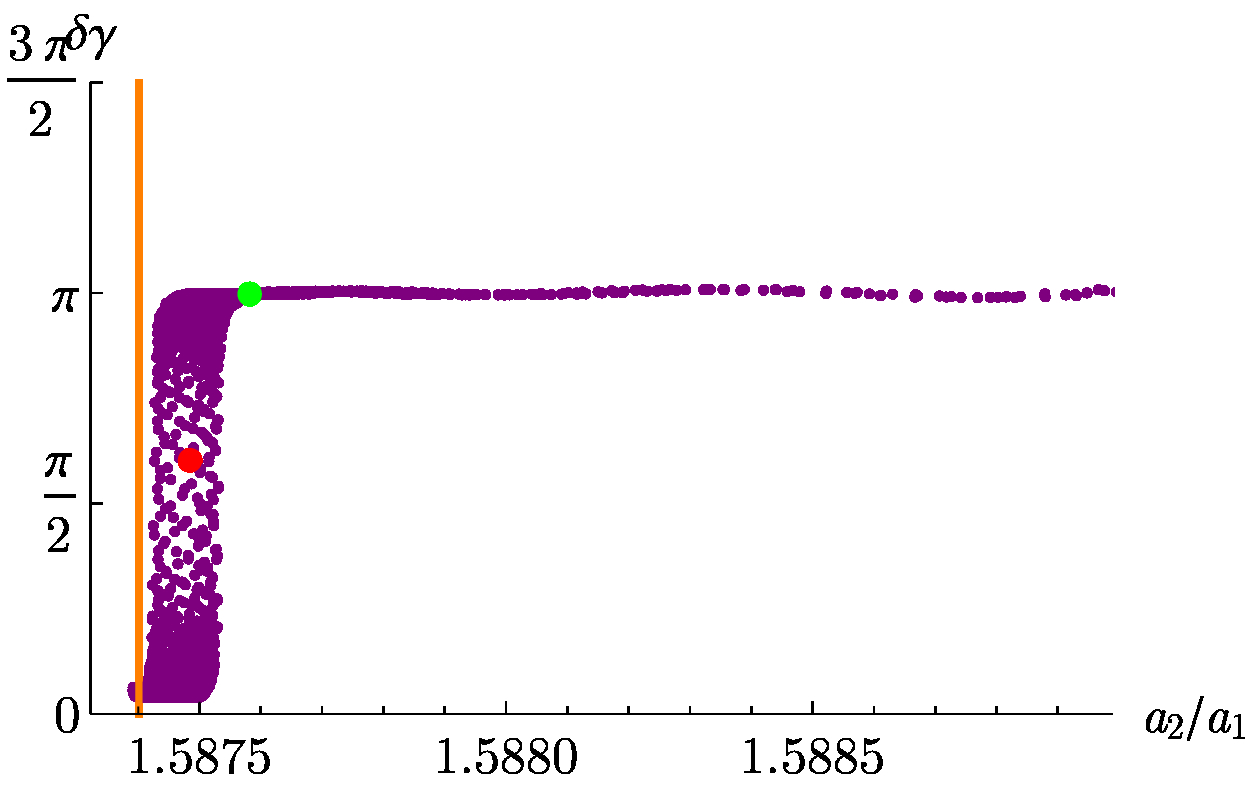}
\caption{Angle $\DELTAGAMMA$.}
\label{fig:EvolutionFor2-1.subfig:deltavarpi}
\end{subfigure}

\caption{
Result of a numerical simulation for two planets in the 2-1 mean motion resonance, with planetary masses
$m_1 = m_2 = m = 10^{-5} M_*$. In panel (a) we show both eccentricities $e_1$ (in blue) and $e_2$ (in black); 
in panel (b) the resonant angles $\PSI_1 = \theta + \gamma_1$ (in blue) and $\PSI_2 = \theta + \gamma_2$ (in black);
in panel (c) the angle $\DELTAGAMMA$.
In all panels the quantities are given in terms of the semi-major axes ratio $a_2/a_1$, to easily compare the results 
with the panels in Figure \ref{fig:EqCurvesForAllResonances-WithExpansions.subfig:2-1}; 
in panel (a) we also superimpose the equilibrium curves, shown in dotted grey lines, for $\DELTAGAMMA=\pi$ and $\DELTAGAMMA=0$.
We again indicate in all plots the location of exact Keplerian resonance with an orange vertical line.
The green points correspond to the equilibrium configuration of this system when $e_2 \simeq 0.006$ is maximal; 
the red points correspond to the equilibrium configuration of the system when $e_2$ has then reached the value 0.
We observe that the evolution of the orbital parameters is very well described by our analytical curves;
the large oscillations, visible especially in panel (a), are short-periodic, 
due mainly to the fast synodic angle $\LAMBDA_1-\LAMBDA_2$, which is averaged out in the analytical model.
We notice that when $e_2$ reaches 0 (red point) the value of $\DELTAGAMMA$ 
is changing from $\pi$ to 0. 
This happens without triggering large oscillations as the system is still smoothly following 
the curve of stable equilibrium points, see text for details. 
}
\label{fig:EvolutionFor2-1} 
\end{figure}

\subsection{Frequencies in the limit of small amplitude of libration}\label{subsec:FrequenciesOfLibration}
In this section we calculate the frequencies of the system around an equilibrium point
assuming small amplitude of libration by considering the linearised system near the equilibrium point. 
As we will see in the next Section, we expect that in our numerical simulations the planets will be very close to the equilibrium in the variables \eqref{eq:FinalCoV}, and will move from an equilibrium corresponding to some value 
of the constant of motion $\OMEGONA$ to the next while preserving a small amplitude of libration.
We then discuss how we can check numerically the validity of our calculations.

Near the equilibrium point $\mathbf x_{eq}$ the Hamiltonian 
$\bar\Ha(\PSIONA_1,\PSIONA_2,\PSI_1,\DELTAGAMMA) = \bar\Ha(\mathbf x)$ can be approximated as
\begin{equation}
\bar\Ha(\mathbf x) = \bar\Ha(\mathbf x_{eq}) + \bar\Ha_{lin}(\mathbf x) + \bar\Ha_{quad}(\mathbf x) + \mathcal O(\mathbf x^3).
\end{equation} 
The linear part $\bar\Ha_{lin}(\mathbf x) \equiv 0$ by definition of equilibrium point, and the quadratic part is given by
\begin{equation}
\bar\Ha_{quad}(\mathbf x) = \frac{1}{2} (\mathbf x - \mathbf x_{eq})^{\intercal} C (\mathbf x - \mathbf x_{eq}),
\end{equation} 
where $C := \mathbb H(\bar\Ha(\mathbf x_{eq}))$ is the Hessian of $\bar\Ha$ at the equilibrium point 
$\mathbf x_{eq}$.
Dropping the unimportant constant term $\bar\Ha(\mathbf x_{eq})$ and ignoring the higher order terms, 
the linearised Hamiltonian system of equation then becomes 
\begin{equation}\label{eq:LinearisedEquationsAroundEquilibriumPointForAveragedHamiltonian}
\frac{\D{}}{\D t}{(\mathbf x-\mathbf x_{eq})} = J\nabla\bar\Ha_{quad} = J C (\mathbf x-\mathbf x_{eq}),
\end{equation} 
where $\nabla=\nabla_{\mathbf x}$, and $J$ is the symplectic matrix
\begin{equation}
J = \left(
\begin{matrix}
\mathbf 0 & -\mathbb I \\
\mathbb I & \mathbf 0
\end{matrix}
\right)
.
\end{equation} 
The study of the stability of the equilibrium then reduces to writing the matrix $J C$ and finding its eigenvalues. 
Moreover, given that the system is Hamiltonian, it is well known that the four purely imaginary eigenvalues 
come in pairs, $(+i \omega_1, -i \omega_1)$ and $(+i\omega_2, -i\omega_2)$, with $\omega_{1,2}>0$. 
These $\omega_1$ and $\omega_2$ are the two characteristic frequencies of the system at vanishing amplitude of 
libration around the equilibrium point:
they are associated respectively with the (faster) libration of the 
resonant pair $(\PSIONA_1,\PSI_1)$, and with the (slower) secular libration to which the pair of variables 
$(\PSIONA_2,\DELTAGAMMA)$ is subjected. 
We expect that $\omega_1$ will be much higher than $\omega_2$, except at vanishing eccentricities, where the 
system exhibits a fast precession of the perihelia.

We check that our analytical calculations of the frequencies are correct as follows. 
We first take a system of two planets well in resonance, e.g.\ in the 3-2 mean motion resonance, 
$\dot\LAMBDA_1 \simeq \frac{3}{2}\dot\LAMBDA_2$, but not exactly on the equilibrium point.
Here we take $m_1 = m_2 = 10^{-5} M_*$. 
We then observe the evolution of the orbital elements $a$ and $e$, from which we obtain that of the 
four actions, and we record $\bar\PSIONA_1$, $\bar\PSIONA_2$, $\bar\KAPPONA$, $\bar\OMEGONA$ their mean values. 
Note that the mean values are needed because the system is undergoing a fast evolution due to the 
non-resonant angles, which have been averaged out in our analytical model. 
In particular, e.g.\ in Figures \ref{fig:aExcitedSystemIn3-2Resonance-Frequencies.subfig:a1ShortPeriod} 
we notice the prominent effect of the harmonic relative to the circulating angle $\LAMBDA_1-\LAMBDA_2$, 
with a frequency that can be calculated as 
$\omega_{\LAMBDA_1-\LAMBDA_2} = (\dot\LAMBDA_1-\dot\LAMBDA_2)=\frac{1}{3}\dot\LAMBDA_1 = \frac{1}{3} 2\pi/(a_1^{3/2})\simeq 65.4$, 
for the actual value of $a_1 = 0.1008$  AU and assuming $\GravC M_* = (2\pi)^2$, 
that is a period $T_{\LAMBDA_1-\LAMBDA_2}\simeq 0.096$ years for $M_* = M_\odot$. 
We then look at the two angles, checking that the resonant angle 
$\PSI_1$ is librating (around 0) and noticing that $\DELTAGAMMA$ librates (around $\pi$); we therefore 
fix $\bar\PSI_1=0$ and $\bar\DELTAGAMMA=\pi$.
Using the values for $\bar\KAPPONA$, $\bar\OMEGONA$ and of the two angles $\bar\PSI_1 = 0$, 
$\bar\DELTAGAMMA = \pi$, 
we calculate analytically an equilibrium point $\mathbf x_{eq}$ as explained above. 
This equilibrium point well represents the state of the system, with $\PSIONA_{1,eq}$ and $\PSIONA_{2,eq}$ differing 
from the observed mean values $\bar\PSIONA_1$, $\bar\PSIONA_2$ by less that 0.03\%.
For this equilibrium point we calculate the two frequencies $\omega_1 \simeq 0.62$ and $\omega_2 \simeq 0.23$, 
i.e.\ periods of $T_1 = 2\pi/\omega_1 \simeq 10.5$ years and $T_2 = 2\pi/\omega_2 \simeq 26.9$ years. 
In order to clearly see these two frequencies in a numerical simulation, 
we excite the system's initial condition, in the semi-major axes ratio 
and in eccentricity respectively, thereby increasing the amplitude of librations relative to the resonant angle 
$\PSI_1$ and the angle $\DELTAGAMMA$.
In practice, we first take the same initial conditions of the original unexcited system, 
and slightly excite the value of $R = a_2/a_1$, 
e.g.\ by forcingly change the initial value $a_2(0)$ of $a_2$ to $(1+\epsilon) a_2(0)$, 
where $\epsilon$ is a small number. 
We plot the resulting evolution of the semi-major axis and eccentricity for the inner planet in Figure 
\ref{fig:aExcitedSystemIn3-2Resonance-Frequencies}, where we see clearly an oscillation with period 
$T_1 \simeq 10.5$ years (panels (b), (d)).
Similarly, we take again the same initial condition of the unexcited system and slightly excite the value of $e_2(0)$ 
to $(1+\tilde\epsilon) e_2(0)$, where $\tilde\epsilon$ is a small number.
We plot the resulting evolution of the semi-major axis and eccentricity for the inner planet in Figure 
\ref{fig:eExcitedSystemIn3-2Resonance-Frequencies}, where we now also see an oscillation with period 
$T_2 \simeq 26.9$ years on top of the one with period $T_1 \simeq 10.5$ years 
(panel (d)).
In both Figures \ref{fig:aExcitedSystemIn3-2Resonance-Frequencies} and \ref{fig:eExcitedSystemIn3-2Resonance-Frequencies} we overplot the result of the analytical explicit integration of the linearised equations of motion \eqref{eq:LinearisedEquationsAroundEquilibriumPointForAveragedHamiltonian} around the equilibrium point. These follow very closely the evolution of the $3$-body integrations.

\begin{figure}[!ht]
\centering
\begin{subfigure}[b]{0.3 \textwidth}
\centering
\includegraphics[scale=0.43
]{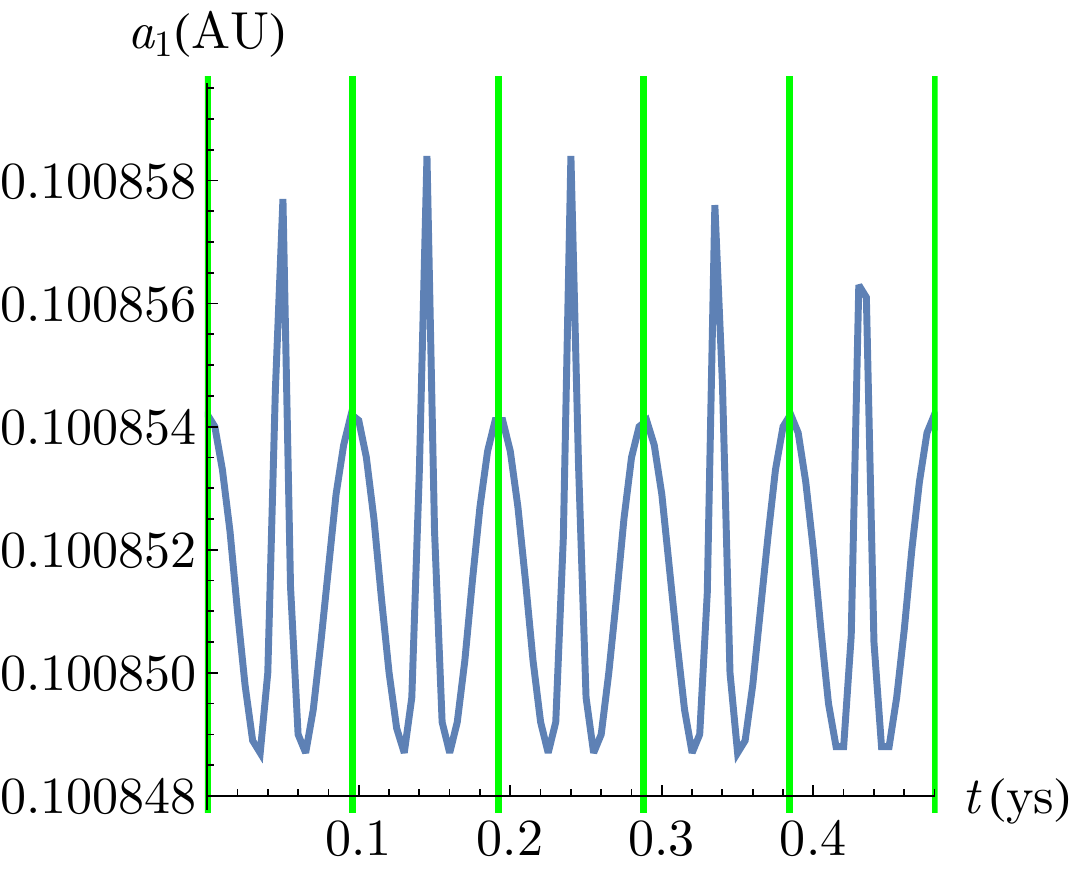}
\caption{$a_1$, short period oscillations.}
\label{fig:aExcitedSystemIn3-2Resonance-Frequencies.subfig:a1ShortPeriod}
\end{subfigure}
\begin{subfigure}[b]{0.69 \textwidth}
\centering
\includegraphics[scale=0.43
]{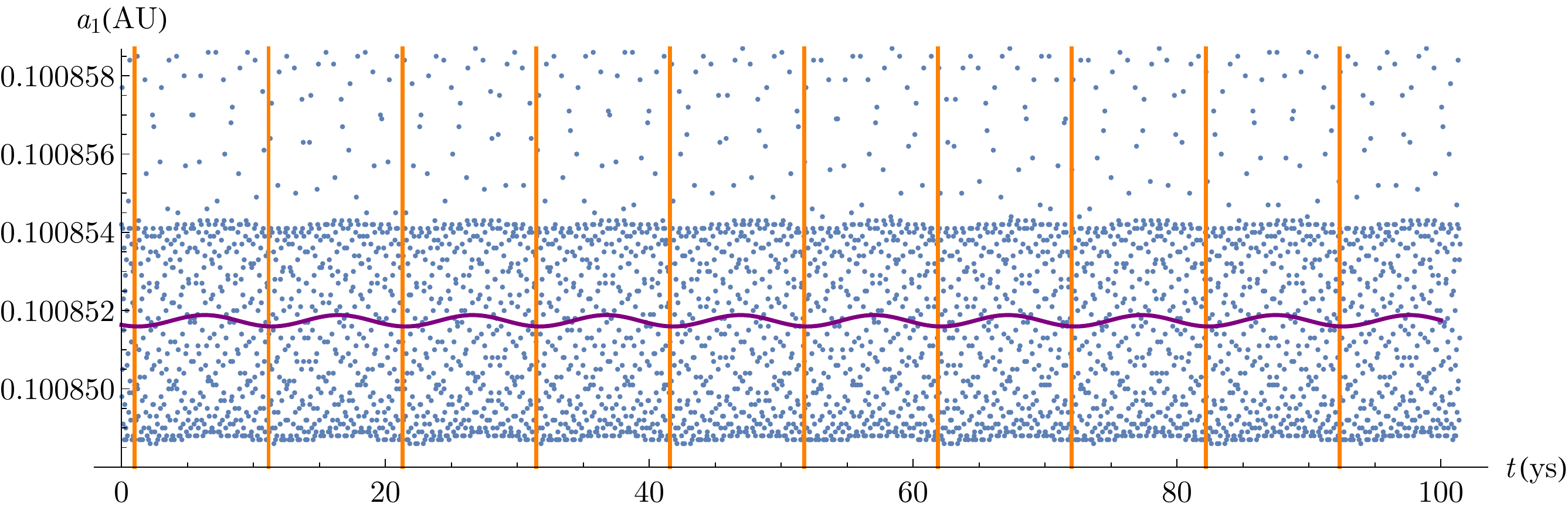}
\caption{$a_1$, long period oscillations.}
\label{fig:aExcitedSystemIn3-2Resonance-Frequencies.subfig:a1LongPeriod}
\end{subfigure}

\centering
\begin{subfigure}[b]{0.3 \textwidth}
\centering
\includegraphics[scale=0.43
]{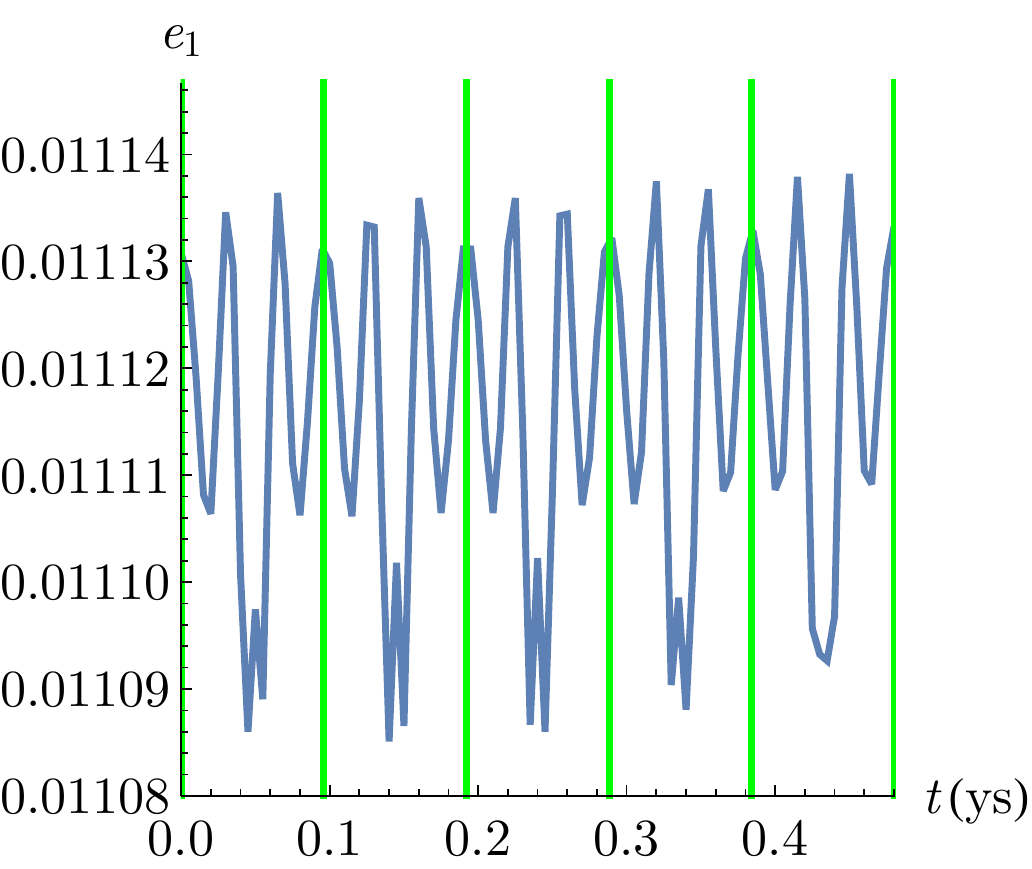}
\caption{$e_1$, short period oscillations.}
\label{fig:aExcitedSystemIn3-2Resonance-Frequencies.subfig:e1ShortPeriod}
\end{subfigure}
\begin{subfigure}[b]{0.69 \textwidth}
\centering
\includegraphics[scale=0.43
]{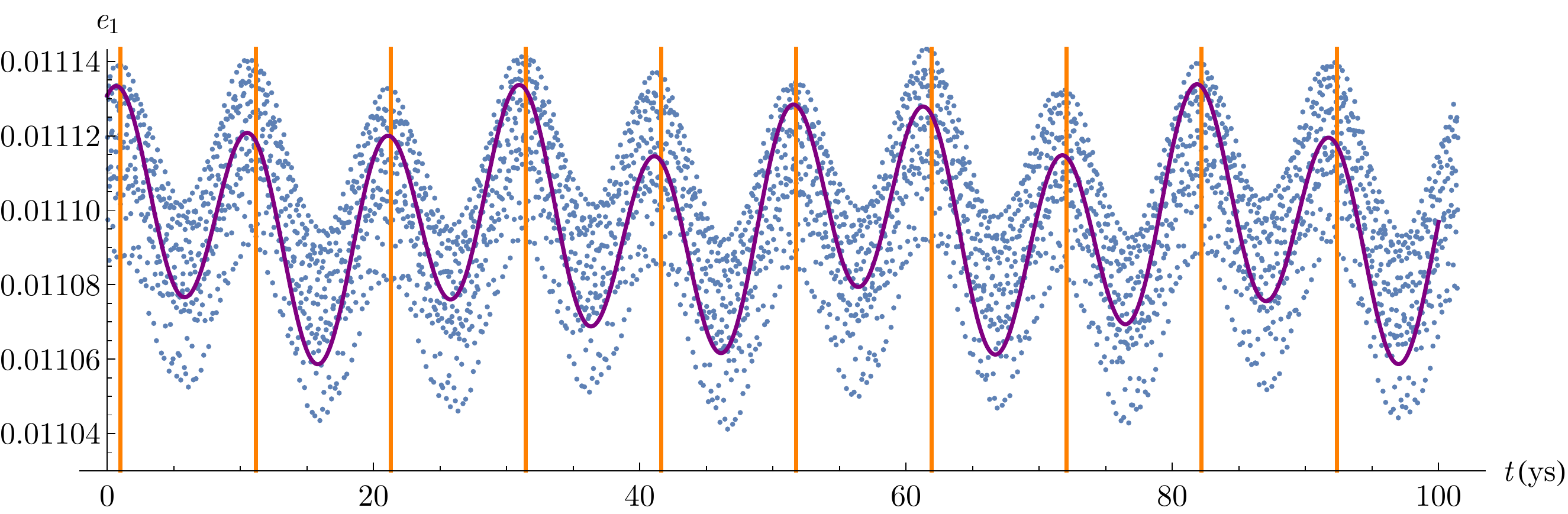}
\caption{$e_1$, long period oscillations.}
\label{fig:aExcitedSystemIn3-2Resonance-Frequencies.subfig:e1LongPeriod}
\end{subfigure}
\caption{Evolution of $a_1$ and $e_1$ on different timescales for a system in the 3-2 mean motion resonance, 
after a forced small excitation of $R=a_2/a_1$. 
Here $m_1 = m_2 = 10^{-5} M_*$, $M_* = M_\odot$, in units where $\GravC M_* = (2\pi)^2$.
We notice the fast evolution on the two left panels due to the synodic period 
$T_{\LAMBDA_1-\LAMBDA_2}\simeq 0.096$ years (green vertical lines).
On the right panels, we notice the oscillations with a longer period of $T_1 = 2\pi/\omega_1 \simeq 10.5$ years 
(orange vertical lines), 
as predicted by the analytical calculation of $\omega_1$, see text.
The thick purple curve is the result of direct integration of the linearised equations of motion around the equilibrium point for the averaged Hamiltonian, equation \eqref{eq:LinearisedEquationsAroundEquilibriumPointForAveragedHamiltonian}; the initial conditions are the same as those for the numerical simulations. One sees that the analytical model follows very closely the averaged evolution obtained via $3$-body numerical integration.
}
\label{fig:aExcitedSystemIn3-2Resonance-Frequencies}
\end{figure}

\begin{figure}[!ht]
\centering
\begin{subfigure}[b]{0.3 \textwidth}
\centering
\includegraphics[scale=0.43
]{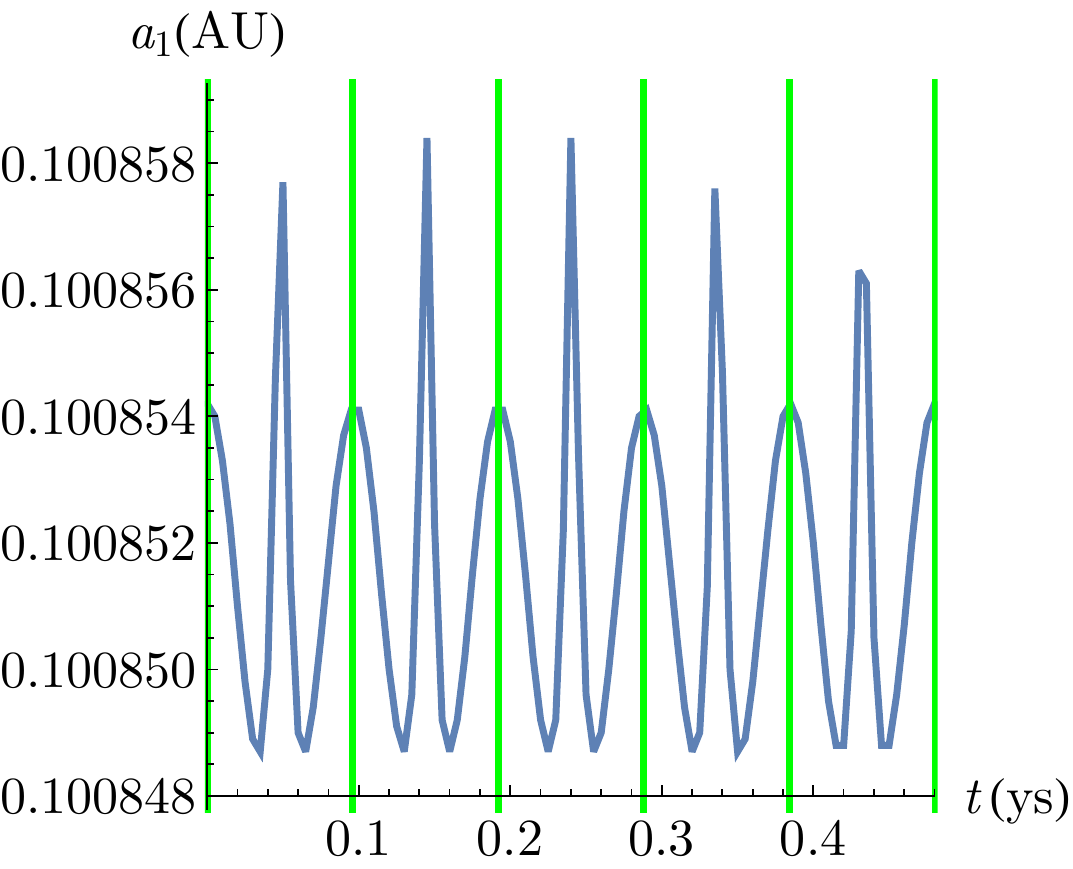}
\caption{$a_1$, short period oscillations.}
\label{fig:eExcitedSystemIn3-2Resonance-Frequencies.subfig:a1ShortPeriod}
\end{subfigure}
\begin{subfigure}[b]{0.69 \textwidth}
\centering
\includegraphics[scale=0.43
]{3-2excitee-a1-longperiod-WithAnalyticalSol}
\caption{$a_1$, long period oscillations.}
\label{fig:eExcitedSystemIn3-2Resonance-Frequencies.subfig:a1LongPeriod}
\end{subfigure}

\centering
\begin{subfigure}[b]{0.3 \textwidth}
\centering
\includegraphics[scale=0.43
]{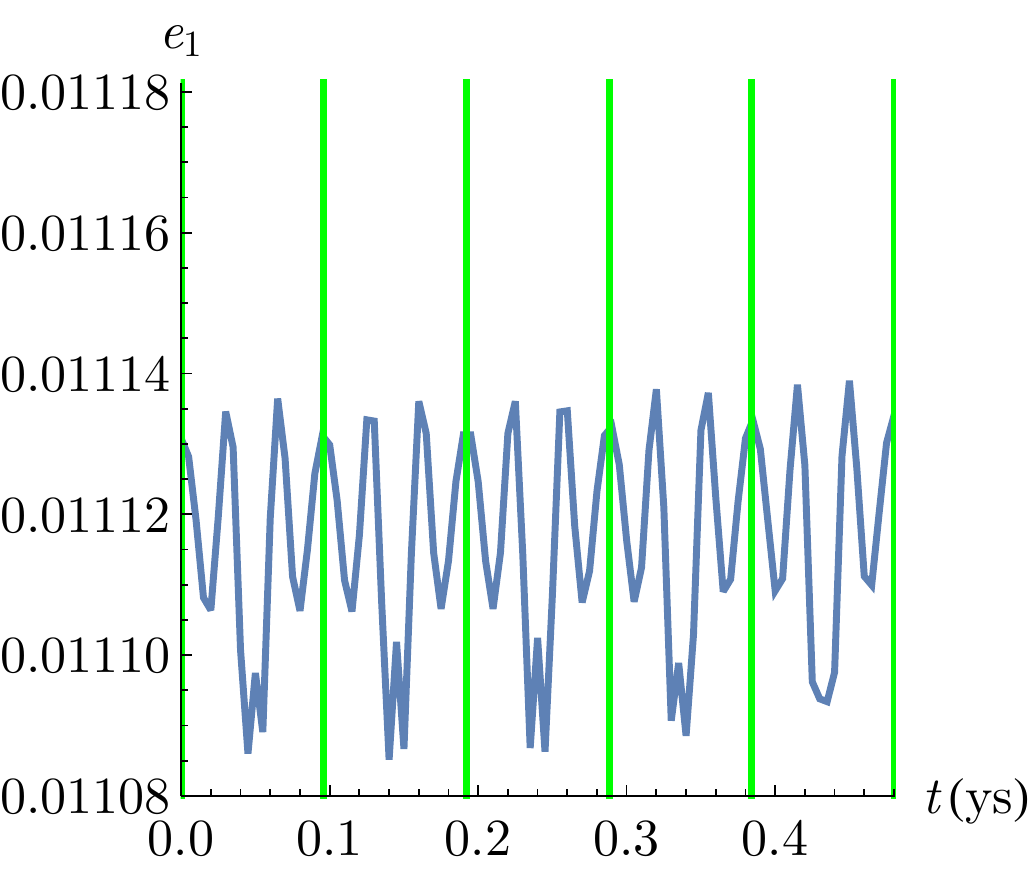}
\caption{$e_1$, short period oscillations.}
\label{fig:eExcitedSystemIn3-2Resonance-Frequencies.subfig:e1ShortPeriod}
\end{subfigure}
\begin{subfigure}[b]{0.69 \textwidth}
\centering
\includegraphics[scale=0.43
]{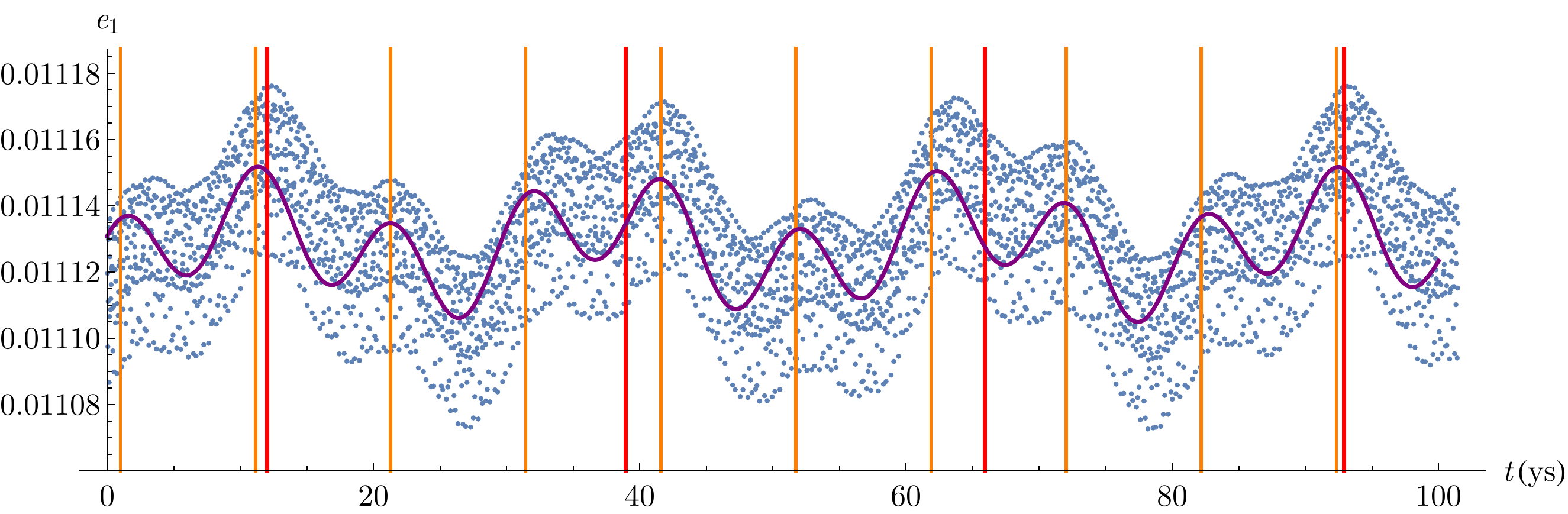}
\caption{$e_1$, long period oscillations.}
\label{fig:eExcitedSystemIn3-2Resonance-Frequencies.subfig:e1LongPeriod}
\end{subfigure}
\caption{
Evolution of $a_1$ and $e_1$ on different timescales for a system in the 3-2 mean motion resonance, 
after a forced small excitation of $e_2$. 
Here $m_1 = m_2 = 10^{-5} M_*$, $M_* = M_\odot$, in units where $\GravC M_* = (2\pi)^2$, 
as in Figure \ref{fig:aExcitedSystemIn3-2Resonance-Frequencies}.
We notice again the fast evolution on the two left panels due to the synodic period 
$T_{\LAMBDA_1-\LAMBDA_2}\simeq 0.096$ years (green vertical line).
On the right panels, we notice still the oscillations with a longer period of $T_1 = 2\pi/\omega_1 \simeq 10.5$ years 
(orange vertical lines); 
in addition, in panel (d) we notice how $e_1$ is now also effected by libration of $\DELTAGAMMA$, with 
characteristic period of $T_2 = 2\pi/\omega_2 \simeq 26.9$ years (red vertical lines), as predicted by the analytical calculation 
of $\omega_2$, see text.
The thick purple curve is again the result of direct integration of equation \eqref{eq:LinearisedEquationsAroundEquilibriumPointForAveragedHamiltonian}, with the same initial conditions as the numerical simulations, showing again good fit.
}
\label{fig:eExcitedSystemIn3-2Resonance-Frequencies}
\end{figure}

\section{Convergent inward migration in disk and resonant capture}\label{sec:ConvergentInwardMigration}
With our resonant model at hand, we now proceed with the study of our first step in our numerical and analytical 
investigations, that of resonant capture in a protoplanetary disk.
This is an efficient method to obtain planets deeply in mutual mean motion resonance 
(e.g.\ \cite{2012Icar..221..624M}, \cite{2017A&A...602A.101R}). 
We start with two planets of equal mass, $m_1=m_2=m$, typically $m/M_*=10^{-5} - 10^{-2}$, 
on coplanar orbits, embedded in a protoplanetary disk. 
We also write $\mu_1=\mu_2=\frac{M_* m}{M_*+m}=:\mu$.
Our numerical simulations consist of the implementation of a symplectic 3-Body integrator ({\codefont swift\_symba}) 
to which fictional analytical dissipative forces are added that describe, in the limit of the purposes of this study, 
the interaction between the planets and a protoplanetary disk. 
In what follows, we describe these forces, dropping for ease of reading the index $i = 1,2$ to denote the planets' 
elements and parameters.

For each planet, the effect of the disk-planet interaction can be viewed as composed of two separate contributions, 
one operating on the eccentricity $e$ and one operating on the semi-major axis $a$. 
Concerning the effect of the gas on the eccentricity $e$, our code implements a damping effect of the disk as
\begin{equation}
\dot{e}_{damp} := -\frac{e}{\tau_e},
\end{equation}
where $\tau_e$ is given, in the limit of vanishing eccentricities, by
\begin{equation}\label{eq:tau_e}
\tau_e \simeq \frac{\tau_{wave}}{0.780},
\end{equation}
and $\tau_{wave}$ is the typical Type-I migration timescale, given by
\begin{equation}\label{eq:twave}
\tau_{wave} = \frac{M_*}{m} \frac{M_*}{\Sigma a^2} \frac{h^4}{\sqrt{\GravC M_*/a^3}},
\end{equation}
see e.g.\ \cite{2006A&A...450..833C}, \cite{2014prpl.conf..667B}.
The parameters $\Sigma=\Sigma(r) = \Sigma_0 r^{-\alpha}$ and $h = h(r) = H/r \propto (r/r_0)^\beta$ are 
the surface density and aspect ratio of the disk respectively and are evaluated at the position of the planet. 
The flaring index $\beta$ is taken to be $\beta = 0.25$ and $H = H(r) =  z_{scale} (r/r_0)^\beta r$ 
is the scale-height.
We take $\left.h\right|_{5.2 \text{AU}} = 5\%$ so that $z_{scale} = 0.05 \times (5.2 \text{ AU}/r_0)^{-\beta}$.
The parameter $\alpha$ sets the surface density profile of the disk; here we take $\alpha = 1$.

Secondly, the disk-planet interaction results in a torque, and therefore in an exchange of angular momentum 
$\ANGMOM$.
For a planet,
\begin{equation}
\ANGMOM = m \sqrt{\GravC (M_*+m) a (1-e^2)}.
\end{equation}
The torque $T:=\dot\ANGMOM$ is taken here to be negative, so that the effect on the semi-major axis $a$ 
is that of inward, Type-I migration. It is modeled in our simulations as
\begin{equation}\label{eq:dotAngMomMig}
\dot\ANGMOM_{mig} = -\frac{\ANGMOM}{\tau_{mig}},
\end{equation}
where $\tau_{mig}$ is given, again in the limit of vanishing eccentricities, by 
\begin{equation}\label{eq:tau_mig}
\tau_{mig} \simeq 2 \frac{\tau_{wave}}{(2.7 + 1.1 \alpha)} h^{-2},
\end{equation}
where again we take $\alpha = 1$.
To calculate the resulting effect on the semi-major axis $a$ due to this planet-disk interaction, we take 
\begin{equation}\label{eq:dotAngMom}
\dot\ANGMOM=\frac{\D\ANGMOM}{\D t} =
   m\sqrt{\GravC (M_* +m)} \left(\frac{\dot a}{2\sqrt{a}}\sqrt{1-e^2} - \frac{\sqrt{a}}{\sqrt{1-e^2}}e \dot e \right),
\end{equation}
and dividing by $\sqrt{a}$ we obtain
\begin{equation}\label{eq:adotovera}
\frac{\dot a}{a} = 2\frac{\dot\ANGMOM}{\ANGMOM} + \frac{2 e \dot e}{1-e^2} 
                 = - \frac{1}{\tau_a} - p \frac{e^2}{\tau_e},
\end{equation}
where $\tau_a=\tau_{mig}/2$ and $p\simeq 2$ for small $e$.
It is customary to introduce the quantity $K = {\tau_{a}}/{\tau_{e}}$ which we call $K$-factor 
(cfr.\ \cite{2017A&A...602A.101R}).
Note that,
\begin{equation}\label{eq:KFactor}
K = \frac{\tau_{a}}{\tau_{e}} \simeq \frac{0.780}{2.7+1.1} h^{-2}:
\end{equation}
given that disks are very thin, e.g.\ here $h = \mathcal O(5 \times 10^{-2})$, we see that the $K$-factor is very large,
of the order of at least $K = \mathcal O(10^{2})$,
meaning that the typical timescale of eccentricity damping is much shorter than that of migration. 
This allows us to assume that the planets approach the resonance on circular orbits, 
as any finite (but relatively small) initial eccentricity would be immediately damped by the disk. 

In order to insure convergent migration and resonant capture, we need to stop the migration of the inner planet, 
since two equally massive planets would migrate inward at roughly the same rate and resonant capture would 
not occur (e.g.\ \cite{2017A&A...602A.101R}).
To do this, we simulate the effect of a disk edge, which corresponds to a sharp drop in $\Sigma$ as $r$ decreases. 
In this conditions, \cite{2006ApJ...642..478M} showed that a coorbital corotation torque is activated, 
which is positive and dominates the inward Type-I torque. 
Thus inward migration stops at the inner edge of the disk. 
\cite{2006ApJ...642..478M} called this a planet trap and we follow this terminology here. 
For simplicity, the trap is modeled here by smoothly reversing the Type-I torque. 
This is not what happens in reality. 
Modeling the real effects would require an appropriate implementation of the corotation torque, 
and that would depend on the $\Sigma$ profile at the edge.
Our recipe, however, is effective to stop the inward migration of the innermost planet and to retain the second planet in
resonance, that is to exhibit the same effects observed in hydrodynamical simulations 
(\cite{2008A&A...478..929M}). 
As we approach the disk edge $d_{edge}$ (at 0.1 AU in our simulations) we implement the planetary trap by smoothly 
reversing the sign of the migration in order to stop the inner planet from migrating all the way into the star.
This is achieved by dividing 
$\tau_{a}$ by a factor $\tau_{a,red}$ given by
\begin{equation}
\tau_{a,red} = 
\begin{cases} 
      1 & a\geq d_{edge} (1+h_{edge}), \\
      5.5 \times \cos\left(\frac{((d_{edge}\times(1.+h_{edge})-a) 2 \pi)}{(4 h_{edge} \times d_{edge})}\right) - 4.5 & d_{edge}(1-h_{edge}) \leq a \leq d_{edge}(1+h_{edge}), \\
      -10 & 0 \leq a\leq d_{edge}(1-h_{edge}),
\end{cases}
\end{equation}
where $h_{edge} = z_{scale} (d_{edge}/r_0)^{0.25}$ is the aspect ratio of the disk at the edge.

As initial conditions in our simulations we first assume circular orbits, $e_{1,init} = e_{2,init}=0$, see above. 
Secondly, we choose the initial semi-major axes to be just outside a specific first order mean motion resonance, 
$a_{2,init} \gtrsim (k/(k-1))^{2/3} a_{1,init}$, $k=2,3,\dots$. 
The two planets will migrate inward at roughly the same rate due to their interaction with the disk; 
the first planet will then reach the disk edge, where our imposed reversal of the sign of migration will cause it
to stop migrating.
The still migrating outer planet approaches the first planet and 
is then automatically locked in the desired mean motion resonance 
as a result of convergent Type-I migration.
The behaviour of the planets as they approach resonance can be understood using adiabatic theory, provided that the
migration timescale is much longer than the resonant libration timescale (see Section \ref{subsec:FrequenciesOfLibration} 
for the latter). 
When the planets are far from resonance, the damping effect of the disk ensures that their orbits are circular. 
But the circular orbit is also the limit of the curve of the resonant equilibria for large $a_2/a_1$ ratio 
(see Figures \ref{fig:EqCurvesForAllResonances-WithExpansions}, \ref{fig:EqCurvesForAllResonances-DifferentMasses}). 
Thus, the planets are very close to the equilibrium in the variables \eqref{eq:FinalCoV} corresponding to their large $a_2/a_1$ ratio. 
If the evolution is adiabatic, the amplitude of libration around the equilibrium point 
(more precisely the value of the libration action - \cite{Arnold1963}) is preserved (\cite{Neishtadt1999}, \cite{Neishtadt&Al2008}, \cite{Henrard1993}). 
Given that initially this amplitude is close to zero, it will remain close to zero throughout the evolution. 
In reality, the application of the adiabatic principle can be done only 
if the non-conservative forces change the parameters of the Hamiltonian, 
and not if they affect directly its variables. 
If there is no damping on the eccentricities but only a drag on the semi major axes, 
\cite{2015ApJ...810..119D} show that, at low-order in $e$, the dissipation only acts 
on the otherwise constant of motion $\OMEGONA$ (see \eqref{eq:FinalCoV}) and does not act 
on the dynamical variables $\PSIONA_1, \PSIONA_2, \PSI_1, \DELTAGAMMA$. 
In this case, the adiabatic principle can be used. 
Thus, as
the planets approach each other, they have to follow the locus of 
equilibrium points computed in Section \ref{subsec:EquilibriumPointsOfAveragedHa} and shown in 
Figures \ref{fig:EqCurvesForAllResonances-WithExpansions}, \ref{fig:EqCurvesForAllResonances-DifferentMasses}. 
This is precisely what we observed in Figure \ref{fig:EvolutionFor2-1} for the 2-1 resonance.
Thus, as the planets approach each other, their eccentricities start to grow. 
As shown in Figure \ref{fig:EvolutionFor2-1}, if there were no eccentricity damping, at least one of the two panets' eccentricities would grow indefinitely. 
However, as discussed above, the disk exerts an eccentricity damping.
This has two effects. 
On the one hand, it stops the eccentricity growth and keeps the planets at a fixed semi-major axes ratio. 
That is, the mutual planet configuration freezes out, as we show in Figure 
\ref{fig:ExampleOfMeanMotionResonanceCapture} for the 3-2 mean motion resonance. 
We discuss how to describe analytically this equilibrium configuration in the the Appendix \ref{subsec:EqInResonantCapture}. 
On the other hand, it breaks the adiabatic approximation. 
The orbit either shrinks towards the equilibrium point, which becomes an attractor, 
or spiral away from the equilibrium, increasing the libration amplitude until it escapes from 
the resonance or reaches a limit cycle (\cite{2014AJ....147...32G}). 
The conditions for one or the other behaviour are quantified in \cite{2015A&A...579A.128D} and 
\cite{2015ApJ...810..119D} as a function of planetary masses, damping 
forces, resonance index $k$. 
We come back to this in the Appendix \ref{subsec:EqInResonantCapture}, 
where we briefly discuss how, for the purposes of this work, we can ensure that the eventual instability would occur 
on very long timescales and by removing the gas early enough we can ignore this complication.

\begin{figure}[!ht]
\centering
\begin{subfigure}[b]{0.32 \textwidth}
\centering
\includegraphics[scale=0.43
]{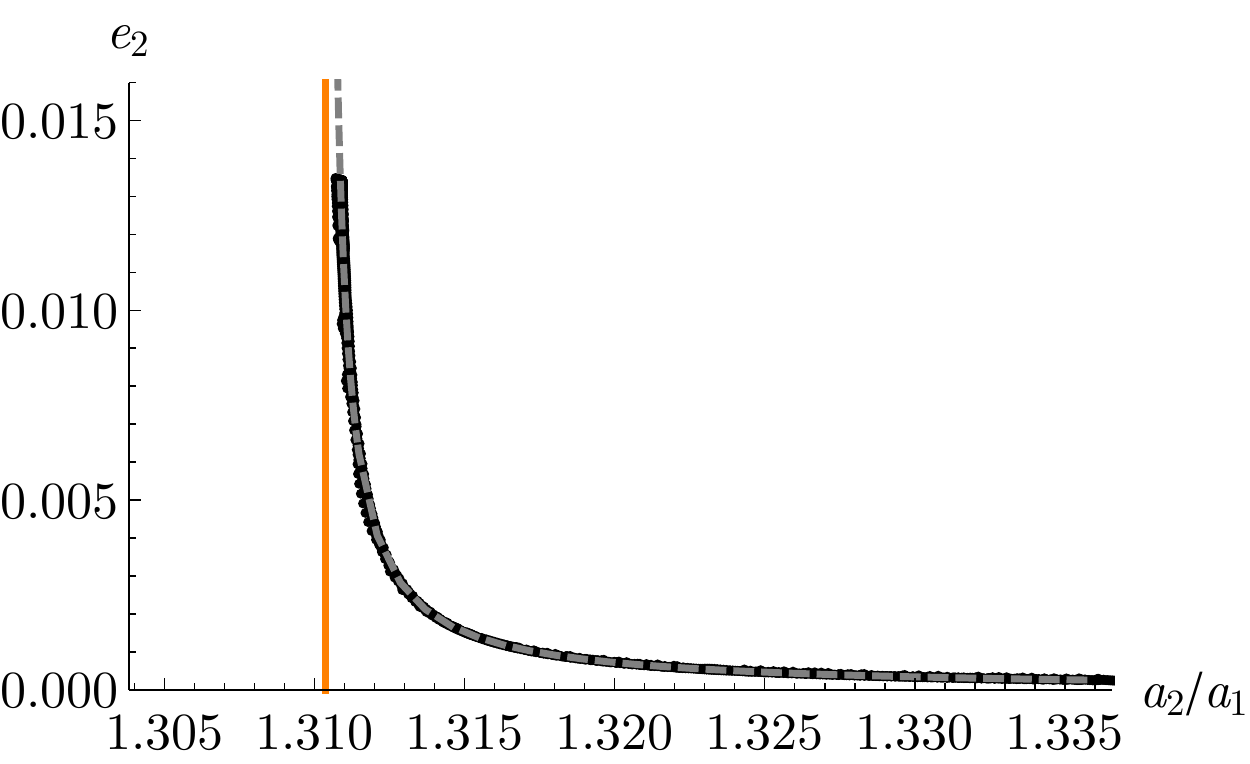}
\caption{$e_2$}
\label{fig:ExampleOfMeanMotionResonanceCapture.subfig:e2}
\end{subfigure}
\begin{subfigure}[b]{0.32 \textwidth}
\centering
\includegraphics[scale=0.43
]{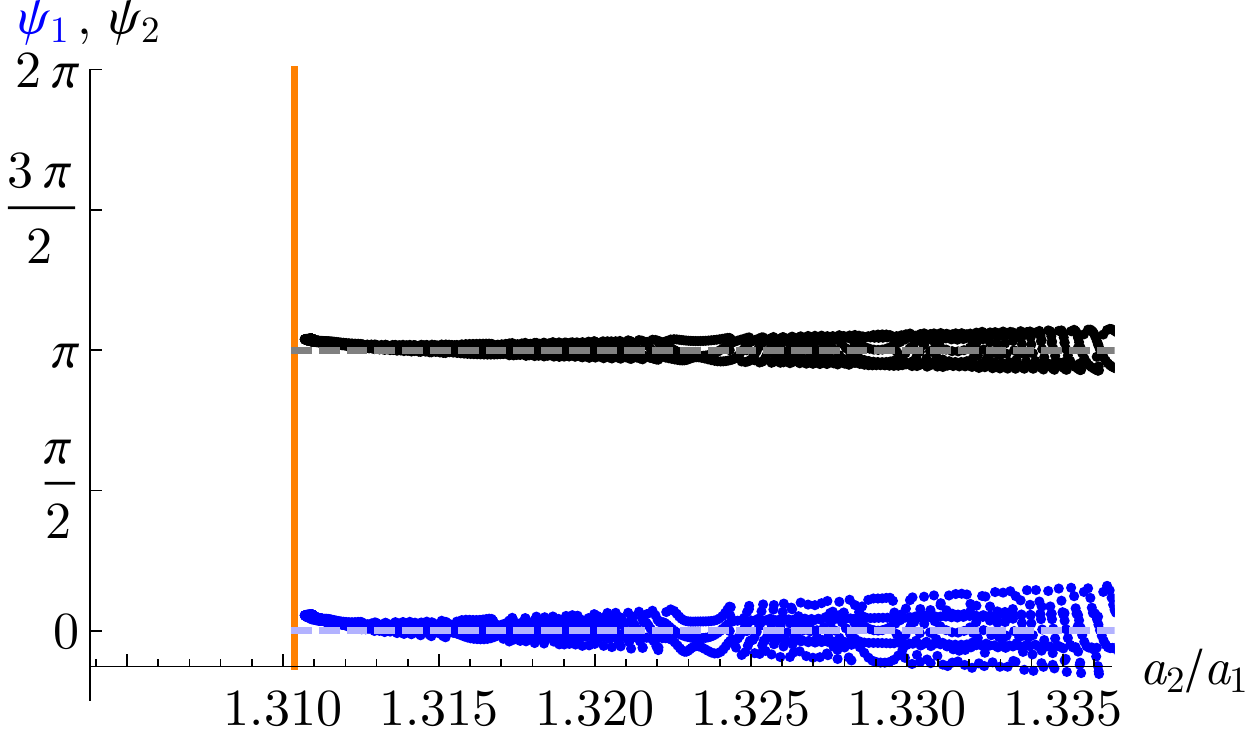}
\caption{Resonant angles $\PSI_i = \THETA+\GAMMA_i$.}
\label{fig:ExampleOfMeanMotionResonanceCapture.subfig:resangles}
\end{subfigure}
\centering
\begin{subfigure}[b]{0.32 \textwidth}
\centering
\includegraphics[scale=0.43
]{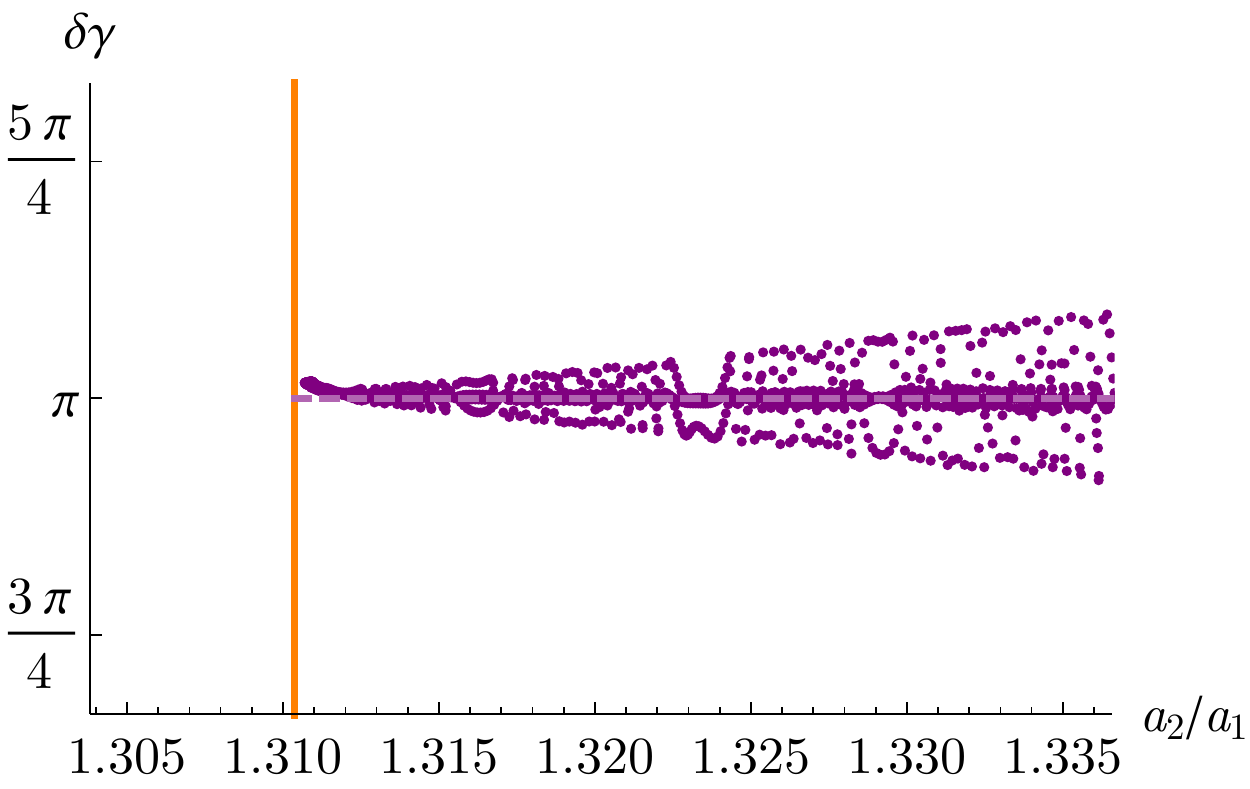}
\caption{Angle $\DELTAGAMMA$.}
\label{fig:ExampleOfMeanMotionResonanceCapture.subfig:deltagamma}
\end{subfigure}
\caption{
Typical evolution of a system during capture into 3-2 mean motion resonance, 
for two planets of equal mass $m_1 = m_2 = m = 10^{-5} M_*$.
All quantities are given as a function of the semi-major axes ratio $a_2/a_1$ in order to compare with the 
analytical calculations carried out in Section \ref{sec:AnalyticalSetup}; to further aid the comparison, we superimposed to all plots the analytical values found using our averaged model with dashed lines.
Note that the initial configuration is at the far right of the plots and with vanishing eccentricity, 
so we are very close to the equilibrium point, i.e.\ in a configuration of small amplitude of libration; 
this property of the system is conserved during its evolution as explained in the text.
In this simulation, both the migration and eccentricity damping effects of the disk on the planets are active, 
so that the system eventually reaches a final configuration of low-amplitude libration 
around an equilibrium point for some value of the angular momentum.
This final configuration is stable, see text for details.
Note that the amplitude of libration of $\PSI_1$, $\PSI_2$ and $\DELTAGAMMA$ shrinks as $a_2/a_1$ decreases. 
This is because initially the eccentricity is very small and therefore even a small oscillation around the equilibrium point 
can cause a large excursion in the angles. 
}
\label{fig:ExampleOfMeanMotionResonanceCapture}
\end{figure}

\section{Limits of stability as a function of planetary mass}\label{sec:MassIncrease}
After the equilibrium configuration is attained, we slowly deplete the gas (that is have $\Sigma$ decrease exponentially in Equation \eqref{eq:twave}). 
This is done slowly enough and the system does not move considerably from the equilibrium configuration calculated 
in the previous Section.
We should only note that the damping in the eccentricities has the effect of changing the equilibrium values of the 
angles $\PSI_1$ and $\DELTAGAMMA$ from the ones which are found in the purely conservative planetary system 
(see e.g.\ \cite{2013AJ....145....1B} for a formula of this shift, linking $\PSI_{1,eq}$, and $\DELTAGAMMA_{eq}$ 
to $\tau_{e}$).
This means that when the latter admits stable symmetric equilibrium points, $\PSI_{1,eq},\DELTAGAMMA_{eq} = 0,\pi$, 
the non-conservative system might seem to contradict this; 
however these are \emph{not} asymmetric equilibrium points, as they are only due to the damping effect: 
when this is removed the system reaches the expected equilibrium values of the angles.

Now that we have an effective method for obtaining numerically a planetary system in mean motion resonance, 
and to describe its properties analytically, 
we intend to investigate its stability.
In particular, we study the stability of pairs of equally massive, $m_1 = m_2 = m$, 
resonant planets by considering their mass as a free parameter. 
We maintain the notation $\mu = m M_*/(m+M_*)$ for the common reduced mass of the planets.

To perform this study, we can take the resonant equilibrium configurations obtained as described in the previous section,  slowly deplete the gas, and then perform long-term integrations with the resulting orbital configuration as initial conditions, checking if the system exits the resonance, in which case the resonant configuration is deemed unstable; this analysis can be then performed for different masses.
One might start with the planets already as massive as desired and repeat the exercise 
of capture in resonance through interaction with a disk of gas and then depletion of the gas (e.g.\ \cite{2012Icar..221..624M}).
However if the region of high amplitude of libration around the equilibrium point is chaotic, 
the capture might not lead to an orbit near the resonant center, 
so that once the gas is removed an instability may develop, 
whereas the orbits might have remained stable if they had had a smaller amplitude of libration.
If instead we take a system of planets deep in resonance and slowly increase their masses 
until the system shows instability, we can ensure that we are indeed probing the region of the phase space 
around the resonant equilibrium point.
For, as long as the rate at which this increase is performed is small enough, the amplitude of libration around the 
equilibrium point will be an adiabatic constant and will be preserved.
For simplicity, we chose a linear law $m(t) = m(0) + M t$, where $M$ is a constant (in practice, for the results shown below, we chose to increase the planetary mass so that it grows by 3 orders of magnitude in $5\times10^4$ years;  
changing $m$ slowly enough, we notice no noticeable difference in the resulting 
evolution if one uses different laws or rates of change for $m(t)$);
in our code, we increase the planetary mass at each integration step.
We should stress here that the increase in the planetary parameter is a purely numerical exercise: 
one should assign no physical meaning to it, and the fact of changing the value of $m$ is just a way to 
explore the stability of deeply resonant systems as a function of planetary masses 
starting from one system that is well in resonance, the configuration of which one can describe analytically. 

Indeed, another advantage of operating this way is that we can follow analytically the evolution of the system 
as the mass increases, at least to a very good approximation. 
To do this, we look at the quantity
\begin{equation}\label{eq:SpecificAngMom}
\ANGMOM_{spec} := \frac{\ANGMOM}{\mu} = 
\frac{m}{\mu}\left(\sqrt{\GravC(m+M_*) a_1 (1-e_1^2)} + \sqrt{\GravC(m+M_*) a_2 (1-e_2^2)}\right),
\end{equation}
which we (improperly) call \emph{specific angular momentum}. 
This quantity is not exactly constant as the planetary mass increases, but its value changes very little up to 
high enough values of $m/M_*$, cfr.\ Figure \ref{fig:3-2MassIncrease.subfig:MvsSpecAngMom}.

In the approximation $\ANGMOM_{spec} = const$, we can follow analytically the evolution of a resonant system in
which the planetary mass parameter $m$ is slowly changing. 
To do this, consider a resonant system in the vicinity of an equilibrium point 
$(\PSIONA_{1,eq}', \PSIONA_{2,eq}', \PSI_{1,eq}', \DELTAGAMMA_{eq}')$ for some value $m'$ of $m$ and some value 
of the integrals of motion $\KAPPONA'$ and $\OMEGONA'$. 
Note now that $\ANGMOM = \frac{m}{\mu}(\KAPPONA + \OMEGONA)$, i.e.\ 
$\ANGMOM_{spec} = \frac{m}{\mu^2}(\KAPPONA + \OMEGONA)$. 
We can then obtain the values of these actions when we change $m$ to $ m''$, by setting 
\begin{equation}
\KAPPONA'' = \frac{m'/(\mu')^2}{m''/(\mu'')^2} \KAPPONA',\quad 
\OMEGONA'' = \frac{m'/(\mu')^2}{m''/(\mu'')^2} \OMEGONA',
\end{equation}
where $\mu'$ and $\mu''$ are the reduced masses relative to the planetary masses $m'$ and $m''$ respectively.
Finally we find the new equilibrium point $(\PSIONA_{1,eq}'', \PSIONA_{2,eq}'', \PSI_{1,eq}'', \DELTAGAMMA_{eq}'')$ 
with the new planetary mass $m''$ and these two actions $\KAPPONA''$ and $\OMEGONA''$ 
in the same manner as in Section \ref{subsec:EquilibriumPointsOfAveragedHa}.
We can then closely follow the evolution of the system as we show in Figure \ref{fig:3-2MassIncrease}, 
where we have superimposed the results 
of a numerical simulation in the case of the 3-2 mean motion resonance and our analytical predictions.
At the same time we plot the real evolution of $\ANGMOM_{spec}$, against the fixed value used for the analytical 
calculations.

\begin{figure}[!ht]
\centering
\begin{subfigure}[b]{0.32 \textwidth}
\centering
\includegraphics[scale=0.43
]{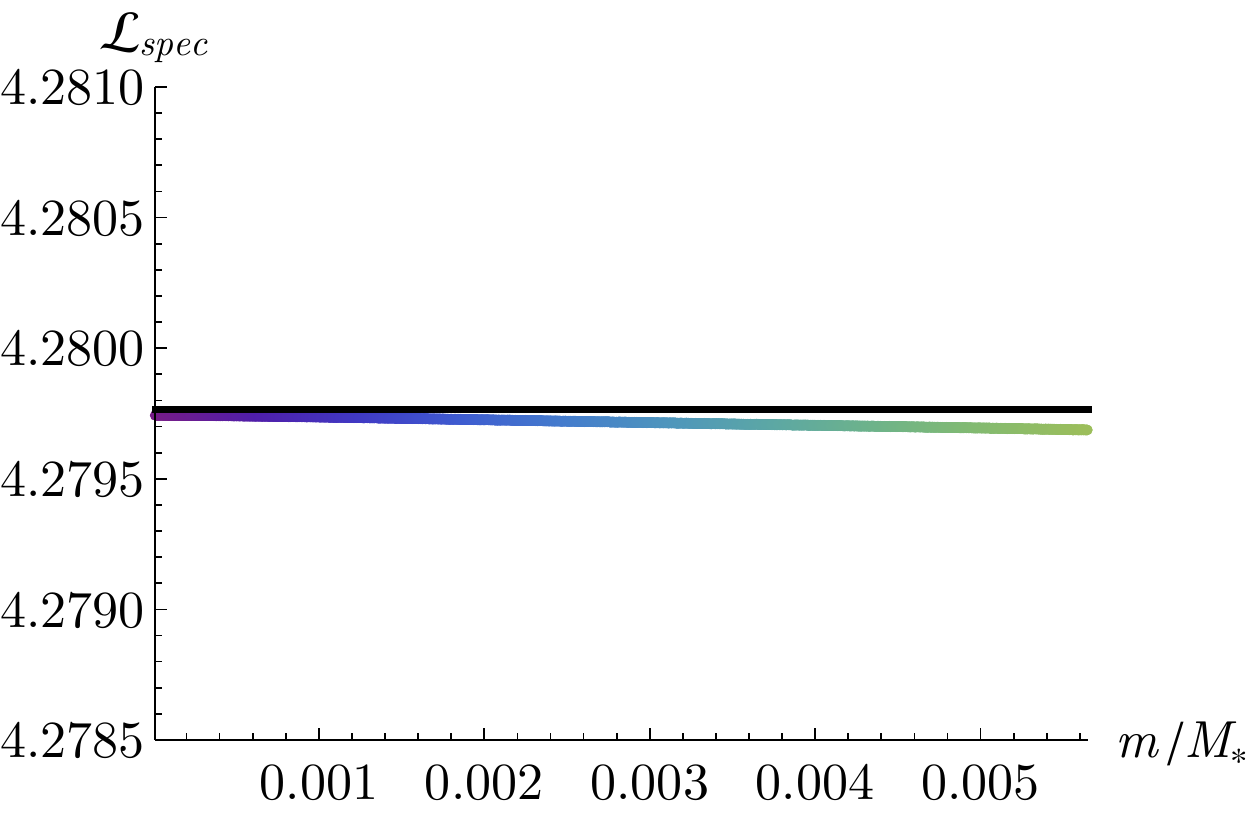}
\caption{$m/M_*$ vs.\ $\ANGMOM_{spec}$}
\label{fig:3-2MassIncrease.subfig:MvsSpecAngMom}
\end{subfigure}
\centering
\begin{subfigure}[b]{0.32 \textwidth}
\centering
\includegraphics[scale=0.43
]{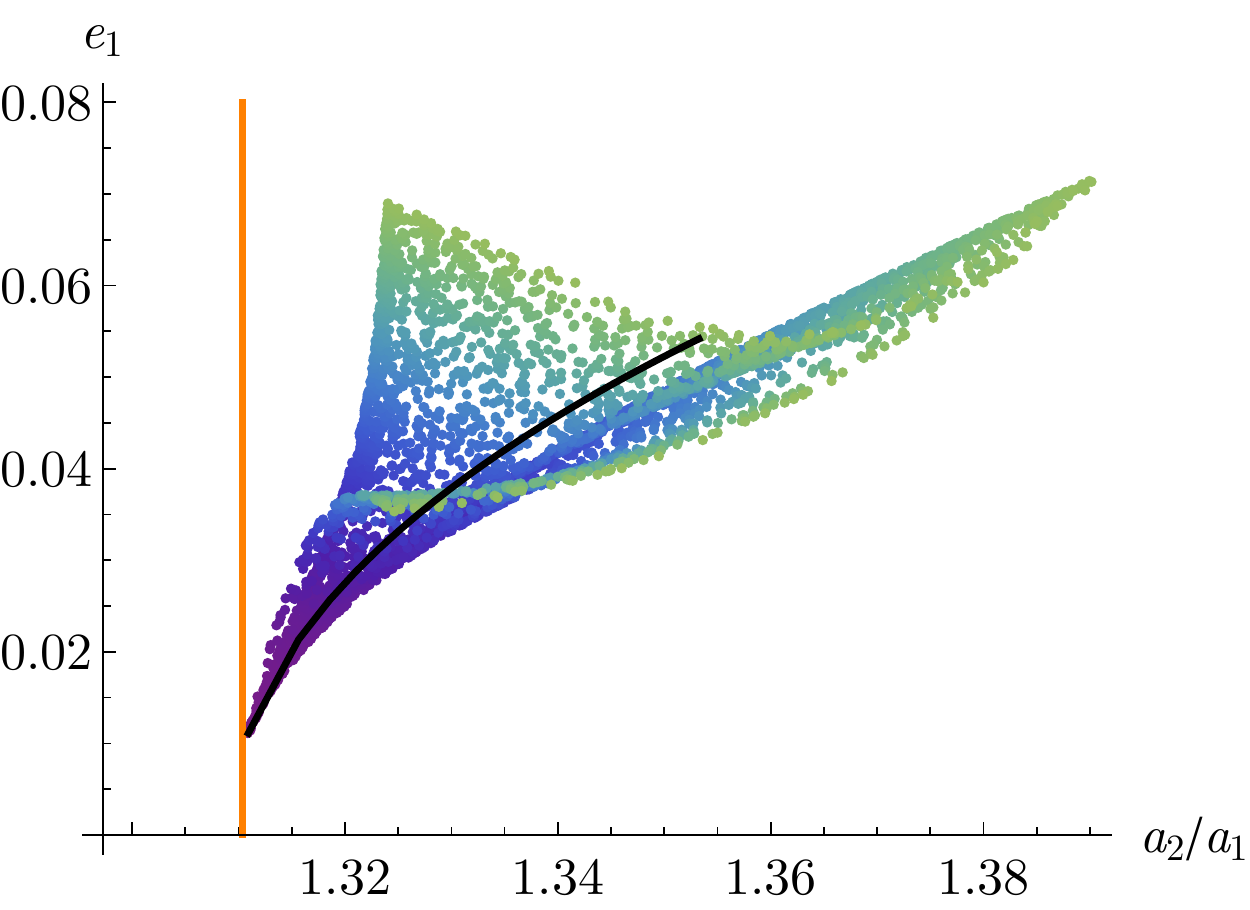}
\caption{$a_2/a_1$ vs.\ $e_1$}
\label{fig:3-2MassIncrease.subfig:a2overa1vse1}
\end{subfigure}
\begin{subfigure}[b]{0.32 \textwidth}
\centering
\includegraphics[scale=0.43
]{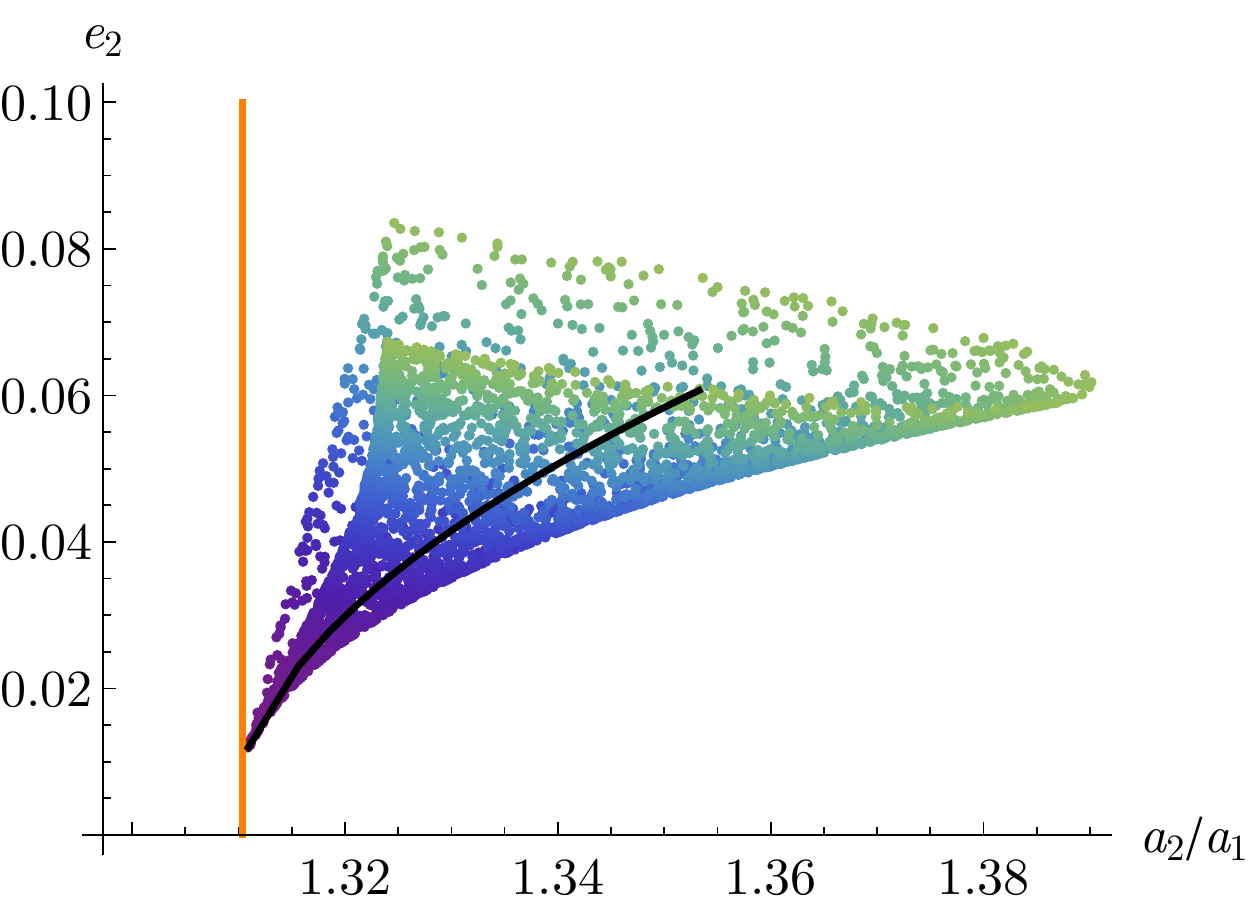}
\caption{$a_2/a_1$ vs.\ $e_2$}
\label{fig:3-2MassIncrease.subfig:a2overa1vse2}
\end{subfigure}

\begin{subfigure}[b]{0.32 \textwidth}
\centering
\includegraphics[scale=0.43
]{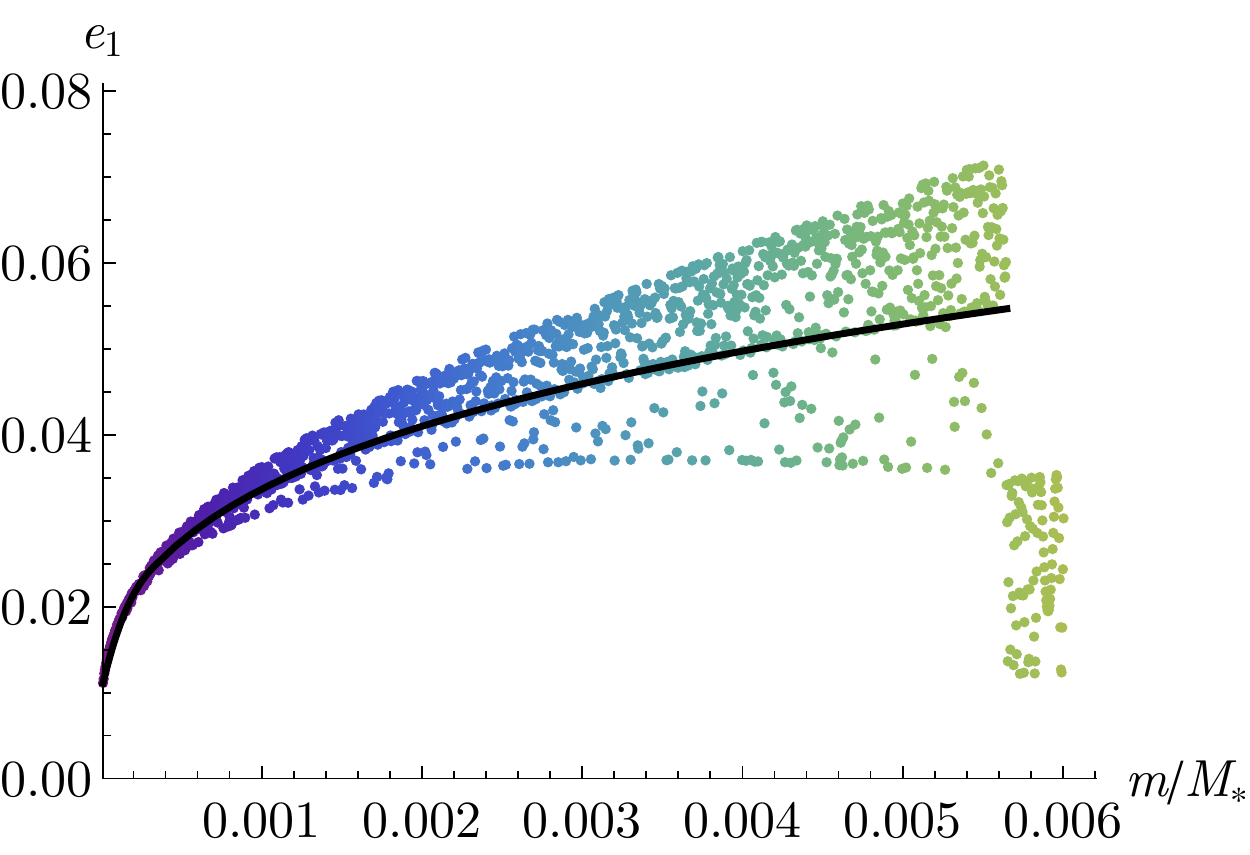}
\caption{$e_1$ vs.\ $m/M_*$}
\label{fig:3-2MassIncrease.subfig:e1vsMploverM0}
\end{subfigure}
\begin{subfigure}[b]{0.32 \textwidth}
\centering
\includegraphics[scale=0.43
]{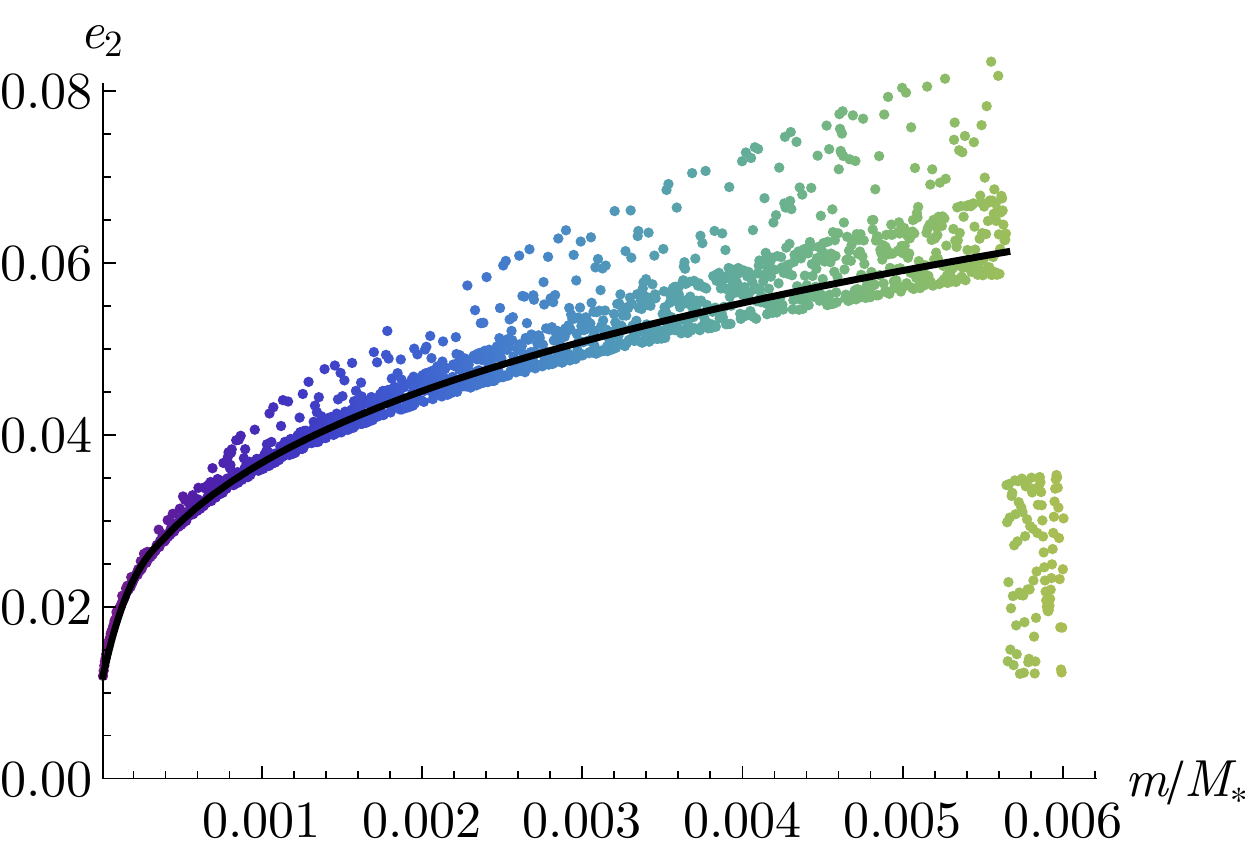}
\caption{$e_2$ vs.\ $m/M_*$}
\label{fig:3-2MassIncrease.subfig:e2vsMploverM0}
\end{subfigure}
\begin{subfigure}[b]{0.32 \textwidth}
\centering
\includegraphics[scale=0.43
]{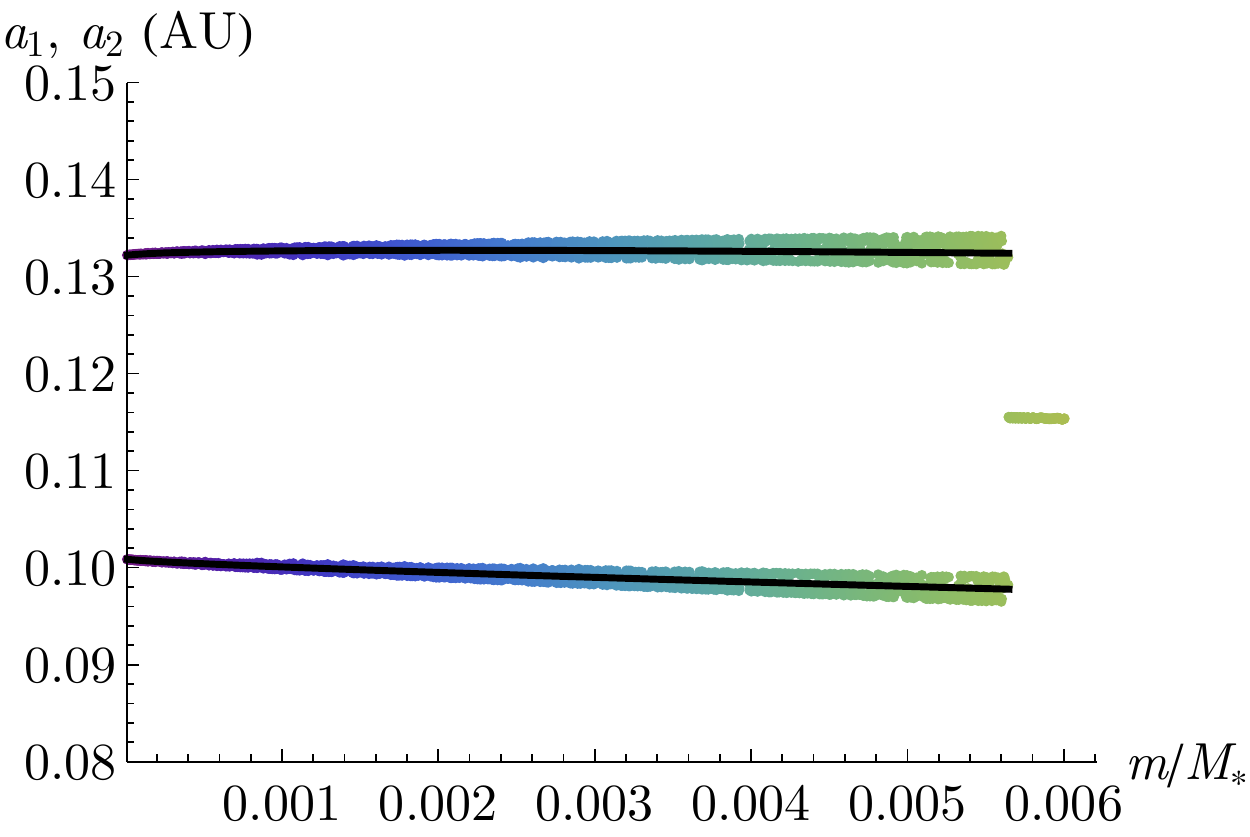}
\caption{$a_1$, $a_2$ vs.\ $m/M_*$}
\label{fig:3-2MassIncrease.subfig:a1anda2vsMploverM0}
\end{subfigure}

\caption{
Evolution of a system deep in the 3-2 mean motion resonance as the planetary mass $m_1 = m_ 2 = m$ increases.
The initial configuration of the averaged system is
$a_2/a_1 = 1.31093$, $e_1 = 0.01112 $, $e_2 = 0.01195$ and $m/M_* = 1 \times 10^{-5}$. 
The true evolution of $\ANGMOM_{spec}$ along the simulation is plotted in panel (a) as a function of $m/M_*$, 
see the coloured line (the colour-coding is reproduced only to indicate the value of $m$ in panels (b) and (c)).
The black line represents the approximation $\ANGMOM_{spec} = const$ used in the analytical calculations, 
showing relatively good agreement up to high values of $m/M_*$.
The plot is interrupted at $m/M_* \simeq 5.64 \times 10^{-3}$, at which point the system goes unstable. 
In panels (b) and (c) we plot both eccentricities as a function of the semi-major axes ratio, 
as they evolve while $m$ increases.
We colour-code the points based on the value of the planetary mass 
(with the same colours used in panel (a)).
We superimpose, with a black line, the result of an analytical calculation aimed at reproducing the evolution 
of the system as explained in the text, assuming $\ANGMOM_{spec} = const$.
Note that the simulation follows closely the analytical prediction. 
The oscillations around the equilibrium points become larger and larger as $m$ increases, 
but they are short periodic ones, i.e.\ they are due to the evolution of the fast angles (the same as those shown in Figures \ref{fig:eExcitedSystemIn3-2Resonance-Frequencies.subfig:e1ShortPeriod} and \ref{fig:aExcitedSystemIn3-2Resonance-Frequencies.subfig:e1ShortPeriod}) which are averaged out in the 
analytical model and are not linked to a growth in the amplitude of resonant libration, 
which is conserved adiabatically until the system becomes unstable.
Panels (d), (e) and (f) show the evolution of the orbital elements as the mass increases, with again a black line being the result of analytical calculations; since we imposed a linear increase of the mass with time, this can be seen as an evolution in time. Notice that in this case the outcome of the instability is a collision, as the two planets eventually merge.
}
\label{fig:3-2MassIncrease}
\end{figure}
At this point, a remark is in order. 
The eccentricity of the equilibrium configuration\footnote{
We are of course referring to the eccentricity in the averaged system, 
in the full one $e$ would oscillate due to the fast 
evolving angles.} 
grows with the planetary masses, as shown in Figure \ref{fig:3-2MassIncrease}, following roughly a line of constant specific angular momentum. 
Instead, the equilibrium eccentricity of planets captured in resonance by planet-disk interaction is independent of the planetary mass (see equation \eqref{eq:e2EquilibriumWithTrap}). 
This means that capturing planets in the resonance with a mass $m'$ or capturing them with a smaller mass $m''$, which is then grown to $m'$ after capture, leads to two different configurations.  
In other words, the two processes of a) first capturing the planets in mean motion resonance and then increasing 
their masses, and b) first increasing their masses and then putting them in resonance, do not commute.
Nevertheless, by assuming different scale-heights of the disk when the planets are captured and then growing the planetary masses, we can explore numerically the full $m$, $e_2$ parameter space characterising the resonant equilibrium. 
We will take initial values of $m$ ranging from $10^{-5} M_*$ to $10^{-4} M_*$, 
and initial values of the eccentricities up to $\sim 0.2$. 
Higher values of $e$ are physically unrealistic as $e_{eq,2} \propto h$ (cfr.\ equation \eqref{eq:e2EquilibriumWithTrap}) 
and disks with high aspect ratios are not expected.

The coloured dots in Figure \ref{fig:FreqWithSimulations} and \ref{fig:MinimalDistanceWithSimulations} 
show the evolution of $e_2$ as the planetary mass grows, 
starting from different initial values, for systems in the 3-2 mean motion resonance. 
We let the masses grow until an instability occurs. 
Denoting by $m_{crit}$ the mass at which the discontinuity happens, we do a 
long-term simulation, over $3\times 10^{7}$ revolutions of the inner planets, with a fixed mass $m= 0.995\times m_{crit}$ to check that the dynamics was still stable up to that point. 
Simulations with higher planetary masses go unstable immediately, after $\simeq 125$ revolutions of the inner planet. This is shown in Figure \ref{fig:StabilityVsInstability}.

\begin{figure}[!ht]
\centering
\begin{subfigure}[b]{0.45 \textwidth}
\centering
\includegraphics[scale=0.43
]{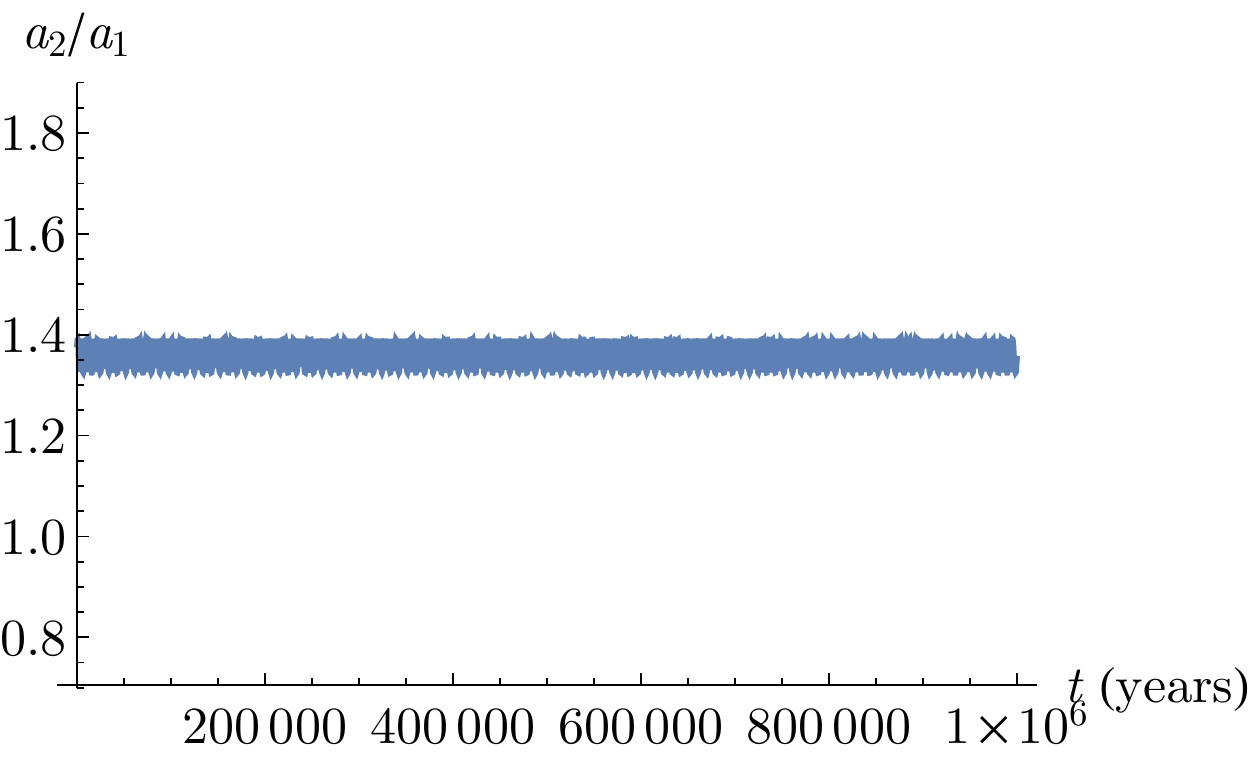}
\caption{$m = 0.995~m_{crit}$.}
\label{fig:StabilityVsInstability.subfig:MplsmllrMcrit}
\end{subfigure}
\centering
\begin{subfigure}[b]{0.45 \textwidth}
\centering
\includegraphics[scale=0.43
]{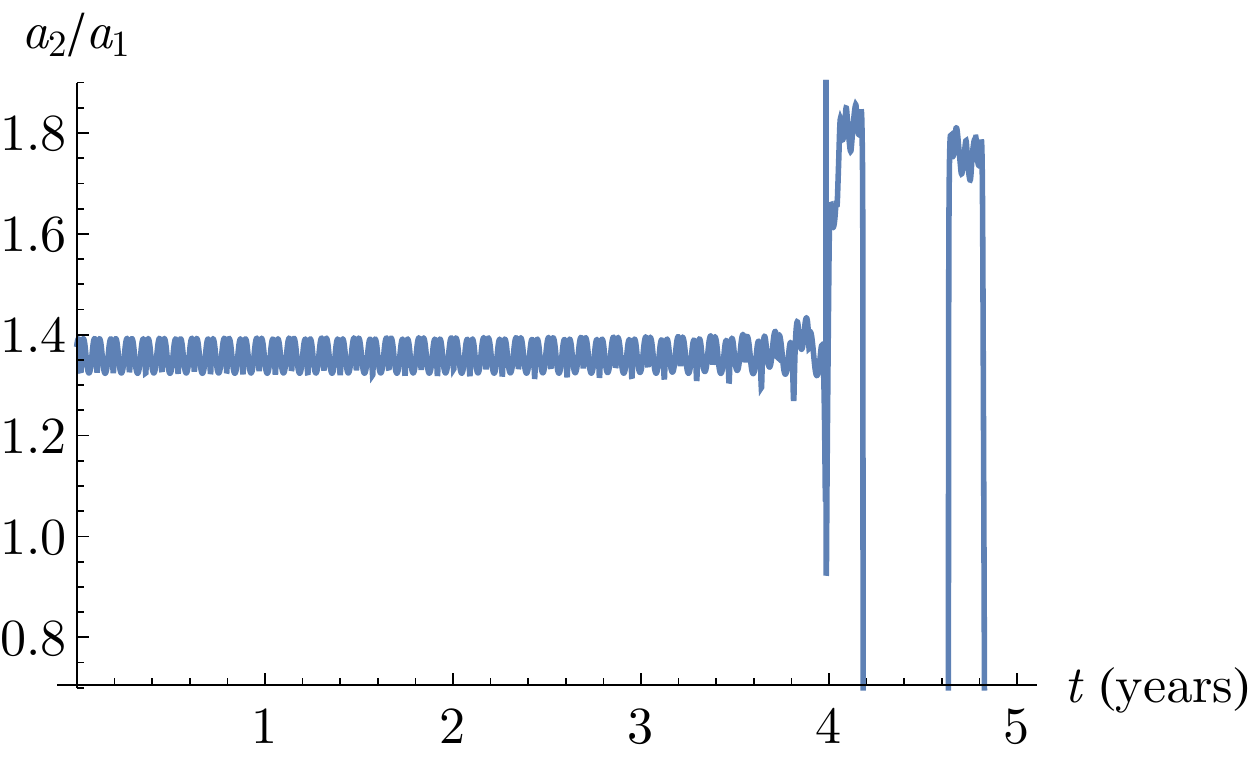}
\caption{$m = m_{crit}$.}
\label{fig:StabilityVsInstability.subfig:MplgrtrMcrit}
\end{subfigure}

\caption{
Two different evolutions of semi-major axis ratio $a_2/a_1$ for a system which starts in a 3-2 mean motion resonance, where the planetary masses $m_1=m_2=m$ are kept constant. In panel (a) $m =0.995~m_{crit}$ and the system shows long-time stable quasi-periodic evolution; in panel (b), $m = m_{crit}$ and the system immediately proves unstable (note that the timescales reported on the horizontal axes differ by 6 orders of magnitude).
}
\label{fig:StabilityVsInstability}
\end{figure}

We test two possible origins of the instability of the system. 
The first one is that of a low-order secondary resonance between the frequency of libration of the resonant angles 
and that of the synodic angle $\LAMBDA_1-\LAMBDA_2$,
which has the most noticeable effect on the faster, short period dynamics of the system 
(cfr.\ Figures \ref{fig:aExcitedSystemIn3-2Resonance-Frequencies.subfig:a1ShortPeriod}, 
\ref{fig:eExcitedSystemIn3-2Resonance-Frequencies.subfig:a1ShortPeriod}).
Note that as the planetary mass increases the frequency $\omega_{\LAMBDA_1-\LAMBDA_2}$ of 
$\LAMBDA_1-\LAMBDA_2$
does not change considerably, as it is fixed by the resonance index $k$, 
the location of the planetary system and stellar mass; 
only the amplitude of this frequency grows with $m$. 
This is visible already in Figure \ref{fig:3-2MassIncrease} and shown again in Figure \ref{fig:FreqWithSimulations}.
Instead, the frequency of libration around the equilibrium point increases with $m$, 
so that for high enough planetary masses it can reach a $l$-$(l-1)$ 
resonance with $\omega_{\LAMBDA_1-\LAMBDA_2}$,
where $l\geq 2$ is an integer, and this might destabilise the system.
To check this first hypothesis we build a map of the libration frequencies as a function of 
the planetary mass and the eccentricity.
To do this, we first fix a planetary mass $m$ and obtain equilibrium points for different values of the constant of motion 
$\OMEGONA$, as detailed in Section \ref{subsec:EquilibriumPointsOfAveragedHa}, 
and for each point we calculate the frequencies of libration $\omega_1$, 
$\omega_2$ as explained in Section \ref{subsec:FrequenciesOfLibration}; 
finally we change the value of $m$.
When we do this, the value of $\KAPPONA$ is adjusted in order to keep fixed the location of the exact 
Keplerian resonance.
For e.g.\ the 3-2 mean motion resonance, for each fixed value of $m$, each equilibrium point is univocally 
characterised by the value of $e_2$, so we can write $\omega_i (m,e_2)$, $i=1,2$. 
We then compare it with the frequency $\omega_{\LAMBDA_1-\LAMBDA_2}$ of the synodic angle $\LAMBDA_1-\LAMBDA_2$.
We show in Figure \ref{fig:FreqWithSimulations} a contour plot of $\omega_1$ and $\omega_2$ as the background 
of the aforementioned numerical simulations.
Since $\omega_1>\omega_2$, as we saw in Section \ref{subsec:FrequenciesOfLibration}, we can focus on 
secondary resonances between $\omega_1$ and $\omega_{\LAMBDA_1-\LAMBDA_2}$.
We notice that some systems become unstable before the frequency $\omega_1$ reaches the 2-1 resonance with 
$\omega_{\LAMBDA_1-\LAMBDA_2}$, while others pass through this low-order secondary resonance unaffected.
We therefore conclude that these secondary resonances do not play a role in the dynamics of the system, 
at least at small libration amplitudes.

\begin{figure}[!ht]
\centering
\begin{subfigure}[b]{0.45 \textwidth}
\centering
\includegraphics[scale=0.7
]{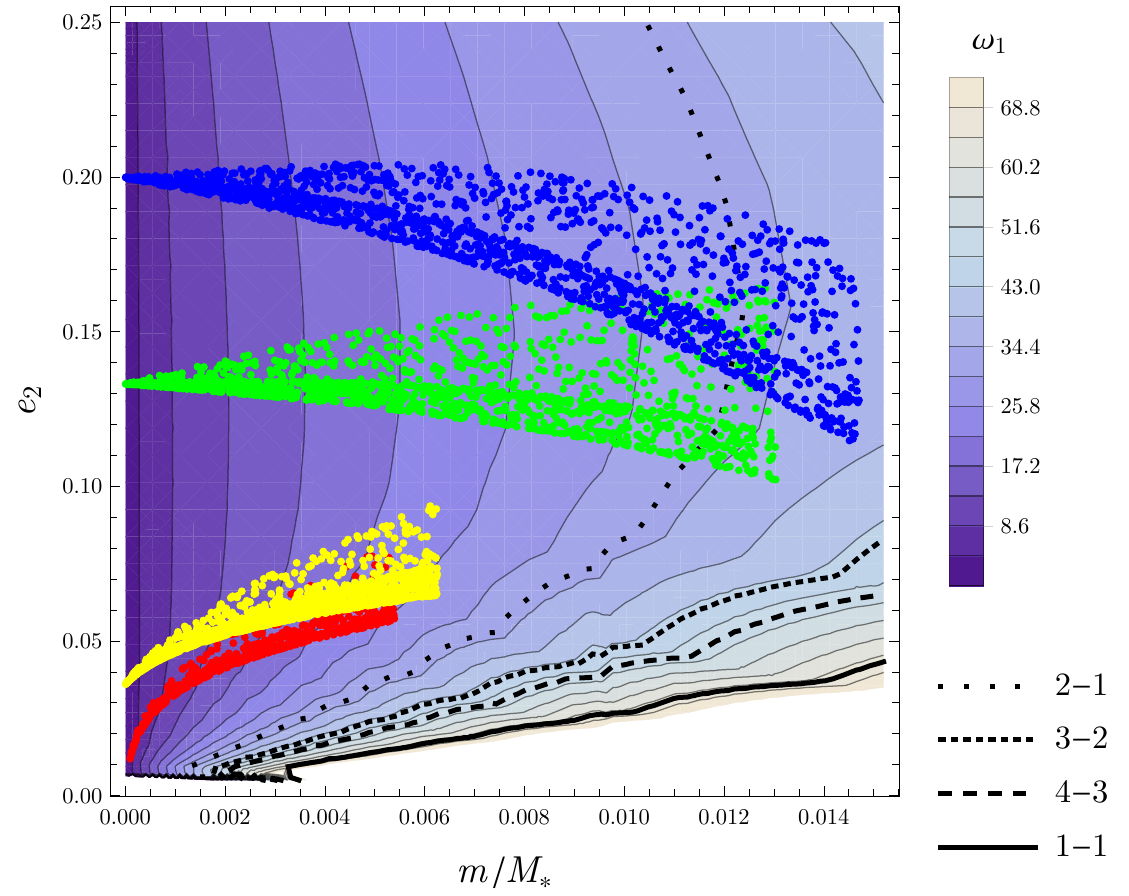}
\caption{$\omega_1$.}
\label{fig:FreqWithSimulations.subfig:omega1}
\end{subfigure}
\centering
\begin{subfigure}[b]{0.45 \textwidth}
\centering
\includegraphics[scale=0.7
]{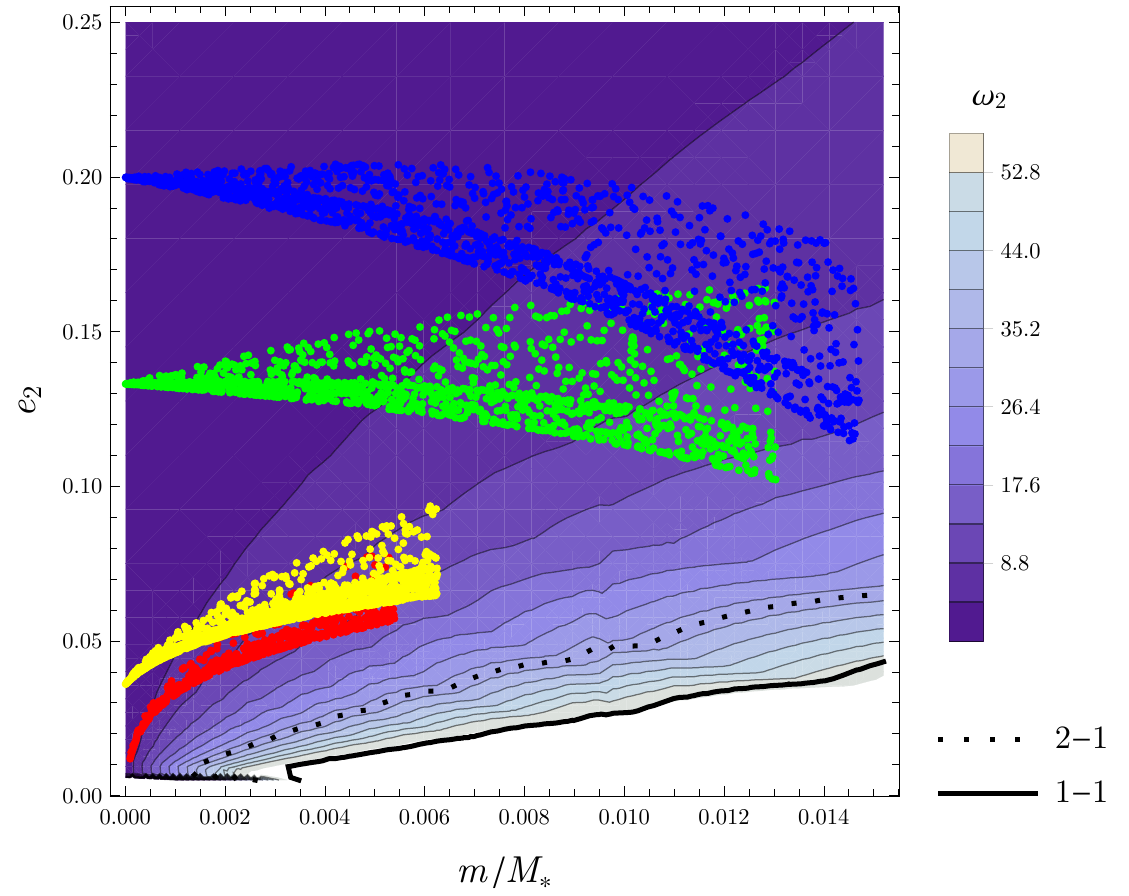}
\caption{$\omega_2$.}
\label{fig:FreqWithSimulations.subfig:omega2}
\end{subfigure}

\caption{
We show contour plots for the frequencies $\omega_1$, panel (a) and $\omega_2$, panel (b), 
as functions of the planetary mass $m$ and the eccentricity $e_2$ in the case of the 3-2 mean motion resonance.
Lighter colours indicate a higher value of the frequencies.
The black lines indicate respectively a 2-1, 3-2, 4-3 and 1-1 resonance with the 
fast synodic frequency $\omega_{\LAMBDA_1-\LAMBDA_2}$, see the legend.
As expected, the frequency $\omega_2$ is smaller than $\omega_1$, except for extremely small values of $e$; 
we can therefore concentrate on $\omega_1$, panel (a), when looking for secondary resonances in the system.
The dots represent the result of the numerical simulations carried out as explained in the text.
The simulations are interrupted when the system becomes unstable.
Note that the red and yellow simulations go unstable even before encountering the first resonance with $\omega_{\LAMBDA_1-\LAMBDA_2}$ (green line).
Instead, the green and blue simulations cross it undisturbed and do not go unstable near any particular resonance.
}
\label{fig:FreqWithSimulations}
\end{figure}

The second hypothesis for the onset of instability is inspired by the criterion of minimal distance between the planets, first proposed by \cite{1993Icar..106..247G}, then revised (see e.g.\ \cite{2017Icar..293...52O}).
These studies show that two non-resonant planets go unstable when their orbital configuration is such that 
they come closer to each other than a critical distance 
\begin{equation}\label{eq:mutualHillRadiusCriterion}
d_{crit} = 2\sqrt{3} r_{H}, 
\end{equation}
where
\begin{equation}
r_H = \left(\frac{m_1 + m_2}{3 M_*}\right)^{1/3} \frac{a_1+a_2}{2}
\end{equation}
is the mutual Hill radius. 
Note that for non-resonant configurations, the closest distance of approach between the two planets coincides with the orbital distance, but for resonant ones this is not the case. Therefore in this case we consider applying Gladman's criterion not to the orbital distance, but to the actual closest approach of the two planets during the evolution in the resonant configuration. 
We can estimate this closest distance analytically as follows. 
As before we find for a fixed planetary mass and a fixed value of $\OMEGONA$ an equilibrium point, which can again 
be identified in terms of the eccentricity. 
We then evaluate the real minimal distance $d$ of the two planets in such orbital configuration, 
by sampling the distance between the planets at different values of $\LAMBDA_1$ in $[0, 2 k \pi]$, 
(recall that the full Hamiltonian is periodic in this angle with period $2 k \pi$), 
and taking the minimum of these distances.
We thus plot in the background of Figure \ref{fig:MinimalDistanceWithSimulations} the value $d/d_{crit}$ 
in the case of the 3-2 resonance, 
as a function again of $m$ and $e_2$, where $d_{crit}$ is given by \eqref{eq:mutualHillRadiusCriterion}.
In Figure \ref{fig:MinimalDistanceWithSimulations.subfig:unexcited} we superimpose the same numerical simulations
as in Figure \ref{fig:FreqWithSimulations}.
We observe that the planetary systems reach the critical distance $d_{crit}$ (black dashed line) 
without displaying any instability.
Therefore, we see that resonant are more stable against close encounters than systems with randomly chosen 
angular parameters.
It is well known that given values of $a_1$, $e_1$, $a_2$, $e_2$ a pair of resonant planets has a minimum approach 
distance which is larger than if the planets are not in resonance. 
Here we show, in our knowledge for the first time, that the center of a resonance is more stable given an actual 
minimum approach distance (not an orbital distance), than a non-resonant configuration.
In fact, we notice that the instability occurs when the planets reach an analytically estimated closest distance $d$ such that $d/d_{crit} \simeq 0.78$, see the black continuous line.
We should note however that as the planetary masses increase and the planets reach $d/d_{crit}\sim1$, one is approaching a singularity (a collision) so that the remainder of the averaged Hamiltonian grows (e.g., \cite{2017CeMDA.128..383P}), meaning that the closest approach calculated along the trajectories of the averaged model might be incorrect. However we checked against the actual minimal approach distance that is obtained along a simulation and we saw that at $d/d_{crit}\simeq1$ the analytically calculated distance is slightly bigger than the real one but correct within an error of $\sim3\%$, and even close to the instability, i.e.\ for $d<d_{crit}$ but $m \lesssim m_{crit}$, it is again slightly bigger than the real minimal distance but correct within a $\sim6\%$ error. The actual minimal distance at which planets in a 3-2 mean motion resonance go unstable is therefore $d/d_{crit} \simeq 0.74$; this is slightly smaller than the number obtained analytically and well smaller than 1.
We repeated the analysis for the 4-3 resonance, and we find that the instability occurs when $d/d_{crit} \simeq 0.6$.
We also run simulations where we took systems initially deep in resonance and slightly excited their amplitude of 
libration of the resonant angles, as we did in Section \ref{subsec:FrequenciesOfLibration}.
With these systems, we repeat the numerical exercise of increasing the planetary mass, see the resulting evolutions for two of them in Figure \ref{fig:MinimalDistanceWithSimulations.subfig:excited}. 
We see that the instabilities occur now closer and closer to the usual criterion, where the $d/d_{crit} = 1$.
This indicates that as the mass increases 
the stable region of stability around the equilibrium point shrinks. 
We further test this explanation by taking a system that is deep in resonance, with low amplitude of libration 
of the resonant angles, and with a mass that is just below the critical mass $m_{crit}$.
Recall that such a system was long-time stable.
We then perturb the system to sightly excite the amplitude of libration, as explained before.
We see that the system immediately goes unstable after $\sim 150$ revolutions of the inner planet, 
indicating that at values of $m \sim m_{crit}$ 
the whole stable region of stability has shrunk to the equilibrium point itself. 
This behaviour is similar to what is shown in Figure \ref{fig:StabilityVsInstability}. The sharp transition between stability and a short instability timescale should not surprise. In a planar model, the closest approach distance is achieved very soon. This is true for both the stable and the unstable case. The difference is that in the first case the closest approach distance is large enough not to destabilise the orbit. The closest encounters can then repeat every few years, but the orbit remains stable forever. In the second case, instead, the first closest encounter makes the semi major axes of the planets jump out of resonance.\\

\begin{figure}[!ht]
\centering
\begin{subfigure}[b]{0.48 \textwidth}
\centering
\includegraphics[scale=0.7
]{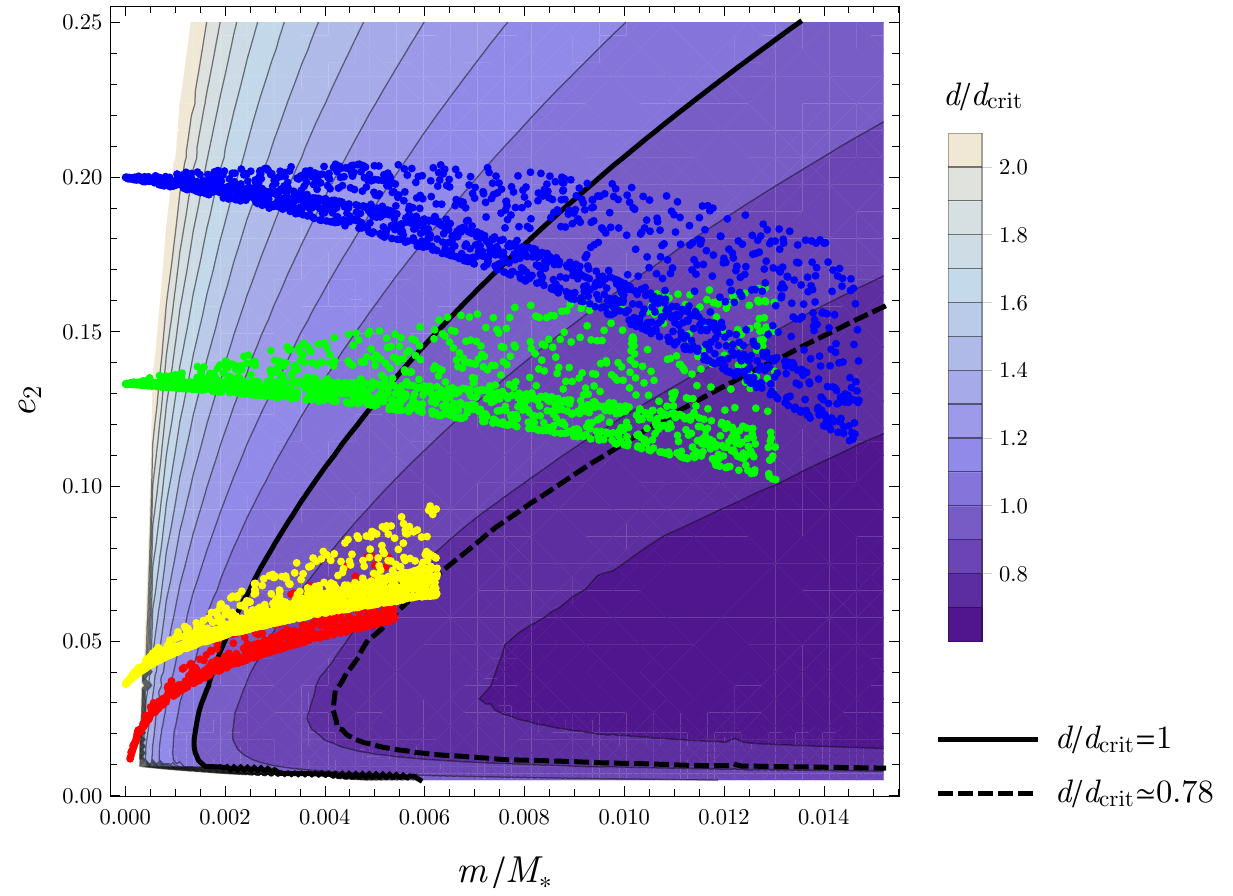}
\caption{Small libration amplitude.}
\label{fig:MinimalDistanceWithSimulations.subfig:unexcited}
\end{subfigure}
\centering
\begin{subfigure}[b]{0.48 \textwidth}
\centering
\includegraphics[scale=0.7
]{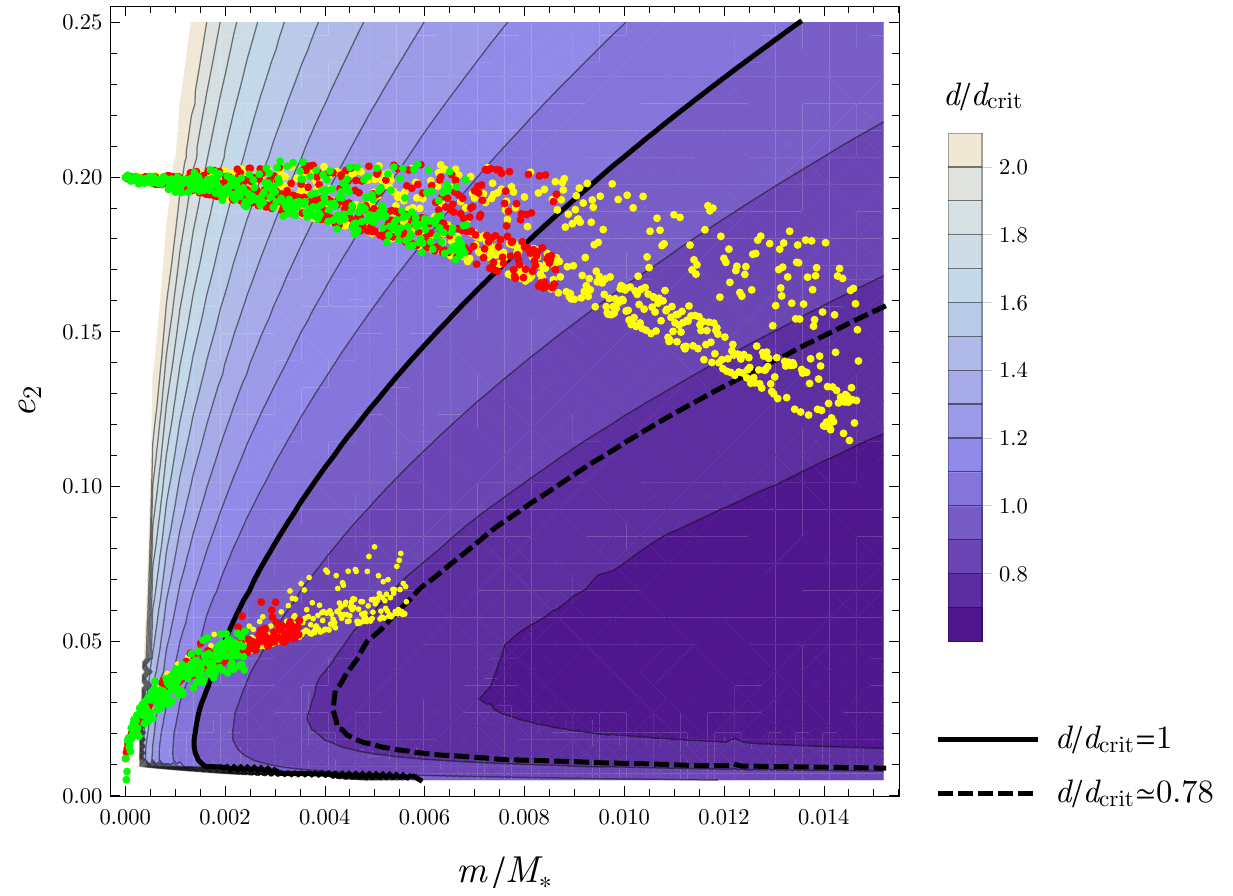}
\caption{Different libration amplitudes.}
\label{fig:MinimalDistanceWithSimulations.subfig:excited}
\end{subfigure}

\caption{
Contour plots of $d/d_{crit}$, where $d$ is the minimal distance between the planets on their orbits around the star, 
and $d_{crit}$ is the critical distance defined in \eqref{eq:mutualHillRadiusCriterion} in terms of the mutual Hill radius.
In both panels, lighter colours indicate higher values of $d/d_{crit}$, and 
the dashed black line indicates the level $d/d_{crit} = 1$.
In panel (a) we superimpose the same numerical simulations as in Figure \ref{fig:FreqWithSimulations}.
We show that the systems cross this line undisturbed and reach configurations where $d/d_{crit} < 1$.
We note however that the instability occurs roughly at the same level, indicated by a solid black line, and corresponding 
to $d/d_{crit} \simeq 0.78$.
In panel (b), se choose two simulations, but we also use initial conditions where we have excited 
the amplitude of libration of the resonant angles.
In both cases, the yellow dots indicate the unexcited case, and the red and green dots show increasing excitation 
of the libration.
We see that with higher degrees of excitation the instabilities occur closer and closer to the usual condition 
$d = d_{crit}$.
As the area enclosed by the libration around the equilibrium is an adiabatic constant with respect to the slowly changing 
parameter $m$, as soon as the stable region becomes too small when increasing the planetary mass 
the system exits the resonance.
}
\label{fig:MinimalDistanceWithSimulations}
\end{figure}

\section{Summary}\label{sec:Conclusions}
In this work, we investigated the dynamics of resonant planetary system, from the capture in mean motion resonance 
via convergent migration in a protoplanetary disk, to the stability of systems with low-amplitude libration of the 
resonant angles.
We treat the simple case of the planar three-body problem, with two equally massive planets.

We present the analytical techniques needed to describe the system in Section \ref{sec:AnalyticalSetup}.
There, we develop the theory for unexpanded Hamiltonians and find semi-analytically the equilibrium points; we validate numerically the analytical calculations, showing perfect agreement.
We compare these with equilibrium points resulting from low-order expansions in the 
eccentricities, showing that the latter they do not capture qualitatively or quantitatively the simulations. 
Since we are interested in the dynamics in the region of the phase space around the equilibrium points, 
we calculate the frequencies of librations in the regime of vanishing amplitude of libration, 
and check again the results with numerical simulations.

In Section \ref{sec:ConvergentInwardMigration} we describe the forces which result from the interactions between the planets and a disk of gas, which is used in the numerical simulations in order to capture the planets in first-order mean motion resonance. 
These interactions include a damping in the eccentricity and a torque which results in an inward Type-I migration.
To ensure convergent migration and resonant trapping, a planetary trap (\cite{2006ApJ...642..478M}) is implemented at the edge of the disk of gas.
These dissipative forces are implemented in our code using simple analytical formul\ae\ which simulate the disk-planet interactions of real hydro-dynamical simulations (e.g.\ \cite{2006A&A...450..833C}).
We present in the Appendix \ref{subsec:EqInResonantCapture} an analytical description of the capture in mean motion resonance following a general approach, and derive formul\ae\ to calculate analytically the final equilibrium configuration. We compare our formul\ae\ with similar ones from previous works, and validate our results with numerical simulations, showing perfect agreement.

In Section \ref{sec:MassIncrease} we investigate the stability of resonant systems at low amplitude of libration. We describe our numerical experiments where we fictitiously increase 
the planetary mass to follow the low-amplitude regime until the onset of instability. 
At the same time, we detail how one can follow analytically the evolution of the system to a good approximation up to 
high value of the planetary masses.
We test against two possible reasons for instability. 
The first is that of a secondary low-order resonance between the frequency of libration of a resonant angle 
(which grows with the planetary mass, while maintaining adiabatically the same amplitude around the equilibrium point) 
and the frequency of the fast synodic angle $\LAMBDA_1-\LAMBDA_2$ 
(which is constant with the planetary mass), 
where $\LAMBDA_i$ is the mean longitude of a planet.
We construct a map of the frequency of libration of the resonant angles as a function of the planetary mass and the 
eccentricity, and compare the calculated values with the frequency of the synodic angle.
We see that some systems become unstable before reaching the 2-1 resonance between the libration and the synodic 
frequency, while others cross it unaffected. 
We therefore conclude that these secondary resonances do not play a significant role in the instability of 
pairs of planets in first order mean motion resonance at low amplitude of libration.
The second hypothesis is that of instability due to close encounters between planets, 
inspired by the mutual Hill radius stability criterion \cite{1993Icar..106..247G}.
In this case we build a map of the minimal distance reached by the planets in their orbital configuration as a function of 
the mass and the eccentricity.
We see that resonant planetary systems are more stable than those with randomly chosen orbital parameters, 
as they can reach a minimal distance that is smaller then the critical distance $d_{crit} = 2 \sqrt{3} r_H$, where 
$r_H$ is the mutual Hill radius, which is the usual critical distance below which two planets go unstable (cfr.\ \cite{1993Icar..106..247G}, \cite{2017Icar..293...52O}).
We find nonetheless that there is a critical distance after which the system goes unstable, 
which is a fraction of the usual $d_{crit}$. 
We see that for systems with bigger amplitude of libration of the resonant angles this critical distance approaches 
more and more the usual $d_{crit}$.
This indicates that the region of stability around the equilibrium point shrinks as the mass increases,  
until the point itself becomes unstable and the system exits the resonance.


\newcommand{\aap}{A\&A}
\newcommand{\aj}{AJ}
\newcommand{\apj}{ApJ}
\newcommand{\apjs}{ApJS}
\newcommand{\mnras}{MNRAS}
\newcommand{\icarus}{ICARUS}
\newcommand{\apss}{APSS}

\begin{appendices}
\section{Disk-planets interactions and evolution in mean motion resonance}\label{subsec:EqInResonantCapture}
The value of the equilibrium eccentricity (and hence the semi-major axes ratio $a_2/a_1$) for two planets
embedded in a protoplanetary disk in the phase of resonant orbital configuration has been computed 
in a number of works (e.g.\ \cite{2005MNRAS.363..153P}, \cite{2008A&A...483..325C}, \cite{2014AJ....147...32G}). 
We propose here a different formulation, consistent with the Hamiltonian resonant description provided in 
Section \ref{sec:AnalyticalSetup} and the adiabatic principle.

For simplicity, we first discuss the case in which the gas only interacts with the outer planet, 
and finally we add the condition that there is a planetary trap at the disk edge so that the inner planet stops migrating.
The first assumption does not lead to any loss in generality, as the main idea will be to work in 
rescaled variables, putting $R=a_2/a_1$, and following the evolution of \emph{this} quantity rather than each 
semi-major axis. 
We will then compare the analytic results with the equilibrium values obtained in numerical simulations.
To simplify the calculation we choose here units in which $m \sqrt{\GravC (M_*+m)}=1$.
We also assume small $e$ to simplify the formul\ae\, but the method is indeed general.

The idea is to start with two fundamental equations. 
The first states that the derivative of the angular momentum of the system is equal to the torque:
\begin{equation}\label{eq:dotL=Torque}
\frac{d\ANGMOM}{dt} = \dot{\ANGMOM} = 
T;
\end{equation}
the second states that the derivative of the energy of the system is equal to the work
\begin{equation}\label{eq:dotE=Work}
\frac{dE}{dt} = \dot{E} = 
W.
\end{equation}
The total torque exerted on the planetary system from the gas is, from equation \eqref{eq:dotAngMomMig}
\begin{equation}\label{eq:T=dotLmig2}
T = \dot\ANGMOM_{mig,2} = -\frac{\ANGMOM_2}{\tau_{mig,2}} = -\frac{\sqrt{a_2 (1-e_2^2)}}{\tau_{mig,2}}.
\end{equation}
Using \eqref{eq:adotovera}, the change in orbital energy is
\begin{equation}\label{eq:dotE_i}
\dot E_i = \frac{\dot a_i}{2 a_i^2} 
       = \frac{1}{a_i} \frac{\dot a_i}{2 a_i} 
       = \frac{1}{a_i} \left( -\frac{1}{\tau_{mig,i}} - \frac{p}{2} \frac{e_i^2}{\tau_{e,i}} \right)
\end{equation} 
where $p\simeq 2$ for small $e$ so that we find that the total work is 
\begin{equation}\label{eq:TotalWork}
W=\frac{1}{a_2} \left(-\frac{1}{\tau_{mig,2}}-\frac{e_2^2}{\tau_{e,2}}\right),
\end{equation}
We now pass in rescaled variables, and write $R=a_2/a_1$. 
Using the expression \eqref{eq:TotalWork} for the work, \eqref{eq:dotE_i} for $\dot E_i$ and multiplying both sides by $a_2$, 
equation \eqref{eq:dotE=Work} reads
\begin{equation}\label{eq:dotE=WorkRescaled}
R\frac{\dot{a_1}}{2a_1}+ \frac{\dot{a_2}}{2a_2} = -\frac{1}{\tau_{mig,2}}-\frac{e_2^2}{\tau_{e,2}};
\end{equation}
using now
\begin{equation}\label{eq:dota2overa2&dota2overa2.wrt.dotRoverR}
\frac{\dot{a_2}}{a_2}= \frac{\dot{R}}{R} + \frac{\dot{a_1}}{a_1}
\end{equation}
this becomes
\begin{equation}\label{eq:dota1Overa1}
\frac{\dot{a_1}}{a_1}=\left[-\frac{2}{\tau_{mig,2}}-\frac{2e_2^2}{\tau_{e,2}} -\frac{\dot{R}}{R}\right]/(R+1).
\end{equation}
Similarly, to rewrite equation \eqref{eq:dotL=Torque} we write
\begin{equation}
\ANGMOM = \sqrt{a_1}\left[\sqrt{1-e_1^2} + \sqrt{R}\sqrt{1-e_2^2}\right];
\end{equation}
then, ignoring in \eqref{eq:dotAngMom} the higher order terms in $e$ in and writing for each planet 
$\dot\ANGMOM_i \simeq \frac{\dot a_i}{2 \sqrt{a_i}}\sqrt{1-e_i^2} - \sqrt{a_i} e_i \dot e_i$, we use 
\eqref{eq:T=dotLmig2} to obtain
\begin{equation}\label{eq:StartOfCalculationToGetEqInR}
\dot\ANGMOM \simeq \frac{\dot a_1}{2 \sqrt{a_1}} \left[\sqrt{1-e_1^2} + \sqrt{R}\sqrt{1-e_2^2}\right] + 
                   \sqrt{a_1} \left[ -e_1 \dot e_1 + \frac{\dot R}{2 \sqrt R} \sqrt{1-e_2^2} - \sqrt{R} e_2 \dot e_2 \right] 
                   = -\frac{\sqrt{a_2 (1-e_2^2)}}{\tau_{mig,2}}. 
\end{equation}
Dividing this equation by $\sqrt{a_1}$ and using \eqref{eq:dota1Overa1} we get
\begin{equation}
\left[-\frac{1}{\tau_{mig,2}}-\frac{e_2^2}{\tau_{e,2}} -\frac{\dot{R}}{2 R}\right] \left[\sqrt{1-e_1^2} + \sqrt{R}\sqrt{1-e_2^2}\right]/(R+1) + 
   \left[ -e_1 \dot e_1 + \frac{\dot R}{2 \sqrt R} \sqrt{1-e_2^2} - \sqrt{R} e_2 \dot e_2 \right] = -\frac{\sqrt{R}\sqrt{1-e_2^2}}{\tau_{mig,2}}.
\end{equation}
We now write $e_i=e_i(R)$ as given by the equilibrium curves shown in Figure \ref{fig:EqCurvesForAllResonances-DifferentMasses}, so that we can write an equation with $R$ as the sole independent variable. 
Using then $\dot e_i = \frac{\D e_i}{\D R}\dot R$ and grouping the terms in $\dot R$ we get
\begin{equation}\label{eq:LawForDotR}
\begin{split}
\left[-\left(\sqrt{1-e_1^2} + \sqrt{R}\sqrt{1-e_2^2}\right)/(2 R (R+1)) - e_1 \frac{\D e_1}{\D R} + \frac{\sqrt{1-e_2^2}}{2\sqrt{R}} -\sqrt{R} e_2 \frac{\D e_2}{\D R} \right] \dot R \\
  = \left[ \frac{1}{\tau_{mig,2}}+\frac{e_2^2}{\tau_{e,2}}\right] \left[ \sqrt{1-e_1^2} + \sqrt{R}\sqrt{1-e_2^2} \right]/(R+1) - \frac{\sqrt{R}\sqrt{1-e_2^2}}{\tau_{mig,2}}; 
\end{split}
\end{equation}
approximating each $1-e^2 \simeq 1$ for small $e$'s one can simplify this equation into 
\begin{equation}\label{eq:LawForDotR-Simplified}
\left[\frac{1}{2\sqrt{R}} - e_1 \frac{d e_1}{dR} - \sqrt{R} e_2 \frac{d e_2}{dR} - \frac{1+\sqrt{R}}{2R(1+R)}\right]\dot{R} = 
\frac{(1-R^{3/2})}{\tau_{mig,2} (1+R)} + \frac{e_2^2 (1+\sqrt{R})}{\tau_{e,2} (1+R)}.
\end{equation}
Such an equation gives the derivative of $R$ as a function of $R$, whereas equation \eqref{eq:dota1Overa1} 
gives the evolution of $a_1$ as a function of $R$. 
The evolution of $e_1$ and $e_2$ is obtained from that of $R$ using the functions $e_1(R)$ and $e_2(R)$.
Together these relationships describe the full evolution of the resonant system as it evolves under the torque and the damping 
caused by the disk. 
If there is no damping (${\tau_{e,2}}=\infty$) then no equilibrium is possible
and $R$ continues to decrease, the right hand side being negative, and the eccentricities keep following the curves 
in Figure \ref{fig:EqCurvesForAllResonances-DifferentMasses}. 
If instead ${\tau_{e,2} \neq\infty}$, the equilibrium point occurs when $\dot{R}=0$, that is, putting the right hand side of e.g.\
the simplified equation \eqref{eq:LawForDotR-Simplified} equal to 0, when
\begin{equation}\label{eq:e2EquilibriumWithoutTrap}
e_2^2 = \frac{(R^{3/2}-1)}{(1+R^{1/2})} \frac{\tau_{e,2}}{\tau_{mig,2}} = \frac{(R^{3/2}-1)}{2 (1+R^{1/2})} K_2^{-1},
\end{equation}
where $K_2 = \frac{\tau_{a,2}}{\tau_{e,2}}$ is the $K$-factor of the outer planet. 
The multiplicative factor multiplying $K_2^{-1}$ can be further approximated by taking $R=\bar R = (k/(k-1))^{2/3}$.

In the case of a trap at the disk edge operating on the inner planet to stop the migration process, 
the requirement is that the torque on the inner planet adapts so that the total torque on the system is 0, 
whatever may be the additional effect of the disk on the inner planet. 
In this case, the first fundamental equation \eqref{eq:dotL=Torque} rewrites
\begin{equation}\label{eq:dotL=Torque=0-WithTrap}
\frac{d\ANGMOM}{dt} = \dot{\ANGMOM} = 0 \quad \text{i.e. } \dot\ANGMOM_1= - \dot\ANGMOM_2.
\end{equation}
This implies that the disk exerts a positive torque on the inner planet 
\begin{equation}\label{eq:dotL1WithTrap}
\dot\ANGMOM_1=+\frac{\dot\ANGMOM_1}{\tau_{mig,1}}
\end{equation}
with
$1/\tau_{mig,1}\simeq\sqrt{R}/\tau_{mig,2}$ (still approximating $1-e^2\sim1$).
The total work on the system is instead not 0. 
Using the torque just computed for the inner planet and \eqref{eq:adotovera}, it is easy to see that the work exerted by the disk on the inner planet is 
\begin{equation}
W_1= \frac{1}{a_2}\left[+\frac{R^{3/2}}{\tau_{mig,2}} - \frac{R e_1^2}{\tau_{e,1}}\right], 
\end{equation}
where we also consider the eccentricity damping on the inner planet 
(on a timescale not necessarily equal to that of the second planet).\footnote{
We stressed the plus sign in the first term in the right hand side of $W_1$, in contrast with the negative sign of the corresponding term in $W_2$, since the effect of the trap on the inner planet is that of outwards migration.}
This allows to rewrite equation \eqref{eq:dotE=WorkRescaled} as
\begin{equation}
R\frac{\dot{a_1}}{2 a_1^2} + \frac{\dot{a_2}}{2 a_2^2} = \left[\frac{R^{3/2}}{\tau_{mig,2}}-R \frac{e_1^2}{\tau_{e,1}}\right]+
\left[\frac{-1}{\tau_{mig,2}}- \frac{e_2^2}{\tau_{e,2}}\right],
\end{equation}
and, using \eqref{eq:dota2overa2&dota2overa2.wrt.dotRoverR}, the equivalent of \eqref{eq:dota1Overa1} becomes:
\begin{equation}\label{eq:dota1Overa1WithTrap}
\frac{\dot{a_1}}{a_1}=\left[-\frac{\dot R}{R}+\frac{2(R^{3/2}-1)}{\tau_{mig,2}} -\frac{2 R e_1^2}{\tau_{e,1}} - \frac{2 e_2^2}{\tau_{e,2}}\right]/(R+1).
\end{equation}
Then, redoing all the calculations as above from \eqref{eq:StartOfCalculationToGetEqInR} to \eqref{eq:LawForDotR-Simplified}, but putting equal to 0 the right hand side of \eqref{eq:StartOfCalculationToGetEqInR} (zero total torque) and using \eqref{eq:dota1Overa1WithTrap} instead of \eqref{eq:dota1Overa1}, the equivalent of equation \eqref{eq:e2EquilibriumWithoutTrap} becomes
\begin{equation}\label{eq:e2EquilibriumWithTrap}
\frac{(R^{3/2}-1)}{\tau_{mig,2}}-\frac{R e_1^2}{\tau_{e,1}} - \frac{e_2^2}{\tau_{e,2}} = 0.
\end{equation}
Notice that if this equation is satisfied, $\dot{a_1}$ in \eqref{eq:dota1Overa1WithTrap} vanishes when $\dot R=0$, i.e.\ the system is at a complete equilibrium, unlike in the previous case where both planets were migrating in resonance, at constant $R$. Indeed, the equilibrium equation could also have been found by imposing directly $\dot{a_1}=0$ and $\dot R=0$ in \eqref{eq:dota1Overa1WithTrap}.
Considering another limiting case as an example, if no damping is applied to planet 1, $\tau_{e,1} = \infty$, 
then the equilibrium in $e_2$ is 
\begin{equation}
e_2^2=(R^{3/2}-1) \frac{\tau_{e,2}}{\tau_{mig,2}} = \frac{(R^{3/2}-1)}{2} K_2^{-1};
\end{equation}
e.g.\ for the 3-2 resonance the multiplicative coefficient, estimated again using $R=\bar R$, is about twice of the one in
\eqref{eq:e2EquilibriumWithoutTrap}, meaning that the higher relative push between the two planets against one another, provided by the trap, has the effect of increasing the equilibrium eccentricity.

Analytical formul\ae\ to calculate the equilibrium eccentricity during the capture in resonance 
and valid in the low-eccentricity regime have already been produced. 
E.g.\ \cite{2017A&A...602A.101R} reproduce a formula which they derive from \cite{2005MNRAS.363..153P}: 
taking these formul\ae\ in the limiting case of $\tau_{e,1}=\infty$ and $\tau_{a,1}=\infty$, 
one obtains our formula \eqref{eq:e2EquilibriumWithoutTrap}. 
Another point of view was adopted in \cite{2008A&A...483..325C}, where the authors obtained the damping time 
$\tau_{e,1}$ needed to reach a given value of eccentricities at the equilibrium configuration. 
Their final formula (16) indeed leads to our formula \eqref{eq:e2EquilibriumWithTrap} by 
using equation \eqref{eq:dotL1WithTrap}
and by 
again replacing $1-e^2$ with 1, their $\eta$ by $\sqrt{R}$ and their $\epsilon$ by $1/R$ 
(note that their $-1/\tau_a$ is defined as $\dot a/a$, while in the present work the latter is expressed 
by $-1/\tau_a - 2e^2/\tau_e$).
\cite{2014AJ....147...32G} derived a formula for the equilibrium eccentricity in the simplified case of the 
planar, circular, restricted three-body problem with a massless inner planet and using equations to first order in $e_1$. 
They found that 
\begin{equation}\label{eq:GoldreichAndSchlichtingFormula}
e_{1,eq} = \sqrt{\frac{\tau_{e,1}}{k ~ \tau_{mig,eff}}},
\end{equation}
where $\tau_{mig,eff}^{-1} = \tau_{mig,2}^{-1} - \tau_{mig,1}^{-1}$.

We now look at numerical simulations to confirm these analytical predictions.
For formula \eqref{eq:e2EquilibriumWithoutTrap}, we consider the case of $\beta = 0$, $\beta$ being the flaring index of the disk.
This is because, even when the equilibrium described by \eqref{eq:e2EquilibriumWithoutTrap} is reached, $\dot R = 0$ but the two 
planets keep migrating due to the torque on the outer one;
now since $h(r) = z_{scale} r^{\beta}$ and the $K$-factor depends on $h$ via \eqref{eq:KFactor}, 
it is convenient to keep $h$ a constant so that the equilibrium eccentricity attained by the system does not evolve as $r=a_2$ does. 
In this case, $K_2 \simeq 82.11$.
We estimate with \eqref{eq:e2EquilibriumWithoutTrap} the equilibrium eccentricity $e_{eq,2} \simeq 0.0311$ for the 4-3 resonance, 
$e_{eq,2} \simeq 0.0377$ for the 3-2 resonance, and $e_{eq,2} \simeq 0.0519$ for the 2-1 resonance. 
We show in Figure \ref{fig:TestingFormulae2EquilibriumWithoutTrap} the result of numerical simulations with the 
described setup, showing good agreement with the predicted values. 
We note that the equilibrium found is always stable, because 
$\tau_{e,1}=\infty$ (\cite{2002ApJ...567..596L}; \cite{2015ApJ...810..119D}).
\begin{figure}[!ht]
\centering
\begin{subfigure}[b]{0.32 \textwidth}
\centering
\includegraphics[scale=0.43
]{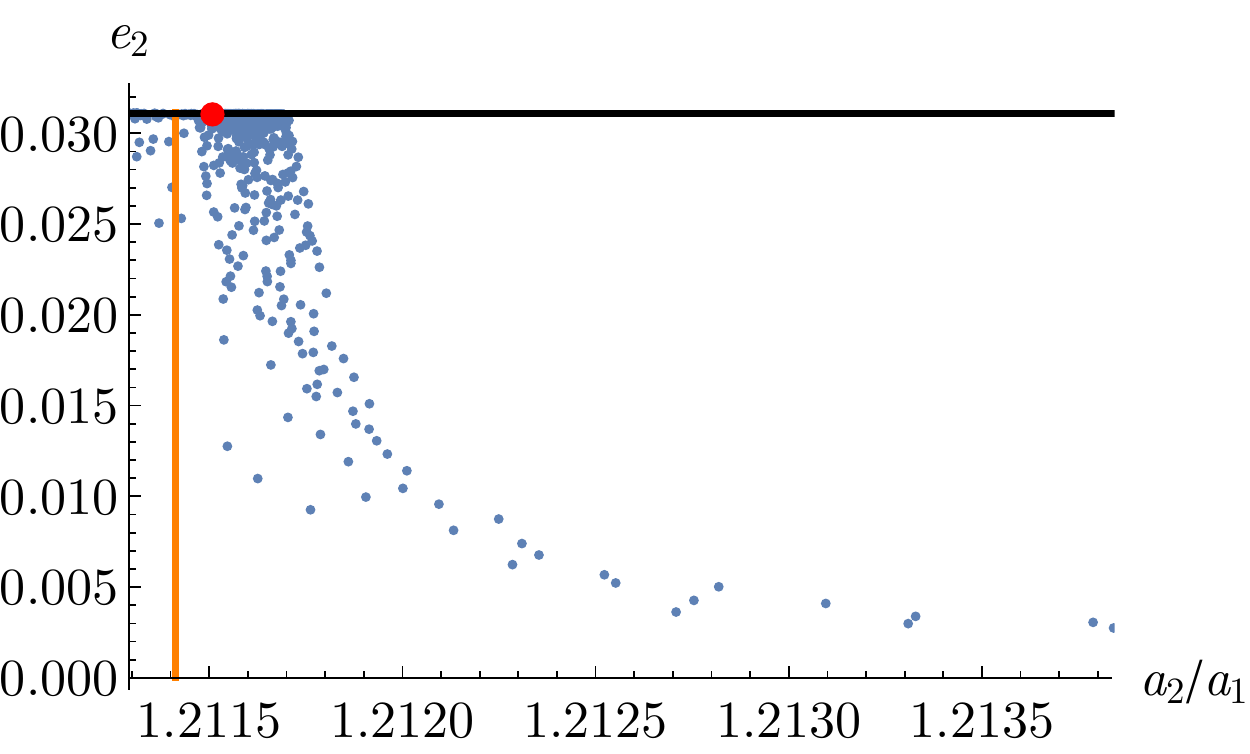}
\caption{4-3 mean motion resonance.}
\label{fig:e2EquilibriumWithoutTrap.subfig:4-3}
\end{subfigure}
\begin{subfigure}[b]{0.32 \textwidth}
\centering
\includegraphics[scale=0.43
]{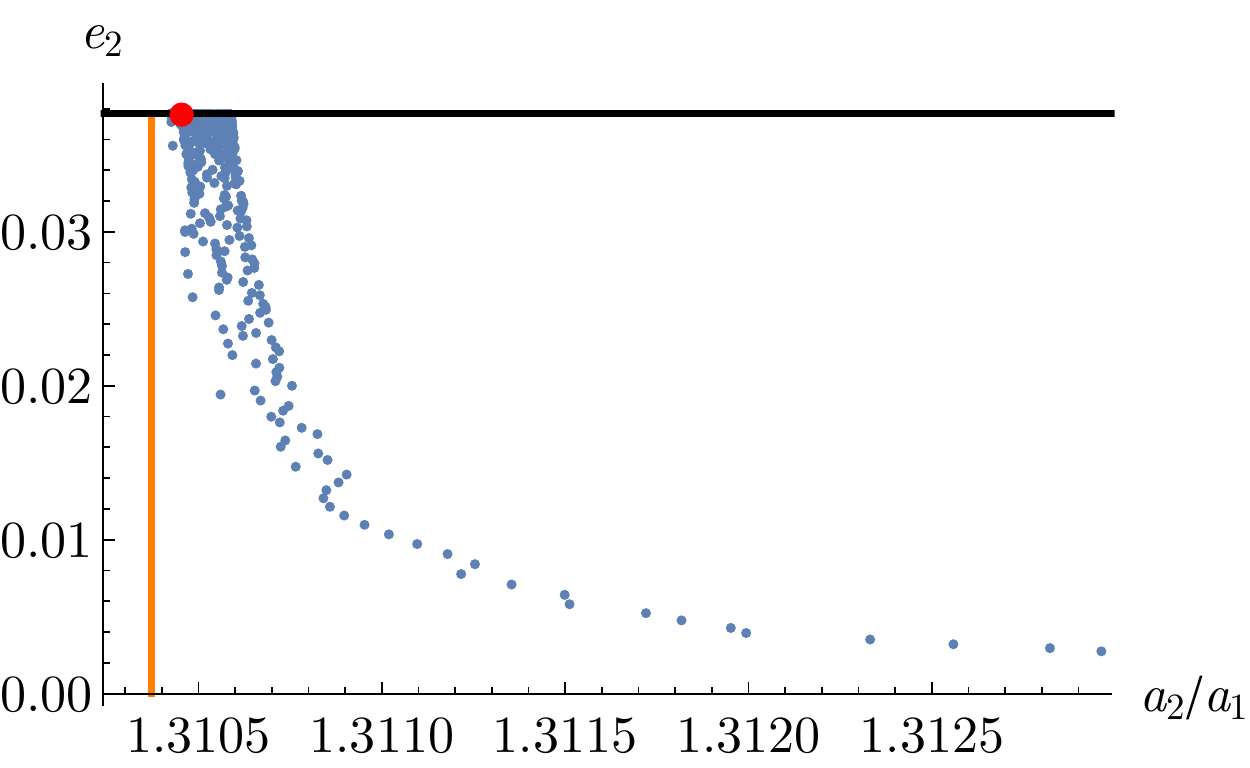}
\caption{3-2 mean motion resonance.}
\label{fig:e2EquilibriumWithoutTrap.subfig:3-2}
\end{subfigure}
\centering
\begin{subfigure}[b]{0.32 \textwidth}
\centering
\includegraphics[scale=0.43
]{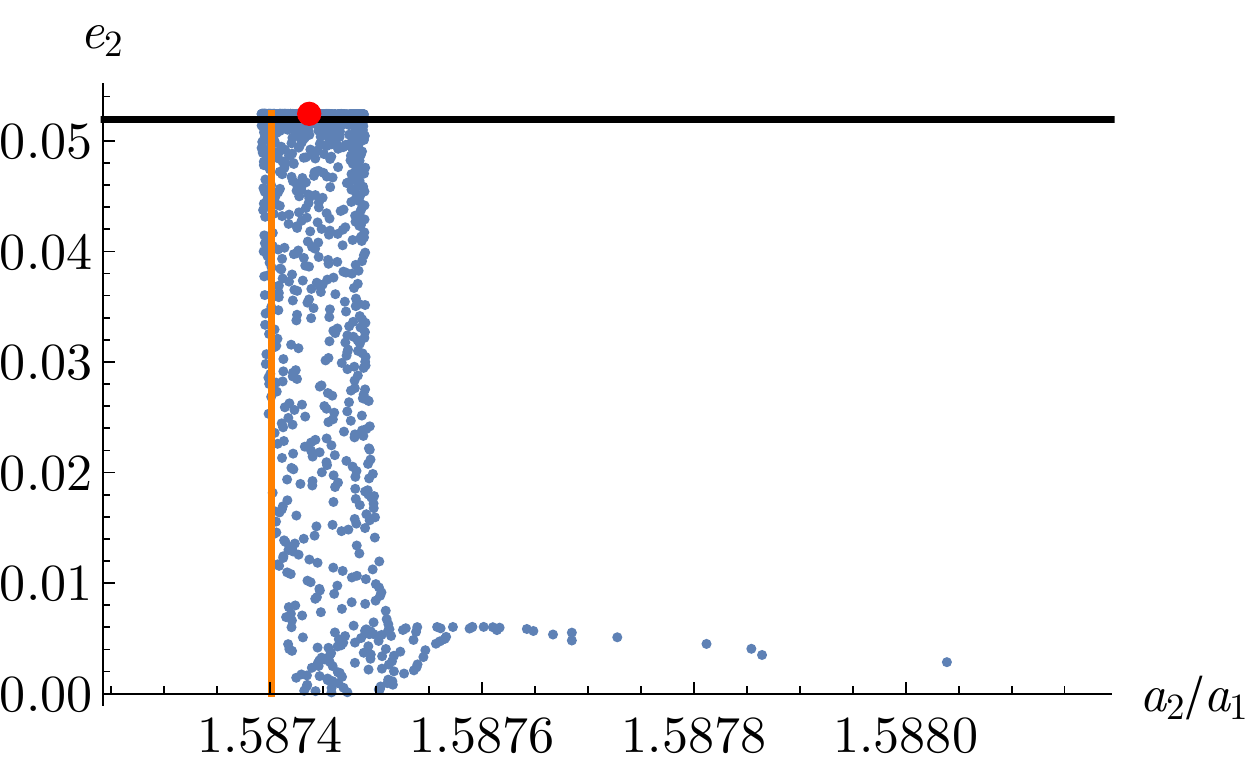}
\caption{2-1 mean motion resonance.}
\label{fig:e2EquilibriumWithoutTrap.subfig:2-1}
\end{subfigure}
\caption{
Confirmation of formula \eqref{eq:e2EquilibriumWithoutTrap} for the equilibrium value of $e_2$ where 
$K_2 = \frac{\tau_{a,2}}{\tau_{e,2}} \simeq 82.11 \equiv const$ (in the case of a disk with flaring index $\beta = 0$), 
in the case of various first order mean motion resonances. 
We show as a function of the semi-major axes ratio $a_2/a_1 = R$ the evolution of $e_2$ in the numerical simulations under the conditions explained in the text. 
The red dot indicates the configuration of the system after the equilibrium is attained.
We indicate with an horizontal line the predicted value for $e_{2,eq}$, showing good agreement.
Note that in the 2-1 mean motion resonance case, we see the same behaviour as shown in Figure \ref{fig:EqCurvesForAllResonances-WithExpansions.subfig:2-1}, 
associated to the smooth change of $\DELTAGAMMA_{eq}$ from $\pi$ to $0$ as described in 
Section \ref{subsec:EquilibriumPointsOfAveragedHa}; 
the goodness of the prediction of $e_{2,eq}$ is unaffected.}
\label{fig:TestingFormulae2EquilibriumWithoutTrap}
\end{figure}

For formula \eqref{eq:e2EquilibriumWithTrap}, we can again consider a flared disk, $\beta = 0.25$. 
To solve that equation, we first need to write $e_1 = e_1(e_2)$ from the equilibrium curves in Figure 
\ref{fig:EqCurvesForAllResonances-DifferentMasses} for the different resonances, 
and then to calculate from \eqref{eq:tau_e} and \eqref{eq:tau_mig} the values for $\tau_{e,1}$, $\tau_{a,2}$ and 
$\tau_{e,2}$. 
Note that we don't need to calculate the value of each $\tau_{wave,1}$ and $\tau_{wave,2}$ 
at the positions $a_1$, $a_2$ of the planets, since we can just factor out one of the semi-major axes 
and easily reduce this factor to a quantity depending only on $R$, 
which we again approximate with $\bar R$.
However since the disk is flared, to obtain $\tau_{mig,2}$ we need to calculate the value of 
$h = h(a_2)$ at the position of the outer planet, and we again write $a_2$ as a function of $e_2$.
We thereby estimate with \eqref{eq:e2EquilibriumWithTrap} a value $e_{2,eq} \simeq 0.0114$ for the 
4-3 resonance, $e_{2,eq} \simeq 0.0134$ for the 3-2 resonance and $e_{2,eq} \simeq 0.0040$ for the 2-1 resonance.  
We show in Figure \ref{fig:TestingFormulae2EquilibriumWithTrap} the result of numerical simulations with this 
setup, showing again good agreement with the analytical predictions. 
To use formula \eqref{eq:GoldreichAndSchlichtingFormula} from \cite{2014AJ....147...32G}, 
we put $|\tau_{mig,1}| = \tau_{mig,2}/\sqrt{R}$ (cfr.\ equation \eqref{eq:dotL1WithTrap}). 
We obtain in the cases discussed above  $e_{1,eq} \simeq 0.019$ for the 
4-3 resonance, $e_{1,eq} \simeq 0.022$ for the 3-2 resonance and $e_{1,eq} \simeq 0.024$ for the 2-1 resonance, 
the real values obtained from the numerical simulations being respectively 
$e_{1,eq} \simeq 0.011$, $e_{1,eq} \simeq 0.012$ and $e_{1,eq} \simeq 0.018$.
This shows that using such an approximated formula one obtains the right order of magnitude but the accuracy may be 
off by a factor of 2. 

We note that for the 2-1 resonance (Figure \ref{fig:e2EquilibriumWithTrap.subfig:2-1}), the case with
$m_1=m_2=10^{-5}M_*$ and $K\sim 100$ should lead to an instability of the equilibrium point.
(see Fig. 3 of \cite{2015ApJ...810..119D}). 
We have checked that this is indeed the case. 
However the growth of the libration amplitudes manifests itself on a timescale $\tau_e$ 
(see Eq. (29) in \cite{2014AJ....147...32G}), 
which is very long given the low surface density of the disk that we assume to ensure a slow evolution. 
We stop the simulation before that the instability produces any noticeable effect. 
This is appropriate for the purposes of our study, which is to place planets deep in resonance 
to study their stability as a function of planetary mass in absence of dissipation (see Section \ref{sec:MassIncrease}).

\begin{figure}[!ht]
\centering
\begin{subfigure}[b]{0.32 \textwidth}
\centering
\includegraphics[scale=0.43
]{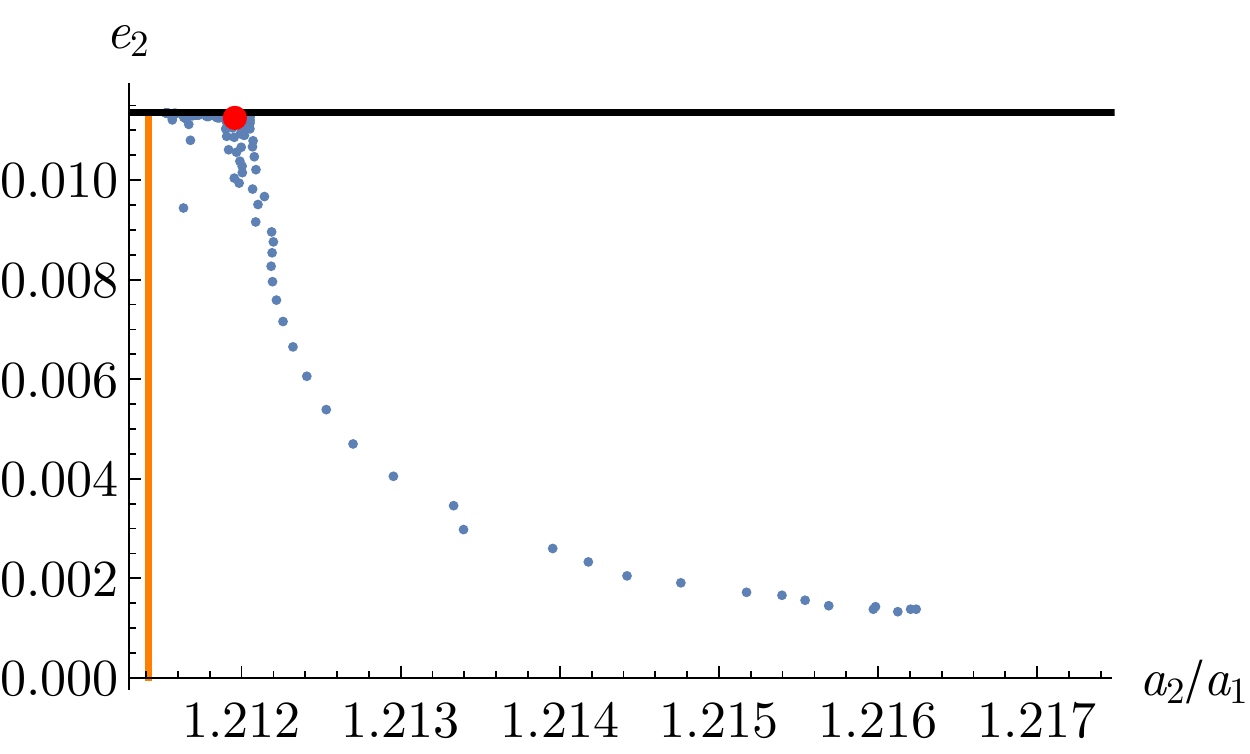}
\caption{4-3 mean motion resonance.}
\label{fig:e2EquilibriumWithTrap.subfig:4-3}
\end{subfigure}
\begin{subfigure}[b]{0.32 \textwidth}
\centering
\includegraphics[scale=0.43
]{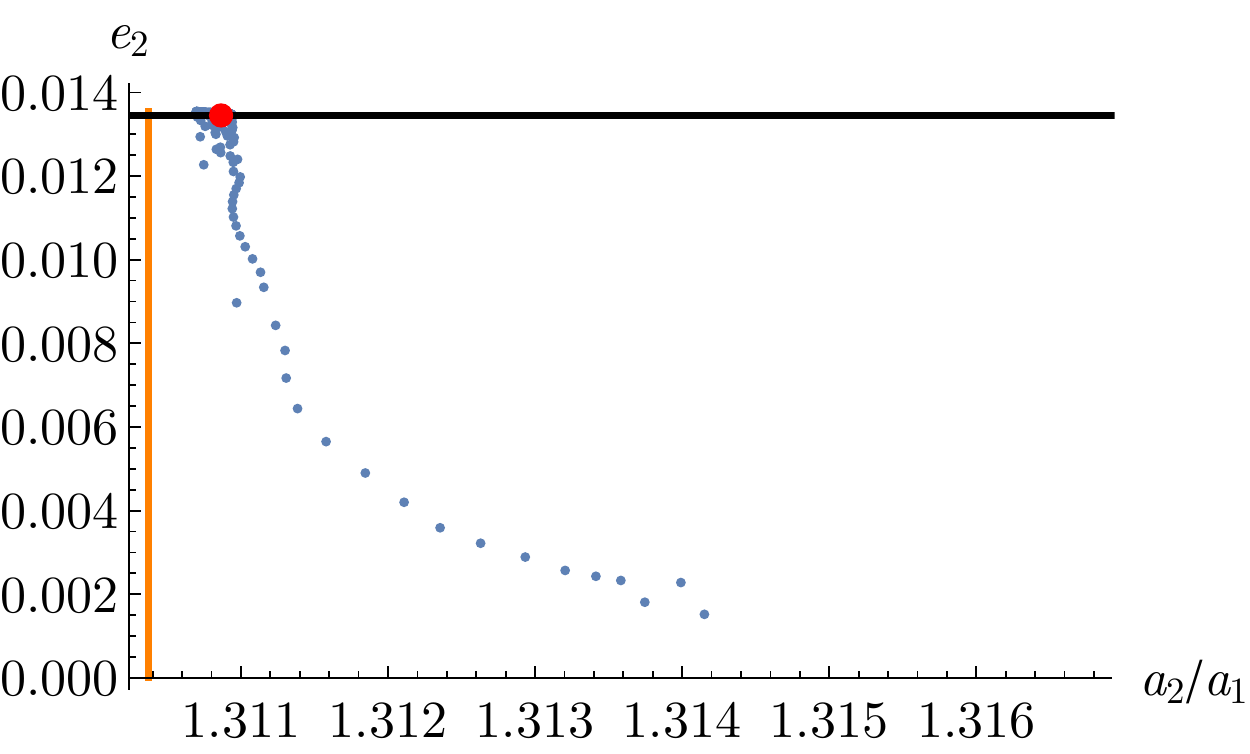}
\caption{3-2 mean motion resonance.}
\label{fig:e2EquilibriumWithTrap.subfig:3-2}
\end{subfigure}
\centering
\begin{subfigure}[b]{0.32 \textwidth}
\centering
\includegraphics[scale=0.43
]{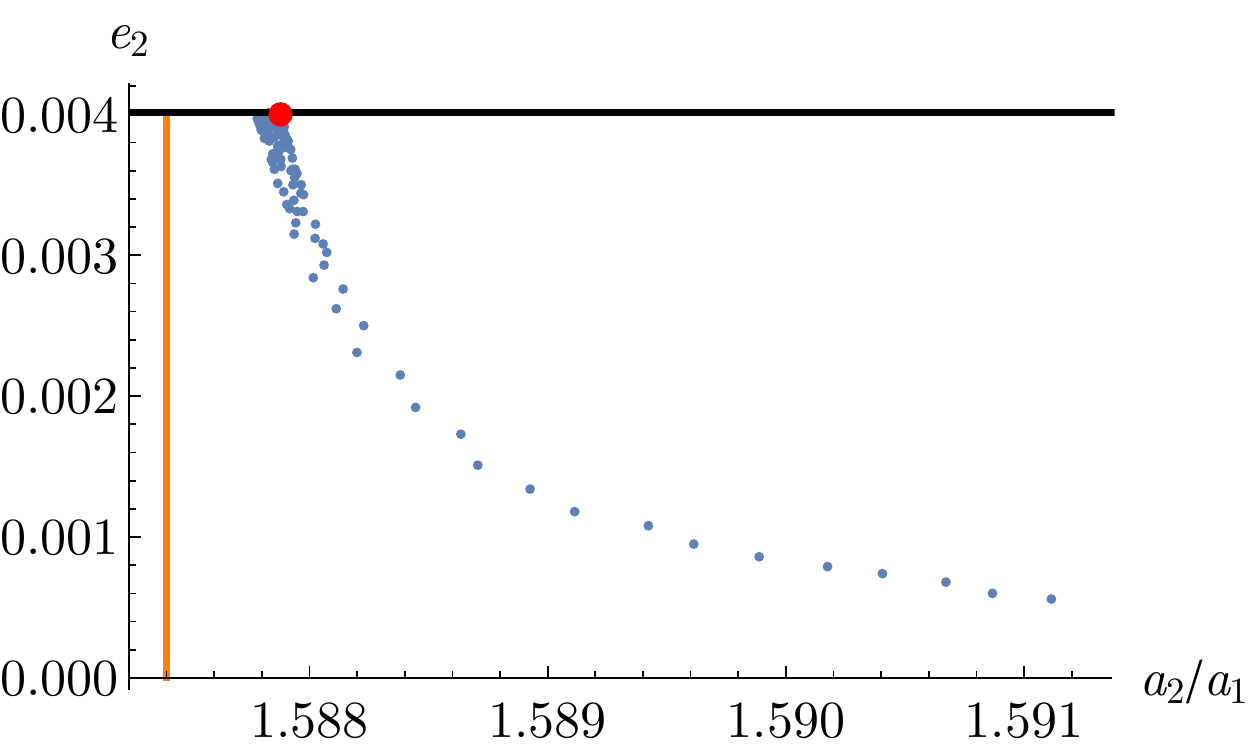}
\caption{2-1 mean motion resonance.}
\label{fig:e2EquilibriumWithTrap.subfig:2-1}
\end{subfigure}
\caption{
Confirmation of formula \eqref{eq:e2EquilibriumWithTrap} for the equilibrium value of $e_2$ in the case of 
a disk with flaring index $\beta = 0.25$, 
for various first order mean motion resonances. 
We show as a function of the semi-major axes ratio $a_2/a_1 = R$ the evolution of $e_2$ in the numerical simulations under the conditions explained in the text. 
The red dot indicates the configuration of the system after the equilibrium is attained.
We indicate with an horizontal line the predicted value for $e_{2,eq}$, showing good agreement.
}
\label{fig:TestingFormulae2EquilibriumWithTrap}
\end{figure}

We conclude this Appendix by noticing that in both equations \eqref{eq:e2EquilibriumWithoutTrap} and
\eqref{eq:e2EquilibriumWithTrap} the coefficient in $\tau_{wave}$ which depends on the planetary mass $m$ 
and of the gas surface density $\Sigma$ is eliminable 
(formula \eqref{eq:e2EquilibriumWithTrap} in principle depends on the planet mass-ratio, here fixed to 1), 
meaning that the final configuration does not depend on these
quantities. This is confirmed by our simulations, as shown in Figure 
\ref{fig:TestingFormulae2EquilibriumWithTrap-DifferentMasses}.
The fact that the final configuration does not depend on the disk surface density means also that we can, 
for the purposes of our study here, let $\Sigma$ be small so to ensure a slow enough change in angular momentum 
and invoke the adiabatic approach mentioned at the beginning of Section \ref{subsec:EquilibriumPointsOfAveragedHa}, 
without affecting the final resonant configuration reached by the system.

\begin{figure}[!ht]
\centering
\includegraphics[scale=0.43
]{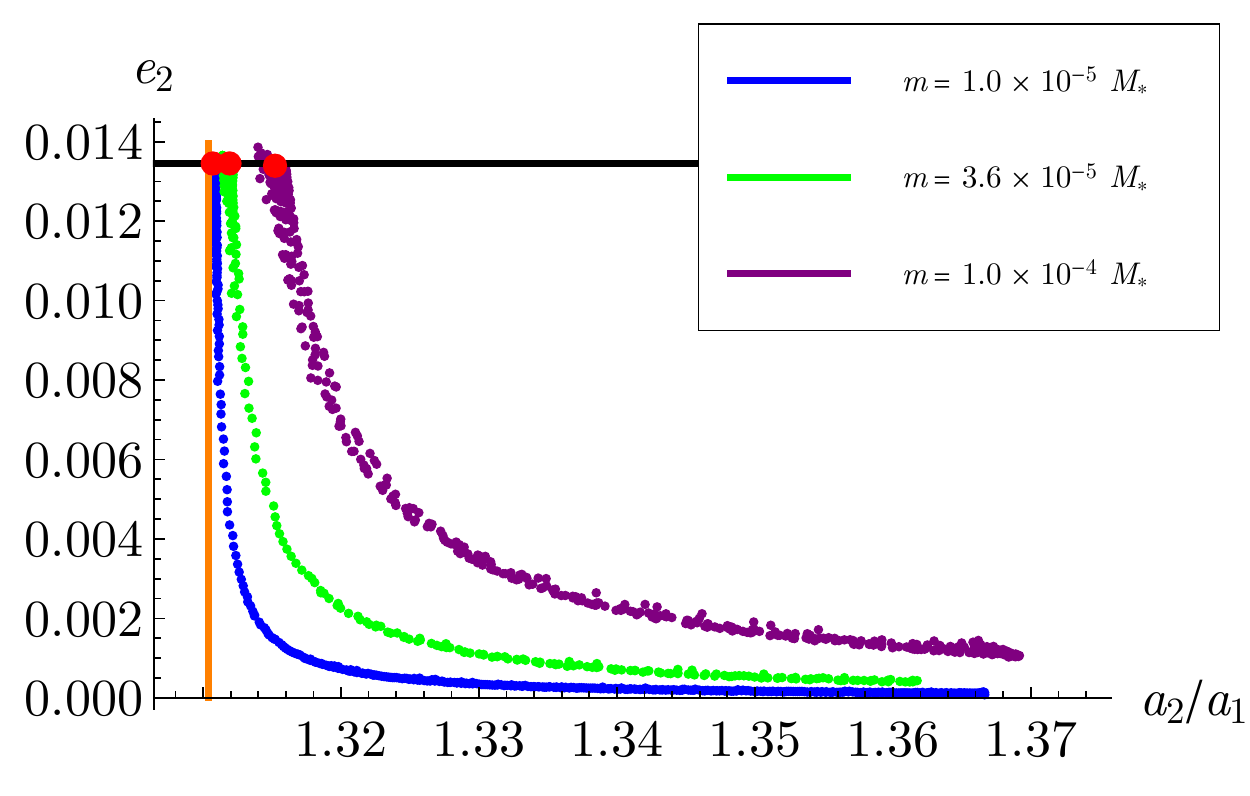}
\caption{Evolution of $e_2$ during the capture in the 3-2 resonance for 
different planetary masses $m_1 = m_2 = m$.
The vertical black line indicates the calculated value of $e_{2,eq}$ obtained with \eqref{eq:e2EquilibriumWithoutTrap}.
The red dots represent the final configurations of the systems after the equilibria are reached, 
showing that the resulting equilibrium value of $e_2$ is independent 
of $m$, as predicted by the analytical formul\ae.}
\label{fig:TestingFormulae2EquilibriumWithTrap-DifferentMasses}
\end{figure}
\end{appendices}

\end{document}